\documentclass[aps, floatfix, twocolumn, 10pt, prx]{revtex4-2}
\usepackage[T1]{fontenc}
\usepackage[utf8]{inputenc}
\usepackage{amsmath, amssymb, mathrsfs, graphicx, bm, enumitem}
\usepackage{soul}
\usepackage[svgnames,dvipsnames]{xcolor}
\usepackage[colorlinks=true, citecolor=NavyBlue, urlcolor=NavyBlue, linkcolor=NavyBlue]{hyperref}

\begin{document}
\title{Forces at the scale of the cell}

\author{K. Vijay Kumar}
\email{vijaykumar@icts.res.in}
\affiliation{International Centre for Theoretical Sciences, Tata Institute of Fundamental Research, Bengaluru North 560089, India.}

\author{Mandar M. Inamdar}
\email{minamdar@iitb.ac.in}
\affiliation{Department of Civil Engineering, Indian Institute of Technology, Bombay, Mumbai 400076, India.}

\author{Pramod A. Pullarkat}
\email{pramod@rri.res.in}
\affiliation{Raman Research Institute, Bengaluru 560080, India}

\author{Gautam I. Menon}
\email{gautam.menon@ashoka.edu.in}
\affiliation{Department of Biology, Trivedi School of Biosciences and Department of Physics, Ashoka University, Sonepat 131029 India}

\date{\today}

\begin{abstract}
The importance of molecular-scale forces in sculpting biological form and function has been acknowledged for more than a century. Accounting for forces in biology is a problem that lies at the intersection of soft condensed matter physics, statistical mechanics, and experimental methods, all adapted to a cellular context. This review surveys how forces arise within the cell. We provide a summary of the relevant background in cell biophysics, of soft-matter systems in and out of thermodynamic equilibrium, and of various force measurement methods. These ideas are then deployed to describe how forces are generated, transmitted, and sensed in specific cell-scale processes. We emphasize active matter descriptions, methodological tools that provide ways of incorporating non-equilibrium effects in a systematic manner into conceptual as well as quantitative descriptions. Our examples include polymerization forces, the motion of molecular motors, the properties of the actomyosin cortex, the mechanics of cell division, and shape changes in tissues. We suggest that a deeper understanding of cell function will necessarily require integrating the action of cell-scale physical forces with the assimilation and processing of information.
\end{abstract}

\maketitle

\tableofcontents

\section{Introduction}

The cell is the basic structural and functional unit of life. Cells are self-organizing structures that continually build, remodel, and dismantle their own internal architecture. This dynamic organization allows them to carry out the processes required for life~\cite{nurse2003great, alberts2014molecular}. 

Sustaining such organization requires forces that act at the scale of the cell~\cite{phillips2012physical}.  At the molecular level these forces arise from configurational changes in biological macromolecules~\cite{HowardBook}, induced by the binding, unbinding, and translocation of molecules at specific sites, or by variations in the local environment such as pH and electrostatics~\cite{dill2010molecular, parthasarathyResourceLetterBP12021}. But sustained force generation that performs net work is impossible in thermodynamic equilibrium. An equilibrium state obeys detailed balance and supports no net currents~\cite{de2013non}. While forces may be present, such as the entropic tension of a stretched polymer or an osmotic pressure, they cannot drive directed motion nor extract work around a cycle.

Directed, sustained force generation therefore requires a persistent energy flow that breaks detailed balance, so that the forces originate in energy-consuming processes. What distinguishes biological systems from other systems with a continuous energy throughput is that force generation in biology typically involves the hydrolysis of small energy-storing molecules, above all adenosine triphosphate (ATP)~\cite{alberts2014molecular}, the energy currency of the cell. This localized, nanoscale energy input is called \emph{activity}~\cite{menon2010activematter}. The ATP concentration is maintained out of equilibrium through metabolism, the biochemical conversion of nutrients into energy~\cite{alberts2014molecular}. Living matter thus occupies one end of a continuum of systems with energy throughput, employing a nearly universal currency in ATP.

Whereas the ability to \emph{sense} force is evolutionarily ancient and near-universal~\cite{kung2005unifying, martinac2003evolutionary}, the machinery that actively \emph{generates} and directs force is more elaborate, and its growing sophistication tracks the growth of cellular complexity. Even the simplest cells, such as mycoplasmas, need force-generating machinery to divide and to replicate their DNA. Swimming bacteria such as \textit{E.~coli} locomote by spinning filamentous appendages driven by rotary, torque-generating protein assemblies.

A qualitative jump comes with eukaryotic cells, which (unlike bacteria) enclose their DNA in a nucleus and partition their interior into membrane-bound organelles that must be actively and dynamically maintained. This maintenance calls for \emph{molecular motors}, specialized protein complexes that convert the chemical energy of ATP into directed motion by stepping along the filamentous tracks of the \emph{cytoskeleton} (both introduced in Section~\ref{sec:basic_biology}), to carry cargo directionally. In addition,  many single-celled eukaryotes have evolved dedicated machinery for locomotion. A further leap accompanies multicellularity, which demands force generation coordinated across many cells to drive tissue assembly, morphogenesis, and whole-organism responses. The advent of processive, track-walking motors may thus mark a threshold in cellular complexity, a theme developed in Section~\ref{sec:molecular_motors}.

This review is premised on the idea that molecular-scale cellular forces, originating in non-equilibrium processes, are a key component of the organization of living systems \cite{DarcyThompson}. The idea is broadly reminiscent of descriptions of ``dissipative structures'' in non-equilibrium thermodynamics~\cite{kondepudi2014modern}. The difference is that it is \emph{active forces} that maintain such structures and underlie the functioning of living systems.
 
 At the purely biophysical level, we can think of a living cell as a complex soft-matter system imbued with activity~\cite{needleman2017active}. Theoretical and experimental models of soft matter systems that incorporate such activity can be shown to share many mechanical features of living systems. The contemporary field of \emph{active matter} emphasizes these connections, although with many applications beyond biology~\cite{marchettiHydrodynamicsSoftActive2013}.

Active matter ideas were initially developed to describe the emergent behaviour of particles capable of autonomous motion~\cite{vicsek1995novel, toner2005hydrodynamics}. An example is flocking, whereby groups of birds, fish, and other organisms locally coordinate their motion, overcoming noise, to establish robust collective patterns at scales much larger than the individual. A convenient description is in terms of self-propelled objects, particles that convert energy from an internal depot into the work needed to sustain motion in a viscous fluid~\cite{menon2010activematter}. Connections to \emph{force-free} active systems, such as swimming bacteria that experience no net external force and move by pushing on the surrounding fluid, were made early on, and the first major conference devoted to driven soft matter and biology brought many of these connections to the fore~\cite{ictp2006}. While much initial work focused on \emph{orientable} active systems, built from aligned rod-like units described by a polar or nematic order parameter (as in a flock or a filament bundle), it soon became clear that this physics was mirrored in the cellular cytoskeleton, cell-surface organization, and bacterial motility. These phenomena could also be reproduced outside the cell, for example in beads that moved spontaneously when coated with polymerizing biological components~\cite{Upadhyaya2003}.

Our focus is on forces that operate at the scale of a single cell. Such forces are crucial to the generation and maintenance of form, broadly the processes of development, homeostasis, and physiological function in living systems. However, we also provide a few examples of individual force-generating units at the subcellular scale, as well as of the behaviour of aggregates of cells. We emphasize the relationships between key ideas from the physics of non-equilibrium systems and descriptions of diverse biological systems at the cellular level. 

Biological form and function, driven by activity, are inseparable from the processing and storage of information. 
This motivates treating mechanics not as a mere downstream output of biochemical signalling but as a signalling channel in its own right: tensorial and long-ranged whereas chemistry is scalar and diffusive. This expanded view of \emph{mechanobiology}~\cite{lim2010mechanobiology} is conceptually central to the review and is developed in the concluding Section~\ref{sec:conclusions}. 

The understanding of forces at the scale of the cell requires some knowledge of what a cell actually is and what processes go on within it~\cite{alberts2014molecular}. We first supply a compact introduction to the relevant biology, aimed at a physics audience. We then summarize those aspects of soft condensed matter physics and non-equilibrium statistical mechanics that are especially relevant to understanding forces in cells. We outline some broad principles of mechanobiology that connect generalized thermodynamic forces to fluxes in biological contexts. We then show how such forces can be experimentally measured, demonstrating how coordinated forces in the cellular cytoskeleton underlie cell division and cell movement. We describe how simple models of tissues can be constructed, bridging the behaviour of individual cells with that of cell collectives. 

This review is aimed at (a) physicists from disciplines other than biological physics who wish to learn about how forces act at the scale of the cell, (b) biologists who wish to understand the motivation behind the theoretical descriptions, and (c) students potentially interested in entering this field. We assume familiarity with introductory physics and mathematics at the undergraduate level, but little more.

A review that attempts to tie all these diverse strands together must, of necessity, make choices as to what to retain and what to omit. The choices here reflect our view of what are likely to be general paradigms. We end by emphasizing the links between the many contexts in which forces are relevant to biological function at the cellular scale, reiterating the need for new tools that bridge function to a general physical description in terms of active matter. A glossary of technical terms that we use in this review is provided at the end.

\section{\label{sec:basic_biology}The cell: a biophysical primer}

Almost all living systems are built from fundamental units called cells. (Viruses, which straddle the living--non-living boundary and are not cellular, are dealt with briefly in Section~\ref{sec:other_examples}). Living organisms can be classified into two broad categories, eukaryotes (such as the cells in our bodies) and prokaryotes (bacteria). Eukaryotes are larger and have enclosed organelles (the nucleus, mitochondria, endoplasmic reticulum etc.), bound by a membrane. Prokaryotes are small, evolutionarily more ancient and lack complex intracellular structuring~\cite{alberts2014molecular, luria36lectures}. 

Cells vary in size and shape. Typical cells range in size from about $1~{\rm \mu m}$ (e.g., bacteria) to about $100~{\rm \mu m}$ (e.g., human egg cells), but the spread is far wider: neurons and skeletal muscle fibres can extend from centimetres up to a metre or more, and some single-celled organisms reach comparable sizes. Eukaryotes contain organelles ranging in size from $\sim 10~{\rm nm}$ to $\sim 1~{\rm \mu m}$. A good approximation for the density of cell constituents is the density of water $\approx 10^3~{\rm kg/m^3}$: a bacterium of about $1~{\rm \mu m}$ in size weighs approximately one pico-gram~\cite{milo2010bionumbers}. 

Life's organization involves the formation of extended structures from constituent molecules. For example, cellular organelles are typically protected by a thin continuous bilayer membrane of thickness $\approx 4-5~\mathrm{nm}$~\cite{milo2010bionumbers}. The development of a complex multicellular organism starting from a single fertilized egg involves shape and size changes at the level of aggregates of cells (tissues). This process is called morphogenesis.

Purely biochemical processes are fast. They occur over time-scales ranging from $10^{-12}~{\rm s}$ to $10^{-9}~{\rm s}$. However, processes at the scale of the cells that are relevant to sustained force generation are typically in the time-range of micro-seconds to seconds or longer~\cite{milo2010bionumbers}. When a mother cell divides into two daughter cells, hereditary information is transferred from the mother to the daughters.  The typical interval between cell division ranges from a few  minutes to several hours~\cite{milo2010bionumbers}. Some cells do not divide at all (e.g., neurons).

There is a fundamental distinction between the physics of assemblages of biological molecules that are studied outside the cell and their physics within a living cell. The former, referred to as \emph{in vitro} studies, typically involves a much smaller set of purified components, ignoring their underlying cellular context. In contrast, the latter, or \emph{in vivo}, studies these components as they present themselves in a living cell with its attendant complications of crowding, confinement, and regulation.

\subsection{Constituents of life}

About $20$ elements suffice to generate all of life's molecules~\cite{nelson2008lehninger}. The variety of organic molecules derives from the many ways complex molecules form through bonding. A convenient measure of energy scales involved with such bonds is $1~ k_BT \approx 4.1~{\rm pN~nm} \approx 1/40~{\rm eV}$, evaluated at $T \approx 300~{\rm K}$; the physiological temperature of $37~{\rm ^{\circ}C}$ corresponds to $\approx 310~{\rm K}$. Covalent bonds are strong, typically taking $10-100~{\rm k_BT}$ to break. The breaking of a triphosphate bond (e.g., hydrolysis of ATP) releases about $20~{\rm k_BT}$. Hydrogen bonds range in strength between $2-12~{\rm k_BT}$. 

There are four major types of significant biological molecules: proteins,  carbohydrates, lipids, and nucleic acids~\cite{lodish2021molecular}. Protein molecules are specified as a sequence of smaller amino acids, drawing from 20 amino acids known to occur naturally. Proteins can be short or long, incorporating between $\sim 10$ and $\sim 10^4$ amino acid units. Proteins are structurally diverse, ranging from collapsed globular structures to long linear ones~\cite{alberts2014molecular}.

Proteins can function individually, or as protein complexes~\cite{nelson2008lehninger}. Protein-protein interactions per amino acid involve energies intermediate between that of a covalent bond ($\approx 100 k_{\rm B}T$) and thermal energies ($\approx k_BT$).

Lipids are small amphiphilic molecules with a typical molecular length of about $2$~nm~\cite{milo2010bionumbers}. They are made of a hydrophilic head group and one or two hydrophobic chains. Lipid molecules are the constituents of cell and nuclear membranes. The self-assembly of lipid molecules into membrane-like structures can occur even in thermodynamic  equilibrium.

\begin{figure}
\centering
\includegraphics[width=\linewidth]{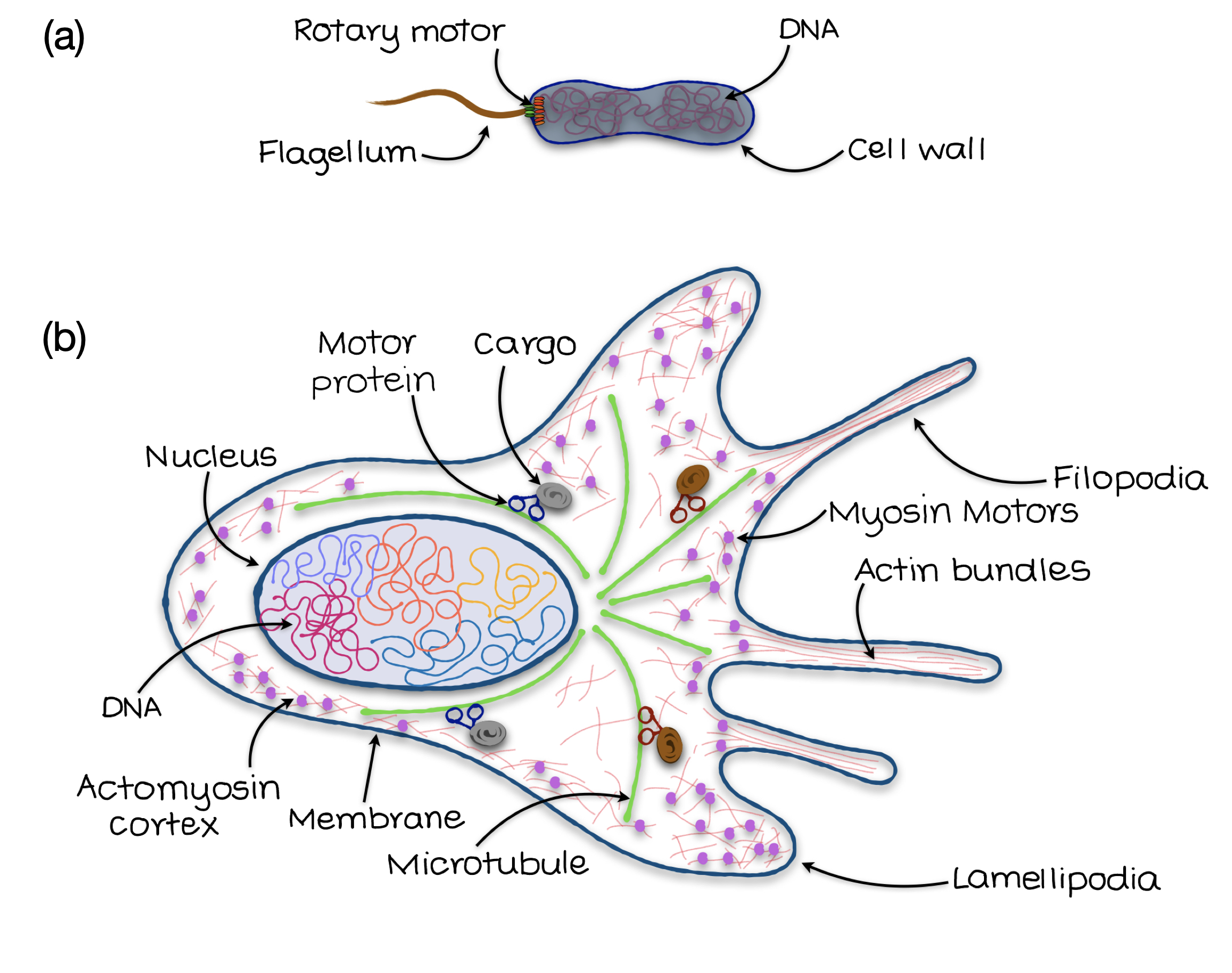}
\caption{Schematic representation of (a) a prokaryotic and (b) a eukaryotic cell (not to scale), showing only those structures whose force-generating or force-bearing physics is discussed in this review; many biochemically important components are omitted. (a) Prokaryotes are typically small ($\sim 1$--$2~\mu$m) and lack internal membrane compartments; the \textit{E.~coli} shown propels itself with a rotary motor that turns helical flagella. (b) Eukaryotes ($\sim 10$--$20~\mu$m) maintain a compartmentalised interior: DNA is packaged into chromosomes enclosed by the nuclear envelope; microtubules serve as tracks along which molecular motors transport cargo; and a cortical meshwork of actin, myosin motors, and crosslinkers (the actomyosin cortex) generates protrusions such as filopodia and lamellipodia for adhesion and motility.}
\label{fig:generic_cell}
\end{figure}

Nucleic acids, both DNA and RNA, are long molecules that embed information required for cell function and replication. The information contained in the DNA sequence is first transcribed into an mRNA molecule and then translated into a protein sequence~\cite{krebs2013lewin}. Both these processes, of \emph{transcription} and \emph{translation}, require energy~\cite{ortega2024minimum}.

Carbohydrates are a prominent energy storehouse for cells. These are broken down in the body into sugars. Further breakdown of a single glucose molecule under aerobic conditions leads to the production of around $30$ ATP molecules~\cite{milo2015cell}. However, most cellular machinery operates by using the energy released by the hydrolysis of predominantly ATP, sometimes GTP. It is estimated that the human body turns over an equivalent of half its weight in ATP molecules each day~\cite{milo2010bionumbers}.

\subsection{Cells and compartmentalization}

The four basic classes of molecules are the building blocks of more complex structures inside living cells.  Figure \ref{fig:generic_cell}(a) shows a simple  bacterium of size $\approx 1 \mu \mathrm{m}$, with a rotary motor embedded in its cell wall. Eukaryotic cells, shown schematically in Figure \ref{fig:generic_cell}(b),  contain multiple intracellular organelles. These use internal membranes to separate the inside from the outside. A detailed glossary of these structures is available in Ref.~\cite{wingreenGlossaryCellularComponents2006}.

The most prominent compartment is the cell itself. A typical eukaryotic cell is contained within a {\it plasma membrane}, an effectively two-dimensional bilayer fluid membrane made of lipid molecules, proteins, and sugars~\cite{alberts2014molecular}. Its mechanics as a fluid film, set by in-plane tension and an out-of-plane bending cost, is developed in Section~\ref{sec:soft_matter_theory}.  The membranes of living cells are in constant flux, exchanging nutrients and signalling molecules with the external environment. 

Compartments provide niches for chemical reactions that require specialized environments, isolating and maintaining spatially heterogeneous regions within a cell. Driven by energy flows, thermodynamically unfavourable processes can proceed in such regions. These regions, e.g., mitochondria, can also act as energy depots that drive cellular processes on time-scales much longer than those of energy production.  

Compartmentalization can also occur without membranes, through phase separation that creates regions enriched in specific biochemical components (e.g., nucleoli and Cajal bodies)~\cite{hymanLiquidLiquidPhaseSeparation2014,berry2018physical}. 

\subsection{Structural framework of cells and tissues}

Living systems have evolved cross-linked polymer networks, collectively called the cytoskeleton, that generate and transmit mechanical forces. Its filaments vary in size, the largest comparable to the cellular scale of $\approx 1-10~{\rm \mu m}$; they are distributed across the cell and especially reinforced just below the plasma membrane (the cortex).

The eukaryotic cytoskeleton is composed of three types of self-assembled polymeric filaments (microtubules, actin filaments, and intermediate filaments) together with the proteins that bind them~\cite{HowardBook, pollard2017cytoskeleton}, including force-generating molecular motors. Microtubules and actin filaments (FIG.~\ref{fig:generic_cell}(b)) have structural polarity~\cite{alberts2014molecular}. These filaments organize into higher-order structures that provide structural integrity, facilitate cell division, attach to substrates, propel cells, and serve as tracks for cargo transport by molecular motors. Intermediate filaments provide tensile strength and mechanical resilience, particularly in epithelial tissues and neurons~\cite{alberts2014molecular}; since they do not directly participate in ATP-driven force generation, we do not discuss them further.

Cytoskeletal filaments differ in their diameter, structural and mechanical properties. Of these, microtubules, with diameter $\approx 25~{\rm nm}$,  are the stiffest. Actin filaments have an overall diameter of $\approx 8 {\rm nm}$.  Bacterial cells possess analogues of tubulin and actin proteins~\cite{busiek2015bacterial}.

Modern approaches, developed over the past few decades, account for the presence of active elements such as pumps on the membrane surface and the underlying meshwork of protein filaments, the actomyosin cortex. This meshwork, ranging from a few tens to hundreds of nanometres in thickness~\cite{clarkMonitoringActinCortex2013}, is a dense crosslinked polymer layer with embedded molecular motors that consume ATP and generate active stresses (FIG.~\ref{fig:generic_cell}(b)).  The coupling between the cell cortex and the plasma membrane regulates important biophysical properties of the cell membrane~\cite{brayMembraneAssociatedCortexAnimal1986, chalutActinCortexBridge2016, chughActinCortexGlance2018, kumarActomyosinCortexCells2021}.

\subsection{\label{sec:neq_active_cells}Non-equilibrium active processes in cells}

The notion of a system being in thermodynamic equilibrium depends on the time-scales of observation~\cite{ma1985statistical}. Typically, fast processes such as the opening and closing of ion channels ($\sim 1 {\rm \mu s}$) and the binding-unbinding of small (ligand) molecules ($\sim 1-10~{\rm \mu s}$) can be understood within the framework of equilibrium statistical mechanics~\cite{phillipsNapoleonEquilibrium2015, phillips2020molecular}. Slower processes such as protein synthesis ($\sim 1-10~{\rm s})$, cell division/locomotion ($\sim 10^2~\rm{s}$), and morphogenesis ($\sim 10^3 - 10^{5}~{\rm s}$) are driven by a throughput of energy. The consumption of energy to drive non-equilibrium processes makes the cell an \emph{active} system.

An exemplar of active processes is the directed transport of material by molecular motor proteins that move unidirectionally on filamentous tracks. They draw energy from ATP hydrolysis, or from the electrochemical potential of proton gradients across cell membranes, and must work against the viscous drag of the crowded cell interior, where the viscosity can be up to $\sim 10$ times that of water.

ATP-driven linear molecular motors include myosins, kinesins, and dyneins. They are indispensable for cellular activities such as muscle contraction and intracellular vesicle transport. Other linear motors take part in processes such as  replication, transcription and genome maintenance (helicases and polymerases).

Rotary motors driven by proton-gradients are found in the ATP synthase~\cite{yoshida2001atp} and the bacterial flagellar motor~\cite{berg2004coli} (shown in FIG.~\ref{fig:generic_cell}(a)). The ATP synthase is a molecular machine, a complex of proteins, in which one part can rotate relative to the other. In a proton gradient, the unidirectional rotation of the ATP-synthase can generate ATP from excess ADP molecules.  In the reverse process, the energy released from the hydrolysis of ATP is used to rotate the ATP-synthase. It rotates at $\approx 4$ revolutions per second while generating torques up to $40~{\rm pN~nm}$~\cite{noji1997direct}.

Molecular motors are an archetypal example of transducing chemical energy into mechanical work. Because no net work can be extracted at thermodynamic equilibrium, sustained force generation requires continuous energy throughput, taking the cell beyond equilibrium frameworks. These non-equilibrium forces drive vital cellular functions. What are their consequences?

\section{Force scales and their characterization}\label{sec:force_scales}

\begin{figure*}[ht]
\centering
\includegraphics[width=0.85\textwidth]{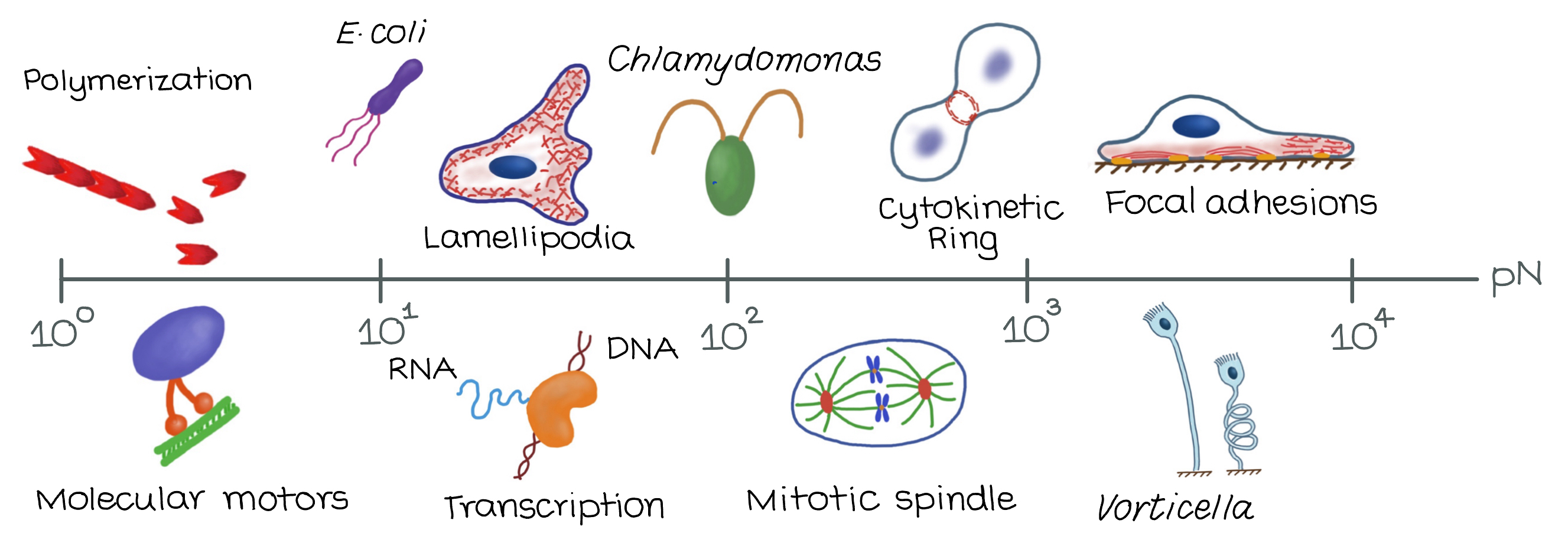}
\caption{This figure illustrates the hierarchy of force scales at the level of a single cell and below. The appropriate force scale is the piconewton ($10^{-12}$ N). Forces from molecular motor translocation are in the regime $1$--$10$ pN while the forces involved in cell adhesion are typically $2$--$3$ orders of magnitude larger. }
\label{fig:force_scales}
\end{figure*}

Force is what force does: forces move, accelerate, deform or rotate the objects they are applied to. FIG.~\ref{fig:force_scales} shows, on a logarithmic scale, the range of forces involved in important cellular biophysical processes. One can roughly estimate the scale of forces required to be sensed, as well as exerted at the single molecule level. A characteristic molecular force scale is $F \sim k_BT/a$, where $a \sim 1~\mathrm{nm}$ is a representative displacement coupled to the force: a force of this magnitude performs work of order $k_BT$ over that displacement. This is not a strict lower limit for force sensing; sensitivity also depends on the sensing mechanism and the observation time. Forces exerted as a consequence of ATP hydrolysis are bounded above by $F \sim 20 k_BT/a$ where $20 k_BT$ is the approximate value of the energy released through the hydrolysis of a single ATP molecule. This yields force scales in the regime $10 - 100~ \mathrm{pN}$ for forces acting at the single molecule scale in biological systems.

At scales relevant to cell biology, all forces operate in a fluid medium. It is often easier to adopt continuum rather than discrete descriptions, which incorporate the surrounding fluid straightforwardly. Newtonian mechanics provides a linear relationship between acceleration and force, $\ddot{\mathbf r} \propto {\mathbf F}$, where an overdot denotes a time-derivative. In a highly overdamped system, however, with friction from an ambient fluid, $\dot{\mathbf r} \propto {\mathbf F}$ gives a more practical description of the long-time dynamics.

The effective interactions between the mesoscopic objects of cell biology are strongly modified by the surrounding solvent. For a solute particle that is uncharged in vacuum, immersing it in a solvent can leave it with a net charge as surface chemical groups dissociate, releasing mobile counter-ions~\cite{israelachvili2011intermolecular}. In a polar solvent, such as water, a network of hydrogen bonds lends water its anomalous properties, including that of being an exceptional solvent for most biological molecules. 

 Added salt screens these charges over a Debye length $\ell_D = \sqrt{\epsilon k_BT/2e^2 n}$, where $\epsilon$ is the permittivity of the medium and $n$ the number density of a monovalent salt (see Glossary): for pure water at room temperature $\ell_D \sim 1~{\rm \mu m}$ while for the cellular cytoplasm $\ell_D \sim 1~{\rm nm}$~\cite{wennerstrom2020colloidal}, reflecting its high salt content. Interactions mediated by water molecules, deriving from the structuring of the hydrogen bond network around protein surfaces, are called hydrophobic interactions. Conformational changes occurring as a result of forces acting on a protein can alter the specificity of its interactions and functions, forming the basis of mechanosensitivity of protein-protein interactions.

Even in the absence of direct electrostatic interactions, effective interactions arising from entropy maximization can be conceptualized as forces. Such forces are conventionally termed as statistical forces. A well known example of these statistical forces, the {\it depletion interaction}, is seen in a system with particles of different sizes immersed in a thermal bath. Clumping the larger particles together increases the configurational space for smaller particles, and thus the overall entropy, resulting in an effective entropic attraction between the larger particles~\cite{asakura1954interaction}. 
Osmotic forces, which arise across a semi-permeable membrane with differential solute concentrations, are partly entropic in origin~\cite{doi2013soft}. Fluctuation-induced forces can also arise in non-thermal situations~\cite{lion2014osmosis}.

In the following sections, we briefly review some of the theoretical frameworks required to describe forces/stresses and their consequences, both in the equilibrium and non-equilibrium contexts. We also discuss appropriate experimental techniques to measure these forces in the context of living systems. 

\subsection{Equilibrium and non-equilibrium forces}

A system in thermodynamic equilibrium minimizes an appropriate free energy. No net currents flow between any of its microscopic states. If a system is perturbed slightly away from its equilibrium state, \emph{thermodynamic forces} restore it to equilibrium~\cite{de2013non}.

Consider now a system which is held in a steady-state but out of equilibrium. It can be maintained in such a state only by a constant flux, typically of energy and matter. For instance, the operation of nano-scale biological machines such as molecular motors and ion-pumps relies on a continuous supply of energy. However, the ``forces'' involved in these processes, typically, do not have a simple and generic relationship with the free energy or any analogue. Much of the cell's functionalities derive from such ``active'' force generating processes. Living systems have evolved metabolic pathways to convert energy input from the external surroundings into a global currency of high-energy molecules, for instance NTPs.

At a microscopic scale, it is plainly impossible to describe the dynamics of the approximately $\sim 10^{15}$ molecules that typically constitute a single cell. Clearly, a reduced description in terms of a much smaller set of important variables is appropriate. Choosing such variables is a problem of coarse-graining. Traditionally, such coarse-graining is in terms of \emph{hydrodynamic variables}~\cite{chaikin2004principles}. These are collective modes (such as the densities of conserved quantities or broken-symmetry variables) whose relaxation slows without bound at long wavelengths. However, given the finite size of the cell and limited time-scale of observation, non-hydrodynamic variables may often be important.

The choice of appropriate coarse-grained variables is as much an art as a science. For example, although a molecular motor protein has several thousand atoms, representing it as a single particle in a suitable potential captures important aspects of its directed motion. What is often done is to use microscopic computational models that are constructed to capture the essential physics in the hydrodynamic limit~\cite{allen2017computer}, to perform computations using simplified kinetic models~\cite{succi2001lattice}, or simply to use general symmetry arguments~\cite{chaikin2004principles} to pick out all possible lowest order relevant terms that should enter the equations of motion.

\subsection{Dimensionless numbers for cellular flows}

Forces can lead to deformations and flows. For fluid flows, an important dimensionless combination of relevant physical quantities is the Reynolds number ($\mathrm{Re} = \rho L v/\eta$, the ratio of inertial to viscous forces for a flow of speed $v$, scale $L$, viscosity $\eta$ and density $\rho$; see Glossary). For example, large scale hydrodynamic flows ($v \approx 10~{\rm \mu m/min}$, $\rho = \rho_{\rm water} = 10^{3}~{\rm kg/m^3}$, $\eta = \eta_{\rm water} = 10^{-3}~{\rm Pa\:s}$, and $L \approx 50~{\rm \mu m}$) in the actomyosin cortex of the {\it C. elegans} embryo correspond to $\mathrm{Re} \approx 10^{-6}-10^{-5}$~\cite{saha2016determining}. For bacteria swimming in water ($v \approx 30~{\rm \mu m/s}$ and $L \approx 1~{\rm \mu m}$), $\mathrm{Re} \approx 10^{-5}-10^{-4}$~\cite{purcell1977life}. Thus, hydrodynamic flows at the cellular level are almost always at low Reynolds number. Inertial forces are negligible at these scales. 

Most biological materials are viscoelastic, i.e., they exhibit a combination of fluid-like and solid-like behaviours. Linear viscoelastic fluids resist deformations induced by applied forces up to a certain fluidization time-scale, known as the Maxwell time, beyond which they flow~\cite{mayer2010anisotropies, saha2016determining}. 

For biological flows in which diffusion and directed transport compete, the appropriate dimensionless quantity is the P\'{e}clet number ${\rm Pe} = vL/D$ (the advection-to-diffusion ratio; see Glossary). In the cellular context, the P\'{e}clet number ranges from values ${\rm Pe} \ll 1$ (e.g., for spatial profiles of signalling molecules established by diffusion), ${\rm Pe} \approx 1$ (e.g. the movement of molecular motors, cytoplasmic streaming, cortical flows), to values ${\rm Pe} \gg 1$ (e.g., in cell motility mechanisms such as crawling and swimming, blood flow). The diffusion coefficients referred to here need not be of thermal origin.

\section{\label{sec:soft_matter_theory}Living matter as active soft matter}

Soft materials in equilibrium are many-body systems in which typical energies for deformation are comparable to the thermal energy. Because of this, thermal fluctuations cannot be neglected. In general, the fluctuations can also be non-thermal, for instance in active systems. Biological materials are archetypal examples of driven soft matter.

At thermodynamic equilibrium, response coefficients such as susceptibilities, conductivities, etc., are intimately connected to spontaneous fluctuations. These constraints are expressed in the form of fluctuation-dissipation relations. However, for a system far away from thermodynamic equilibrium, there are no such constraints between spontaneous fluctuations and response coefficients. Stochastic fluctuations in active materials generically violate detailed balance conditions, discussed in more detail below, leading to non-trivial consequences for their macroscopic descriptions. 

A generic statistical mechanical theory for systems far from thermodynamic equilibrium does not exist, at least as of now. Nevertheless, physical descriptions developed for near-equilibrium systems provide a useful framework, since their predictions can be quantitatively compared to experimental observations. In the next few subsections, we develop a sequence of increasingly coarse-grained descriptions that are valid both at thermodynamic equilibrium and for driven states \emph{close} to thermodynamic equilibrium. After a brief description of various biophysical experimental techniques to measure stochastic fluctuations and response functions, we will use the theoretical formalism developed here to describe several cellular processes. 

\subsection{Stochastic descriptions}

\begin{figure}
\centering
\includegraphics[width=0.9\linewidth]{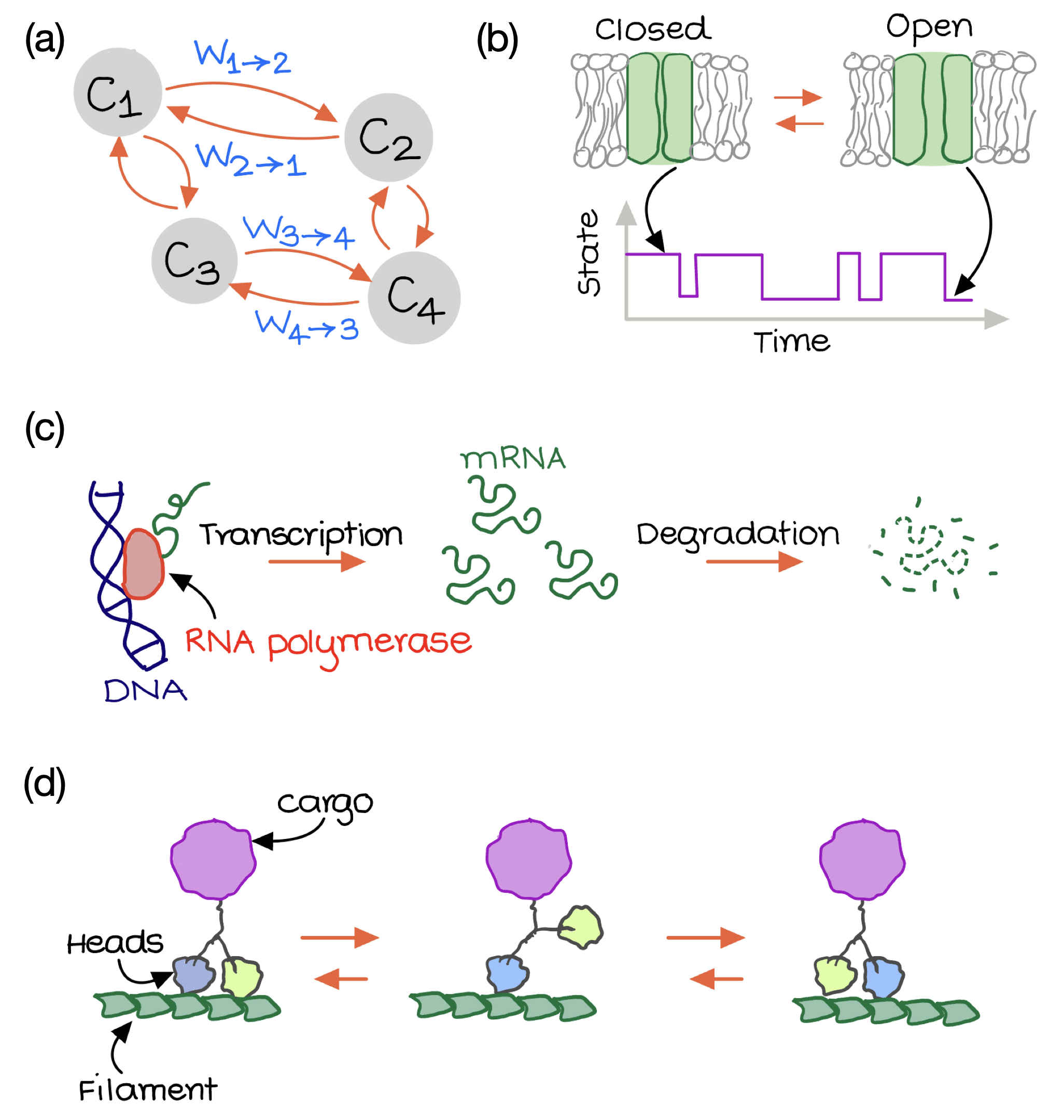}
\caption{(a) A schematic of the master equation \eqref{eq:master_equation} indicating stochastic transitions between the states $C_i$ with transition rates $W_{i \to j}$. (b) Ion-channels are proteins embedded in the cell-membrane which transition between ``open'' and ``closed'' states in a stochastic manner. (c) The production (by transcription) and degradation of the mRNA molecules can be captured within a master equation approach where $P_n(t)$ represents the probability of finding $n$ mRNA molecules at time $t$. (d) The translocation of molecular motors transporting cargo on filaments is another example of a stochastic process with discrete states. Here the internal states of the system are the attachment configurations of the two heads of the motor. The coupling between these transitions and directional asymmetries in the filament leads to motion of the motor.}
\label{fig:masterequation}
\end{figure}

For systems with discrete states such as, for instance, the number of molecules involved in transcription or translation, an appropriate description for the stochastic dynamics is via a master equation~\cite{van1992stochastic}. The probability $P_i$ to find the system in a configuration $i$ evolves in time according to the master equation:
\begin{align}
\frac{dP_i}{dt} = \sum_{j \neq i} \left[ W_{j\to i} \; P_j - W_{i \to j} \; P_i \right]
\label{eq:master_equation}
\end{align}
where $W_{i \to j}$ is the transition rate from a configuration $i$ to $j$. Figure~\ref{fig:masterequation}(a) illustrates the discrete configurations and the transitions between them. The opening and closing of an ion-channel, shown in FIG.~\ref{fig:masterequation}(b), is an example of a two-state stochastic process.

 The master equation description is valid for systems both at equilibrium and those that are out of equilibrium. For equilibrium systems, the detailed balance condition $W_{j\to i} P_j = W_{i \to j} P_i$ (which ensures that there are no net currents) together with a Boltzmann probability $P_{i, \rm eqm} \sim \exp(- \beta E_i)$, with $\beta^{-1} = k_BT$, characterizes the equilibrium state: it fixes the ratios $W_{i\to j}/W_{j\to i} = e^{-\beta(E_j - E_i)}$, but not the rates themselves, which set the relaxation timescales. However, for driven nonequilibrium systems that break time-reversal invariance, there are no such general descriptions. Most cellular processes are driven by energy inputs and, as such, are out of equilibrium.

For instance, in the case of DNA transcription into mRNA, as shown in FIG.~\ref{fig:masterequation}(c) with $P_n$ describing the probability of finding $n$ molecules, the appropriate master equation is
\begin{align}
\frac{dP_n}{dt} = \gamma_d \; (n+1) P_{n+1} + \gamma_p \; P_{n-1} - (\gamma_d \; n + \gamma_p) P_n,
\end{align}
where $\gamma_p$ is the production rate and $\gamma_d$ is the degradation rate. Another example of a stochastic process is the directional stepping of a molecular motor shown in FIG.~\ref{fig:masterequation}(d).

\begin{figure}
\centering
\includegraphics[width=0.9\linewidth]
{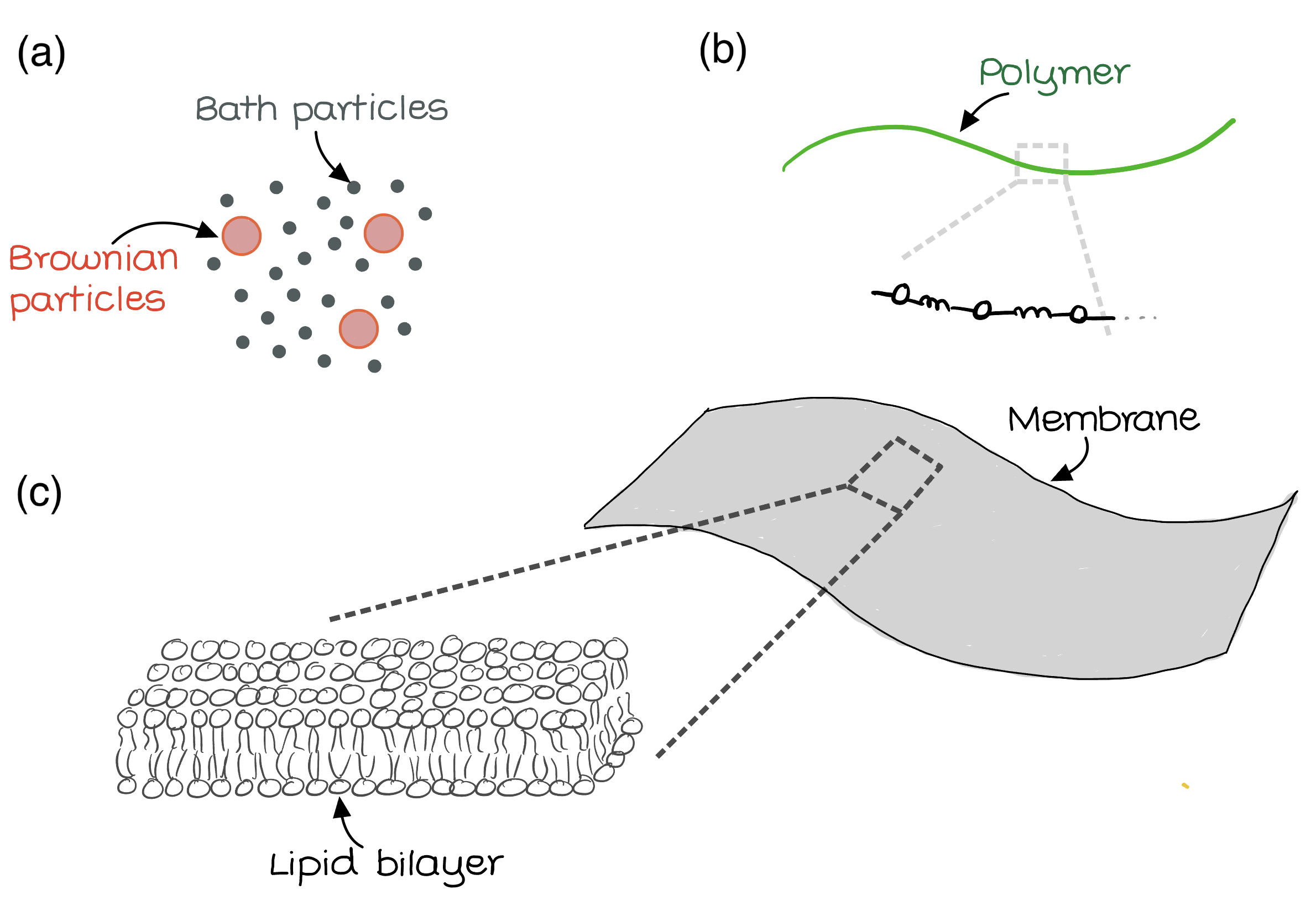}
\caption{Collections of interacting particles driven by stochastic forces arising from the thermal environment. (a) shows a few colloidal scale particles ($10^{-6} \,$m) immersed in a bath of smaller particles. The interactions of the colloidal particles with the heat bath lead to diffusive dynamics at long time. (b) and (c) show extended structures -- one a polymer filament and the other a membrane sheet. The strength of the stochastic forces is comparable to the energies required to cause significant shape change. The extended nature of these structures requires that their energetics be modelled through energy functionals that depend upon shape.
}
\label{fig:langevin_bath}
\end{figure}

In certain limits, one can reduce the master equation description of transitions between states to a set of stochastic differential equations (SDE) for continuous variables. The translocation of a motor protein such as kinesin moving on microtubules, although best described microscopically in terms of transitions between discrete molecular configurations, can be modelled at a coarse-grained level by changes in a continuous position coordinate. Brownian motion is an archetypal example of such a SDE where one considers the stochastic dynamics of a few ``slow'' variables while the effects of the other ``fast'' variables are represented by effective forces arising from a coupling to the medium (FIG.~\ref{fig:langevin_bath}(a)). 

The effects of the medium on an embedded particle come from two sources. One is a drag force, proportional to the velocity of the particle and the other an instantaneous ``noise'' term. For a particle of mass $m$ moving under the influence of a potential $U(\mathbf{x})$, the Langevin equation describing the stochastic dynamics is ~\cite{zwanzig2001nonequilibrium}
\begin{align}
m \frac{d\mathbf{v}}{dt} = -\nabla U(\mathbf{x}) - \int_0^t dt' \; \Gamma(t-t') \, \mathbf{v}(t') + \mathbf{f}(t),
\end{align}
where the scalar friction kernel for an isotropic bath $\Gamma(t)$ represents the drag force and the stochastic term $\mathbf{f}(t)$ represents the noise. The noise term is most often modelled as being drawn from a Gaussian distribution, with zero mean; at equilibrium it obeys the fluctuation-dissipation theorem $\langle f_i(t) f_j(t')\rangle = k_B T \, \Gamma(|t-t'|) \, \delta_{ij}$. Typically, $\Gamma(\tau)$ is sharply peaked around $\tau=0$, is non-zero only in a small region of $\tau$, and can be approximated by $\Gamma(\tau) \approx 2\gamma \, \delta(\tau)$. On timescales $t \gg m/\gamma$, one can also neglect the inertial term, leading to a purely overdamped equation of motion. The position vector $\mathbf{x}$ then evolves according to:
\begin{align}
\frac{d\mathbf{x}}{dt} = -\mu \, \nabla U(\mathbf{x}) + \boldsymbol{\xi}(t),
\label{eq:overdamped_Langevin}
\end{align}
where $\mu = 1/\gamma$ is the mobility and $U(\mathbf{x})$ is a confining potential. The stochastic term $\boldsymbol{\xi}$, representing the effects of the fluctuating environment, is a Gaussian white noise process with zero mean and $\langle\xi_i(t) \xi_j(t')\rangle = 2 D \delta_{ij} \delta(t-t')$ where the diffusion constant $D=\mu k_BT$ is set by the Einstein relation.

In a probabilistic description of states, the master equation \eqref{eq:master_equation} becomes the Fokker-Planck equation. In section~\ref{sec:molecular_motors}, we will use the Fokker-Planck description to illustrate how directed motion occurs in models of molecular motors. The equation governing the time-evolution of the probability density $P(\mathbf{x},t)$, of finding the particle at position $\mathbf{x}$ and time $t$, is governed by the corresponding Fokker-Planck equation (in this context, the Smoluchowski equation):
\begin{align}
\frac{\partial P(\mathbf{x},t)}{\partial t} = -\nabla \cdot \mathbf{J},
\qquad
\mathbf{J} = -\mu \, \nabla U(\mathbf{x}) \, P - D \nabla P,
\end{align}
where $\mathbf{J}$ is the probability current~\cite{risken1996fokker}. At thermodynamic equilibrium, the probability current $\mathbf{J}=0$. This ensures that the Boltzmann distribution $P_{\rm eqm} \sim e^{-\beta U(\mathbf{x})}$ is the asymptotic solution. The stochastic descriptions discussed above can be generalized to include interactions among particles.

\subsection{Polymer mechanics}

Specific cases of interacting particle systems of biological interest include polymers (FIG.~\ref{fig:langevin_bath}(b)) and membranes (FIG.~\ref{fig:langevin_bath}(c)). We can idealize the polymer as a smooth one-dimensional curve and the membrane as a smooth two-dimensional surface. For an inextensible one-dimensional curve of length $L$ with a unit tangent vector $\hat{\mathbf{t}}$, an appropriate elastic energy functional is
\begin{align}
E_{\rm p} = \int_0^L ds \, \frac{\kappa_p}{2} 
c(s)^2,
\label{eq:E_polymer}
\end{align}
where $s$ is the arc-length parameter along the polymer, the curvature $c(s)=|\partial_s\mathbf{\hat{t}}(s)|$ measures the rate at which the unit-tangent vector $\mathbf{\hat{t}}(s)$  changes along the curve, and $\kappa_p$ represents the free-energy cost to bend the polymer -- a bending modulus. This description is also known as the worm-like chain model for the polymer~\cite{doi1988theory}. The stiffness of any polymer is related to the correlation between tangent vectors at different points along the curve. The persistence length is defined by the correlation function $\langle \hat{\mathbf{t}}(s) \cdot \hat{\mathbf{t}}(s') \rangle \sim \exp(-|s-s'|/\ell)$ where $\langle \cdots \rangle$ denotes equilibrium averages. If the environment is a heat-bath at temperature $T$, then $\ell$ is a thermal persistence length $\ell_T$. Typical biopolymers have $\ell_T \sim 50~{\rm nm}$ for double-stranded DNA, $\ell_T \sim 10~{\rm \mu m}$ for actin and $\ell_T \sim 1~{\rm mm}$ for microtubules.

Measurements of the force-extension relationship for DNA/RNA {\it in vitro} attach a colloidal particle to one end of the DNA molecule which is tethered at the other end, and trap this colloidal particle in a movable optical trap~\cite{smithDirectMechanicalMeasurements1992,smithOverstretchingBDNAElastic1996,liphardtReversibleUnfoldingSingle2001}. Exerting a calibrated force on the DNA molecule leads to its extension. Such measurements indicate that DNA at large length scales (greater than $\sim 150$ base-pairs) can best be described as a semi-flexible polymer. A remarkably accurate form for the force-extension relation that only involves the persistence length is~\cite{bustamanteEntropicElasticityLPhage1994} 
\begin{align}
{f \over (k_BT/\ell_T)} = {z \over L} + {1 \over 4 (1 - z/L)^2} - {1 \over 4},
\end{align}
where $f$ is the force required to stretch the polymer of contour length $L$ by an amount $z$ in the direction of the force. The scale of such forces for a $5-10 \%$ strain ($z/L$) is of the order of $10^{-2}~{\rm pN}$.

Two features of this description should be kept in mind, since both are routinely violated in the cellular interior. First, the worm-like chain is an \emph{equilibrium}, single-filament model: the persistence length is defined through Boltzmann-weighted averages, the contour is treated as inextensible, and the surroundings enter only as a thermal heat bath. At the large forces reached in single-molecule pulling, the backbone itself stretches, and the purely entropic force-extension relation above must be supplemented by an enthalpic stretch modulus (the \emph{extensible} worm-like chain) before it reproduces the data~\cite{odijk1995stiff, smithOverstretchingBDNAElastic1996}. In dense or confined environments, steric interactions, entanglements and crosslinking between filaments renormalize the apparent stiffness, so that a single-chain persistence length no longer characterizes the mechanical response of the network~\cite{broedersz2014modeling}. Second, in an environment with active fluctuations, such as in a cell, where the probability distributions will be non-Boltzmann, $\ell$ is no longer a thermal quantity: it is set by the scale of the nonequilibrium fluctuations and can differ substantially from $\ell_T$. For example, measurements of bending fluctuations of microtubules in reconstituted actin networks with embedded myosin motors show that $\ell \sim 1~{\rm \mu m}$ for microtubules~\cite{brangwynneNonequilibriumMicrotubuleFluctuations2008}, orders of magnitude below the thermal persistence length quoted above.

\subsection{Membrane mechanics}

For a two-dimensional fluid membrane, one in which the constituent molecules can diffuse past each other, the corresponding coarse-grained energy functional to lowest order is the Canham-Helfrich energy functional
\begin{align}
E_{\rm m} = \int dS \, \left( \sigma_s + \frac{\kappa}{2} (2H-H_0)^2 + \overline{\kappa} \, \mathcal{K}\right). 
\label{eq:E_membrane}
\end{align}
The effective surface tension $\sigma_s$ (sometimes called the membrane tension), represents the free energy cost to change the area of the membrane by a unit amount. The mean and Gaussian curvatures are defined by $H = (c_1 + c_2)/2$ and $\mathcal{K} = c_1c_2$, respectively, where $c_1$ and $c_2$ are the principal curvatures, taken to be positive for a sphere with outward normal. The spontaneous curvature $H_0$ is non-zero only for an asymmetric bilayer when the two leaflets of the membrane are inequivalent. The free energy cost of bending the membrane is given by $\kappa$ and $\overline{\kappa}$, the appropriate bending rigidities~\cite{boal2012mechanics}. For closed surfaces, as in many biological contexts, the integral $\int dS \, \mathcal{K} = 2\pi \chi_{\mathrm{E}}$, with $\chi_{\mathrm{E}}$ being the Euler characteristic. As such, we can neglect the last term above when the topology remains fixed.

A measure of planarity for a $2$D membrane is the correlation function $\langle \hat{\mathbf{n}}(\mathbf{x}) \cdot \hat{\mathbf{n}}(\mathbf{x}') \rangle$ of the unit normal vector $\hat{\mathbf{n}}$ at two-different points $\mathbf{x}$ and $\mathbf{x}'$. A careful analysis of this correlation function shows that the associated persistence length $\xi_p \sim b \exp(\beta C \kappa)$ where $C$ is a numerical constant, $b$ is a molecular length-scale and $\kappa$ is the bending rigidity~\cite{nelson2004statistical}. 

To find the shape that minimizes the energy $E_m$, we write the Euler-Lagrange equations for membrane shape with appropriate constraints such as a fixed volume and topology. For a closed membrane with spatially constant $\kappa$, $\overline{\kappa}$, $H_0$, and $\sigma_s$, this gives the shape equation~\cite{zhong1989bending, kozlovMembraneShapeEquations2006}
\begin{align}
    &\;2\kappa\,\nabla^2 H \;+\; \kappa\,(2H-H_0)\!\left(2H^2 + H_0 H - 2\mathcal{K}\right) \; \\ \nonumber
   & \qquad \qquad -\; 2\sigma_s H \;+\;\Delta p \;=\; 0\;
\end{align}
where $\Delta p = p_{\rm in} - p_{\rm out}$ is the pressure difference between the inside and the outside of the surface which acts a Lagrange multiplier to conserve the volume. If the bending rigidity can be neglected, then a simple balance of forces in equilibrium leads to the Young-Laplace law:
\begin{align}
2 \sigma_s H = \Delta p,
\end{align}
where $H=1/R$ in the case of a sphere with radius $R$. This relation is used to measure the membrane tension experimentally, see e.g., section~\ref{sec:membrane_measurements}.

The cell membrane is inhomogeneous -- in addition to being composed of various kinds of lipids, the cell membrane contains many transmembrane proteins. It controls the movement of material across it, modulated by selective transport mechanisms such as ion channels and pumps. While pumps consume chemical energy to drive ions against a concentration gradient, the otherwise passive gating of ion-channels can be sensitive to membrane tension. The cell can transport material across the membrane either by fusing lipid vesicles containing internal cargo with the plasma membrane (exocytosis) or by pinching a vesicle from the membrane to encapsulate external cargo (endocytosis). These fusion/fission processes of the membrane are regulated by specialized protein complexes. The forces involved here are $\sim 10 - 100~{\rm pN}$~\cite{schoneberg2018atp}.

Early studies of the shape fluctuations (flicker) in red blood cells observed under a light microscope, assumed that these were purely thermal in nature~\cite{brochard1975frequency}. However, over the past two decades, evidence has accumulated for a strong active component to such fluctuations~\cite{turlier2016equilibrium}. It is possible to write stochastic differential equations at this coarse-grained level using the energy functional~\eqref{eq:E_membrane}. These must include terms representing the active forcing of, e.g., membrane pumps through ATP hydrolysis. For instance, for a nearly flat membrane, the height-field $h(\boldsymbol{x},t)$ describing the membrane position above a $xy-$plane evolves according to
\begin{align}
\frac{\partial h(\boldsymbol{x},t)}{\partial t} = -\mu_m \frac{\delta E_{\rm m}[h(\boldsymbol{x},t)]}{\delta h(\boldsymbol{x},t)} + \xi(\boldsymbol{x}, t) + \xi_{\rm act}(\boldsymbol{x}, t),
\label{eq:active_height_eqn}
\end{align}
where $\mu_m$ is a membrane kinetic coefficient, $\xi(\boldsymbol{x}, t)$ is a thermal noise and the $\xi_{\rm act}(\boldsymbol{x}, t) $ represents active forcing. The active forces can be associated with the density of pumps~\cite{ramaswamy2001physics}. Theoretical analysis where \eqref{eq:active_height_eqn} is supplemented by equations describing fluid flow in the vicinity of the membrane and pump motion within the membrane provides evidence for a fluctuation spectrum that is athermal, as well as for interesting instabilities arising from curvature-pump density couplings~\cite{sankararaman2002two}. These have been studied experimentally~\cite{Lacoste2014}.

Two caveats on the Helfrich description deserve emphasis for living membranes. First, the shape equation and Young-Laplace law above use a single area-conjugate tension $\sigma_s$ within a deterministic continuum description. Once thermal fluctuations are included, one must distinguish the coefficient conjugate to the microscopic membrane area, the mechanical (frame) tension, and the tension inferred from the fluctuation spectrum; how these are related depends on the ensemble and on the coarse-graining scale, and equating them without stating the approximation is a well-documented source of confusion even for passive membranes~\cite{fournier2008direct, diamant2011model}. Second, the functional \eqref{eq:E_membrane} is an equilibrium free energy. Once the membrane is driven (by ion pumps, protein turnover, or an underlying active cortex), no unique thermodynamic tension exists. The effective tension controlling the fluctuation spectrum is then renormalized by the active noise in \eqref{eq:active_height_eqn} and can differ in magnitude, and even in sign, from its passive value; a negative effective tension is the hallmark of the curvature instabilities predicted for active membranes~\cite{ramaswamy2000nonequilibrium, sankararaman2002two}. The Canham-Helfrich energy therefore remains an indispensable organizing framework, but the tension it contains must be interpreted with care whenever the membrane is out of equilibrium.

The nuclear envelope is a double bilayer membrane supported below the inner bilayer by a dense layer of filamentous proteins. Molecules such as mRNA and proteins shuttle across the nuclear envelope through pore complexes. Energy dependent processes are likely involved in the transport of these molecules within the nucleus and across the nuclear envelope~\cite{xie2019mechanisms, vargas2005mechanism}. Moreover, the nuclear envelope also integrates mechanochemical signals, allowing the cell to regulate gene transcription and functionality~\cite{selezneva2022nuclear}. 

\subsection{Hydrodynamic descriptions}

\begin{figure}
\centering
\includegraphics[width=0.9\linewidth]{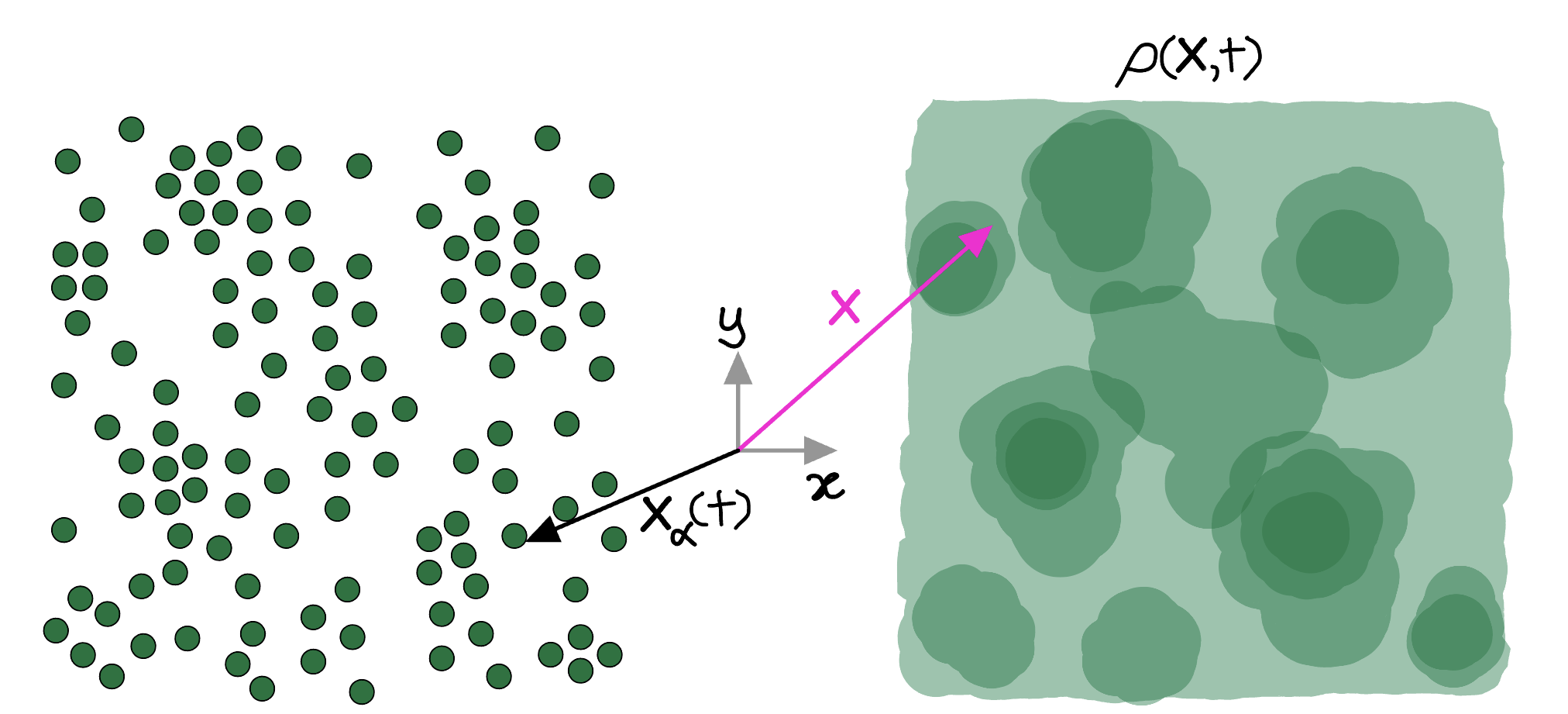}
\caption{For a collection of particles $\alpha$ at positions $\mathbf{x}_\alpha(t)$, we define a coarse-grained density field $\rho(\mathbf{x},t) = \left \langle \sum_\alpha m_\alpha \delta(\mathbf{x}-\mathbf{x}_\alpha(t)) \right \rangle$ by averaging over the particles' degrees of freedom. This leads to a smooth scalar field that varies in space and time.}
\label{fig:stress_tensor}
\end{figure}

The hydrodynamic approach extends the thermodynamic description to include spatiotemporal variations at the longest relevant length- and time-scales~\cite{chaikin2004principles}. The equations that result from such an approach typically involve several phenomenological coefficients. The hydrodynamic description of an isothermal elastic solid or viscous fluid composed of point particles is written in terms of the mass density $\rho(\mathbf{x},t)$ and the momentum density $\mathbf{g}(\mathbf{x},t)$ (FIG.~\ref{fig:stress_tensor}). The hydrodynamic equations for the mass and momentum densities are
\begin{align}
\partial_t \rho = -\nabla \cdot \mathbf{g},
\qquad
\partial_t \mathbf{g} = -\nabla \cdot \mathsf{\Pi},
\label{eq:isotropic_hydrodynamic_equations}
\end{align}
where the momentum flux tensor $\mathsf{\Pi} = \mathbf{g} \otimes \mathbf{v} - \mathsf{\Sigma}$ where $\mathbf{v} = \mathbf{g}/\rho$ the hydrodynamic velocity and $\otimes$ is a tensor-product~\cite{chaikin2004principles}. The generically symmetric second-rank stress-tensor $\mathsf{\Sigma}$ encodes the interactions of the mesoscopic continuum element with its surroundings. The traction $\mathbf{F}$ across a surface with unit normal $\hat{\mathbf{n}}$ is $\mathbf{F} = \mathsf{\Sigma} \cdot \hat{\mathbf{n}}$.

\subsubsection{Simple fluids}
For a simple Newtonian fluid, with a shear-viscosity $\eta$ and a bulk-viscosity $\eta_b$, the stress-tensor is
\begin{align}
\mathsf{\Sigma}_{\rm{f}} = -p \; \mathbb{I} + 2 \eta \left[ \mathsf{E} - \frac{\mathsf{Tr}(\mathsf{E})}{d} \mathbb{I} \right] + \eta_b \mathsf{Tr}(\mathsf{E}) \; \mathbb{I},
\label{eq:isotropic_fluid_stress}
\end{align}
where $p$ is the thermodynamic pressure, and
\begin{align}
\mathsf{E} = {\nabla \mathbf{v} + (\nabla\mathbf{v})^{\mathsf{T}} \over 2}\label{eq:shear_strain_rate}
\end{align}
is the strain-rate tensor. For an incompressible fluid ($\mathsf{Tr}(\mathsf{E}) = \nabla \cdot \mathbf{v} = 0$), this leads to the conventional form of the Navier-Stokes equation (see Glossary) for the velocity field
\begin{equation}
 {\partial }_t {\mathbf v} + {\mathbf v}\cdot \nabla {\mathbf v} = -{1 \over \rho}\nabla p + \nu_{\rm kin} \nabla^2 {\mathbf v}.
\end{equation}
Here $\nu_{\rm kin} = \eta/\rho$ is the kinematic viscosity. 

At the scale of cells and tissues, we can often ignore nonlinearities associated with the equation of motion of the fluid since the Reynolds number ${\rm Re} \ll 1$. A further approximation is to drop the time-derivatives of the velocity field as the accelerations ($\partial_t \mathbf{v} \approx 0$). The velocity can then be assumed to adapt instantly to the applied force. The governing equations of the velocity field are linear and time-reversible. 

The linearity and time-reversibility of the low-Reynolds-number equations have important consequences for the intracellular flows generated by the cytoskeleton and cortex, which we develop in later sections. (Extracellular swimming at low Reynolds number, including the scallop theorem and force-dipole descriptions of self-propulsion, is a large field in its own right; it is treated briefly in Section~\ref{sec:other_examples}, with comprehensive accounts in~\cite{purcell1977life, LaugaBook}.)

\subsubsection{Elastic solids}
 An elastic solid has a stress-free reference state; a material point is displaced by ${\mathbf u}$ from it. When deformations and rotations in the elastic material are small, i.e., $|\nabla {\mathbf u}| \ll 1$, then, analogous to the strain rate~\eqref{eq:shear_strain_rate}, the configurational change in the elastic material is captured by the small deformation strain
\begin{equation}
{\boldsymbol \varepsilon} = {\nabla {\mathbf u} + (\nabla {\mathbf u})^\mathsf{T} \over 2},
\end{equation}
Consequently, similar to \eqref{eq:isotropic_fluid_stress} for a Newtonian fluid, the stress for a linear, isotropic elastic material is given by a generalization of Hooke's law
\begin{equation}
\mathsf{\Sigma}_{\rm s} = 2 \mu_s \left[ \boldsymbol{\varepsilon} - \frac{\mathsf{Tr}(\boldsymbol{\varepsilon})}{d} \mathbb{I} \right] + K_b \mathsf{Tr}(\boldsymbol{\varepsilon}) \; \mathbb{I}, 
\label{eq:elastic_stress}
\end{equation}
where $\mu_s$ and $K_b$ are the shear and bulk moduli, respectively, of the solid. In addition, the equation of motion for the displacement field follows from that for the momentum $\mathbf{g}$, since $\rho \partial_t \mathbf{u} = \mathbf{g}$. 

\subsubsection{Viscoelastic materials}
\begin{figure}
\centering
\includegraphics[width=0.9\linewidth]{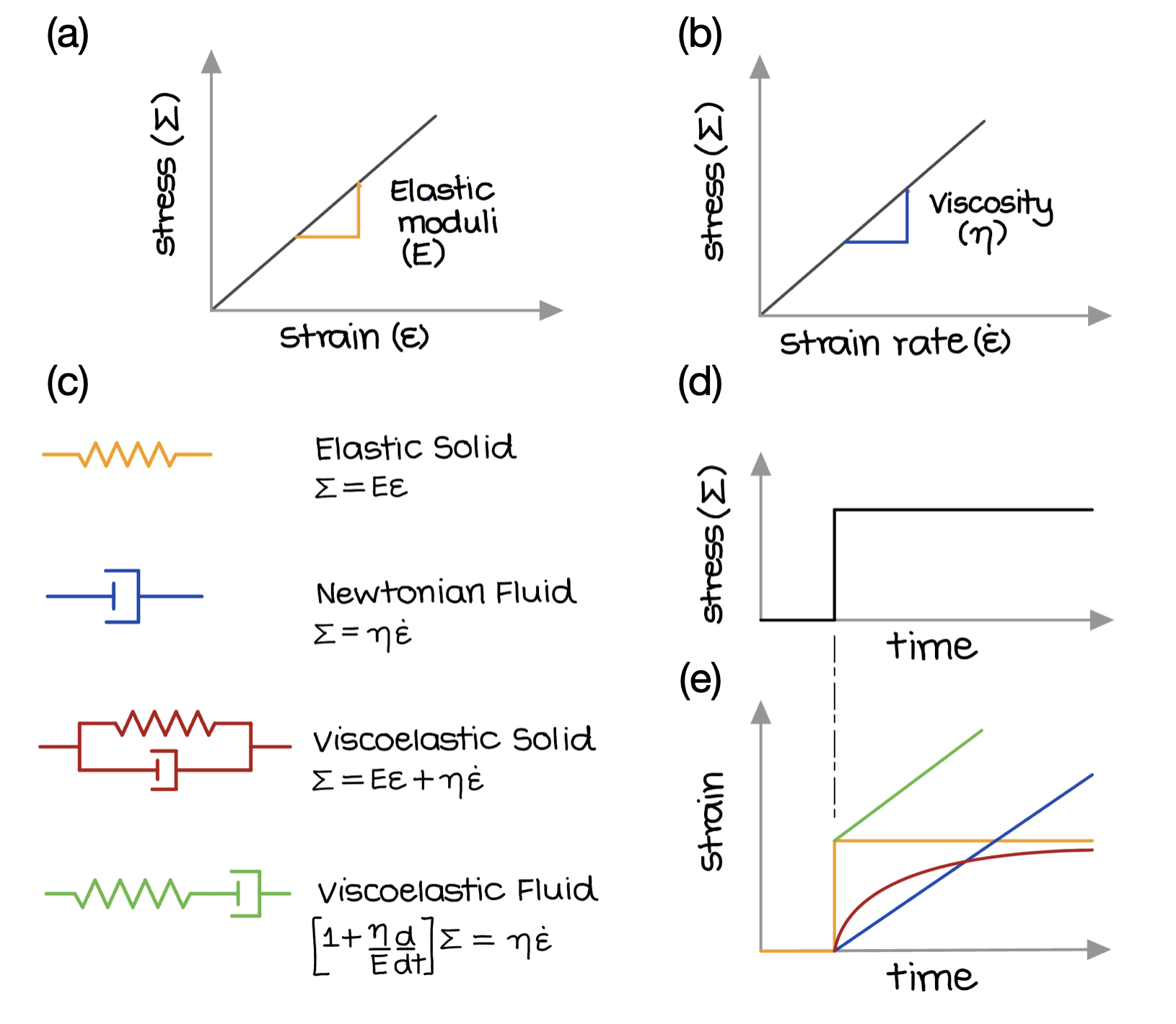}
\caption{The response of macroscopic materials to an applied force can be characterized by its stress $\mathsf{\Sigma}$ versus strain ${\boldsymbol \varepsilon}$ relationship in equivalent one-dimensional models. (a) For a Hookean elastic solid, the stress is linearly proportional to the strain, with the proportionality constant being the elasticity modulus $E$. This is represented by a simple spring. (b) For a simple Newtonian fluid, the stress is linearly proportional to the strain-rate, with the proportionality constant being the shear viscosity $\eta$ as represented by a dashpot. (c) A generic material can display both solid- and liquid-like responses at different timescales. Linear viscoelastic materials can be modelled by serial and parallel arrangements of springs and dashpots. The Kelvin-Voigt viscoelastic solid (red), with the spring and dashpot in parallel, flows like a fluid at short times and is solid-like at long times. The relevant timescale is governed by $\eta/E$. On the other hand, the Maxwell viscoelastic fluid (green), with the spring and dashpot in series, responds as a solid at short times and flows like a fluid at long times. The time-evolution of the strain for the different materials, in response to an imposed step stress (d), are also illustrated in (e).}
\label{fig:rheology_models}
\end{figure}

The ideal Newtonian fluid and the ideal Hookean solid are extreme limits of the mechanical response seen in real biological materials (FIG.~\ref{fig:rheology_models}(a),(b)). Because cellular constituents continuously remodel (through filament turnover, sliding, and transient crosslinking), their mechanical responses are neither those of an ideal solid nor of an ideal fluid~\cite{oswald2009rheophysics}.

An often encountered category of non-Newtonian materials are those that show viscoelastic behaviour. Two broad classes of such materials are the linear viscoelastic solid (Kelvin material) and the linear viscoelastic fluid (Maxwell material) -- see FIG.~\ref{fig:rheology_models}(c)--(e). A Kelvin-Voigt solid flows like a fluid for time-scales $t \ll \tau$, where $\tau$ is the retardation time-scale, whereas for $t\gg \tau$, it responds to mechanical perturbations like an elastic solid. 

On the other hand, a viscoelastic fluid exhibits a short-time elastic response and flows like a fluid at long time scales. As we did for the Newtonian fluid or the Hookean solid, considering a linear relationship between the hydrodynamic stress and mechanical deformations, a simple constitutive model for a Maxwell viscoelastic fluid is
\begin{equation}
\left(1+\tau\,\mathcal{D}_t\right)
\left(\mathsf{\Sigma}+p\mathbb{I}\right)
=\mathsf{\Sigma}_{\rm f}+p\mathbb{I},
\label{eq:Maxwell_Jeffries}
\end{equation}
where $\tau=\eta/E$ is the Maxwell relaxation time with $E$ the elastic modulus of the spring in FIG.~\ref{fig:rheology_models}(c), $\mathsf{\Sigma}$ is the hydrodynamic stress, $\mathsf{\Sigma}+p\mathbb{I}$ is the extra viscoelastic  stress, $\mathcal{D}_t \mathsf{\Sigma} \equiv \partial_t \mathsf{\Sigma} + \mathbf{v} \cdot \nabla \mathsf{\Sigma} + \boldsymbol{\omega} \cdot \mathsf{\Sigma} - \mathsf{\Sigma} \cdot \boldsymbol{\omega}$, is the co-moving, co-rotating derivative with $\boldsymbol{\omega} = (\nabla\mathbf{v} - (\nabla\mathbf{v})^{\mathsf{T}})/2$ being the vorticity tensor, and $\mathsf{\Sigma}_{\rm f}$ is the hydrodynamic fluid stress defined in \eqref{eq:isotropic_fluid_stress}. Typical viscoelastic materials at the cellular scale include the actomyosin cortex (see section \ref{sec:actomyosin_cortex}) and at the larger scale, entire tissues (see section \ref{sec:tissues}). The Maxwell relaxation time for the actomyosin cortex is of the order of a second~\cite{saha2016determining} while that for axons is a few tens of seconds~\cite{bernal2007mechanical}.

The response of a macroscopic system to external perturbations provides insights into its structure and stability. When the magnitude of applied perturbations is comparable to the strength of spontaneous fluctuations in the system, linear response theory provides a good approximation to calculate these responses. Remarkably, linear response theory also connects these response functions to correlation functions of spontaneous fluctuations, as we discuss below. 

\subsection{Linear response}
\label{sec:linear_response}

A fundamental characteristic of many-body systems at (or close to) thermal equilibrium is the relationship between spontaneous fluctuations (as measured by two-point correlation functions, for instance) and the response of the system to the application of a weak external time-dependent perturbation (as measured by response functions). This connection, called Onsager's regression hypothesis, is codified by linear-response theory and leads to the fluctuation-dissipation theorem~\cite{onsagerReciprocalRelationsIrreversible1931,onsagerReciprocalRelationsIrreversible1931a, de2013non, kubo2012statistical}. 

\begin{figure}
\centering
\includegraphics[width=\linewidth]{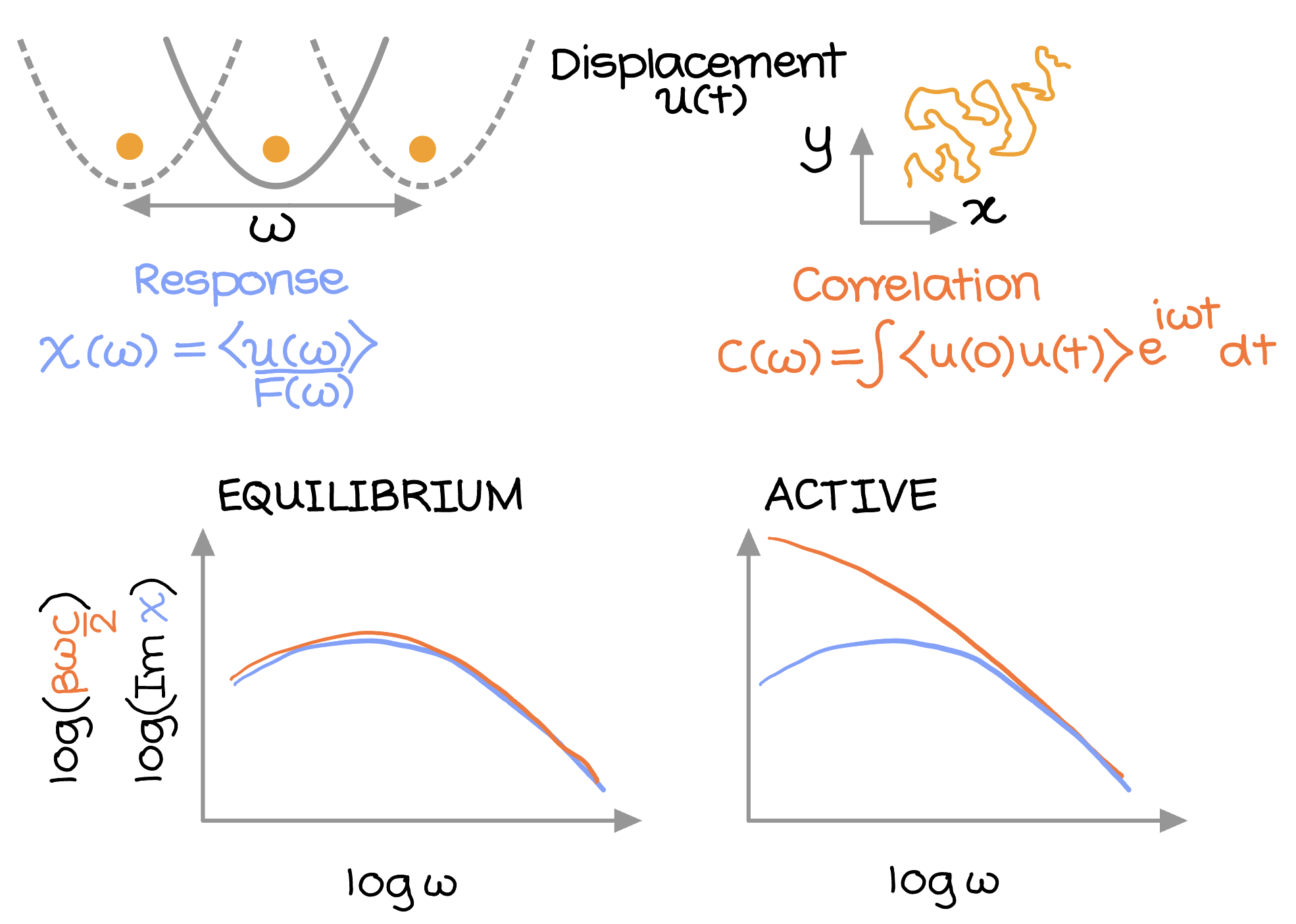}
\caption{The reaction of a macroscopic system to an applied infinitesimal force is captured by generalized response functions $\chi(\omega)$ which depend on the frequency $\omega$ of the applied force $F(\omega)$. For instance, the response function for a Brownian particle can be measured as the ratio of the average displacement $\langle u(\omega) \rangle$ to the applied force (for instance, in an optical trap); in the notation of the main text, $\chi(\omega) \equiv \chi_{uu}(\omega)$ is the displacement response. Spontaneous fluctuations of the displacement $u$, in the absence of external forces, are captured by the Fourier-transform $C(\omega)$ of the two-time correlation function $\langle u(0) u(t) \rangle$. 
For a system at thermodynamic equilibrium, the fluctuation-dissipation theorem \eqref{eq:fdt1} relates the response and correlation functions at all frequencies, a consequence of linear-response theory. For an active system, no such relation exists in general, and the response and correlation functions can be decoupled from each other. 
However, they typically agree at high frequencies, above the inverse of the characteristic timescales of the active processes (e.g.\ motor binding and stroke cycles), where thermal fluctuations dominate. Note that, in principle, either or both the response functions and the correlation functions can deviate from their equilibrium behaviour in the presence of activity.}
\label{fig:correlation_response}
\end{figure}

One way to characterize the behaviour of biological systems is to apply a weak perturbation and study their response. For example, tracking the motion of a small colloidal particle provides insights into the nature of its environment. This can be done in the cytoplasm or the cytoskeleton. For instance, in reconstituted \textit{in-vitro} networks, measuring the deviation between the correlation and response functions provides a characterization of the nonequilibrium nature of the system \cite{mizunoNonequilibriumMechanicsActive2007} as shown in FIG.~\ref{fig:correlation_response}.

Empirically, the response of a system to applied perturbations is probed by measuring the (time-dependent) relaxation of a macroscopic quantity $A$ to its equilibrium value $A_{\rm eqm}$. Consider a generalized force $\mathcal{F}_B$, that couples to another macroscopic quantity $B$, applied on the system. Specifically, we assume that the force $\mathcal{F}_B$ enters the microscopic energy in the form $-\mathcal{F}_B \cdot B$, i.e., we assume that $\mathcal{F}_B$ is the thermodynamic variable conjugate to $B$. If the applied force does not significantly alter the macroscopic thermodynamic equilibrium state of the system, the change $\Delta A(t) \equiv A(t) - A_{\rm eqm}$ is linearly related to the applied force. If $\mathcal{F}_B$ is switched on adiabatically, i.e, much slower as compared to the molecular relaxation time-scales, from $t=-\infty$, then
\begin{align}
\Delta A(t) = \int_{-\infty}^t dt' \, \chi_{AB}(t-t') \; \mathcal{F}_B(t'),
\end{align}
where the response function $\chi_{AB}(t-t')$ captures the coupling between $A$ and $B$. Note that the response $\Delta A(t)$ is, in general, not in phase with the applied force $\mathcal{F}_B(t)$. Causality implies that $\chi_{AB}(t-t')$ is non-zero only for $t>t'$. 

Spontaneous thermal fluctuations of macroscopic quantities are typically captured via two-point correlation functions. For instance, the statistics of the time-dependent fluctuations of macroscopic quantities $A$ and $B$ around their equilibrium values $A_{\rm eqm}$ and $B_{\rm eqm}$ is captured by the correlation function
\begin{align}
C_{AB}(t-t') = \langle \Delta A(t) \Delta B(t')\rangle
\end{align}
where $\langle \cdots \rangle$ denotes an equilibrium average. For a single classical observable ($B = A$), linear response theory connects this correlation function with the response function $\chi_{AA}(t-t')$ via the fluctuation-dissipation relation, which in the time domain reads $\chi_{AA}(t) = -\beta\,\Theta(t)\, dC_{AA}(t)/dt$, where the Heaviside step function $\Theta(t)$ expresses causality, and, in Fourier space [with $f(\omega) = \int dt\, e^{i\omega t} f(t)$],
\begin{align}
\mathrm{Im}[\chi_{AA}(\omega)] = \frac{\omega}{2k_BT} \, C_{AA}(\omega),
\label{eq:fdt1}
\end{align}
where $\mathrm{Im}[\chi_{AA}(\omega)]$ is the imaginary, dissipative part of the response. Systems that are driven out-of-equilibrium by external energy inputs violate this fluctuation-dissipation relation~\eqref{eq:fdt1}~\cite{kubo2012statistical}. Recent years have seen several attempts to generalize linear response theory to situations where the reference state is a non-equilibrium steady state (NESS)~\cite{prostFD2009, marconi2008fluctuation}.

So far, we have considered the response and correlation functions of spatially homogeneous systems that are close to thermodynamic equilibrium. For a system driven out of equilibrium, we do not, as yet, have a generic framework to discuss its thermodynamics. However, a useful approximation is that of local thermodynamic equilibrium. In other words, a spatially heterogeneous system at macroscopic length-scales can be thought to be made up of interacting mesoscopic elements each of which are in \emph{local thermodynamic equilibrium}. These mesoscopic regions reach local equilibrium over time-scales that are much smaller compared to the relaxation time of the entire system. An example of this hydrodynamical approach was discussed in the previous section when we considered the macroscopic description of a simple Newtonian fluid. This framework of generalized hydrodynamics, when extended to complex materials, has turned out to be very fruitful in the description of biological systems from cellular to tissue scales~\cite{marchettiHydrodynamicsSoftActive2013, julicherHydrodynamicTheoryActive2018}. Finally, there is no {\it a priori} justification for the linear response assumption in far-from-equilibrium steady-states. However, in practice, this assumption appears to provide a reasonable description of many biophysical systems.

\subsection{Onsager relations and generalized hydrodynamics}

A systematic way to derive the hydrodynamic equations for the coarse-grained description of a many-body classical system is to develop its linear irreversible thermodynamics. A system at equilibrium is in a state of maximum entropy $S$. Any perturbations in a \emph{generalized displacement} variable $u$ that takes the system away from this state of maximum entropy are countered by a \emph{generalized force} $F$ which tends to restore the equilibrium state. In the ensuing process, the rate at which entropy is generated is proportional to the product of this generalized force and the flux of the generalized displacement $\dot{u}$. For several generalized displacements $u_i$ and corresponding generalized forces $F_i$, at uniform temperature $T$ the internal entropy production rate satisfies
\begin{equation}
 T{dS \over dt} = \sum_i F_i\dot{u}_i.
\end{equation}
If the system is close to equilibrium, the generalized fluxes can be expanded as
a linear combination of the generalized forces, i.e., $\dot{u}_i = \sum_j L_{ij} F_j$, where $L_{ij}$ are called Onsager coefficients. 

Time-reversal invariance constrains the symmetry of $L_{ij}$. For instance, in the case of the simple isotropic fluid discussed earlier, choosing the dissipative stress tensor $\mathsf{\Sigma}_{\rm f} + p\mathbb{I}$ as the generalized force, and the strain rate $\mathsf{E}$ as the generalized flux, gives Onsager coefficients proportional to inverse of viscosities $\eta$ and $\eta_b$. In the case of systems with chemical reactions, the differences in chemical potentials are the generalized forces and the rate of chemical reaction are the fluxes. In general, the chemical and mechanical degrees of freedom are coupled to each other leading to mechanochemical effects. Non-equilibrium chemical reactions within cells can generate active stresses (see section \ref{sec:mechanochemical_transduction}).

To generalize this approach, we write the conservation laws and generic equations of motion for broken symmetry field variables. We then proceed by identifying the generalized forces and fluxes that enter the entropy production rate. Following Onsager \cite{onsagerReciprocalRelationsIrreversible1931,onsagerReciprocalRelationsIrreversible1931a} (and similar to the linear-response theory outlined above), we expand the generalized fluxes as a linear combination of the generalized forces, with appropriate considerations for their behaviour under time-reversal. These lead to the constitutive equations for the system under consideration. Finally, one identifies those terms that survive in the hydrodynamic limit, i.e., long-times and large-lengths compared to the microscopic degrees of freedom. This leads to a closed set of hydrodynamic equations with several phenomenological coefficients. One can also include stochastic components in the hydrodynamic equations as discussed above in the case of an isotropic simple fluid.

The use of this approach in biological systems is illustrated below.

\begin{figure}[ht]
\centering
\includegraphics[width=0.9\linewidth]{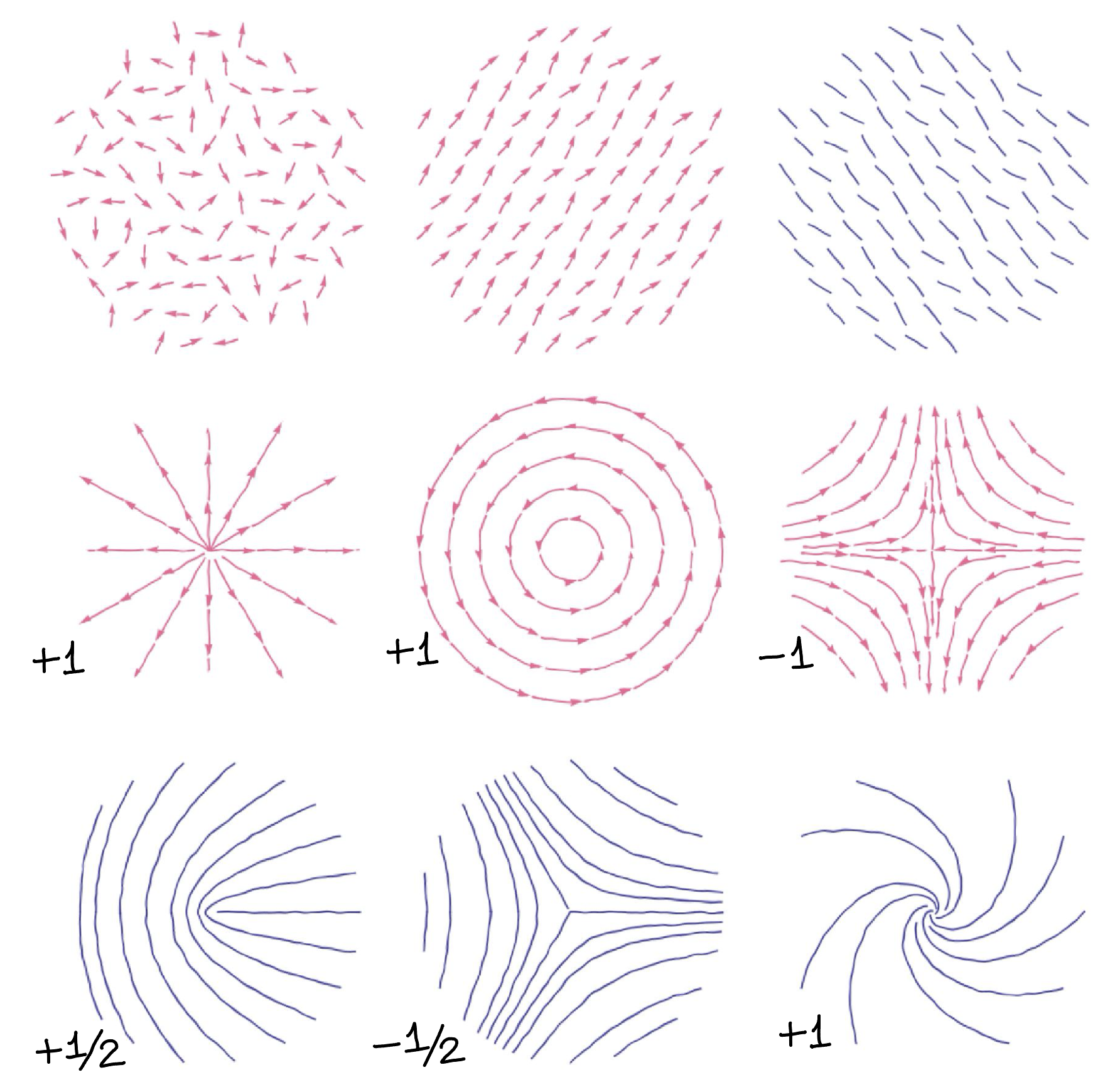}
\caption{Orientational degrees of freedom can have either polar (represented by an arrow) or nematic (represented as an undirected segment) symmetry. The top row of figures shows disordered polar, ordered polar, and ordered nematic structures. This order-disorder transition is typically controlled by temperature and/or density. In the ordered phase, topologically stable excitations appear in the form of defects in the orientational order. Each such singularity is associated with a topological charge that is obtained by measuring the change in the angle of the order parameter around a loop enclosing the defect. The middle row shows examples of $+1$ and $-1$ defects. The bottom row shows half-integer topological defects (which are peculiar to nematics) and a spiral $+1$ nematic defect.}
\label{fig:topo_defects}
\end{figure}

\subsection{Orientational order}

Anisotropic order is often seen in living systems, for example in the cytoskeleton or in cell-ordering in tissues. Understanding the statics and the dynamics of such anisotropic systems requires hydrodynamic descriptions. 
Orientational order can be of multiple kinds. Fore-aft asymmetric filaments (represented by arrows in FIG.~\ref{fig:topo_defects}) can arrange into an isotropic state or a polar state. Fore-aft symmetric filaments (represented by undirected segments in FIG.~\ref{fig:topo_defects}) can arrange into an isotropic state or a nematic state. Nematic ordering can also arise from a symmetric mixture of polar filaments which are aligned but have no net directionality. A vector order parameter $\mathbf{p}$ is relevant for describing orientation order in a polar system. On the other hand, describing local anisotropy in a nematic system requires a traceless and symmetric second-rank tensor $\mathsf{Q}$. One can derive the hydrodynamic equations for an anisotropic system with a polar order-parameter $\mathbf{p}$ and/or a nematic order-parameter $\mathsf{Q}$. 

In the case of polar liquid crystals \cite{deGennesProst}, the simple free energy functional is
\begin{align}
\mathcal{F}_{\rm{p}} &= \int d^d\mathbf{x} \bigg[ A (\mathbf{p} \cdot \mathbf{p}) + B (\mathbf{p} \cdot \mathbf{p})^2 
+ \overline{K} \: (\nabla \mathbf{p}:\nabla \mathbf{p}) \bigg],
\end{align}
where $A$ and $B$ govern the physics near the order-disorder transition. The transition from a disordered state $\mathbf{p}=0$ to an ordered state $\mathbf{p} \neq 0$ occurs at $A=0$, while stability demands $B>0$ always. The gradient term governs the orientational elasticity of the polar order parameter $\mathbf{p}$, with $\overline{K}$ an elastic modulus within the one-constant approximation. The polar order-parameter generates orientational stresses in the system. The stress-tensor therefore has an additional component 
\begin{align}
\mathsf{\Sigma}_{\rm{p}} = \frac{\nu}{2} \left[\mathbf{p} \otimes \mathbf{h} + \mathbf{h} \otimes \mathbf{p} - \frac{2}{d} (\mathbf{p} \cdot \mathbf{h}) \, \mathbb{I} \right] + \overline{\nu} (\mathbf{p} \cdot \mathbf{h}) \; \mathbb{I},\label{eq:polar_passive_stress}
\end{align}
where $\mathbf{h} = -\delta\mathcal{F}_{\rm{p}}/\delta \mathbf{p}$ is the molecular field conjugate to the polarity, and $\nu$ and $\overline{\nu}$ are the alignment parameters. The evolution equation for the order parameter $\mathbf{p}$ is $D_t \mathbf{p} = \mathbf{h}/\overline{\gamma} - \nu \, \tilde{\mathsf{E}} \cdot \mathbf{p} - \overline{\nu} \, \mathsf{Tr}(\mathsf{E}) \, \mathbf{p}$, where $\overline{\gamma}$ is a kinetic coefficient, $\tilde{\mathsf{E}}$ is the traceless part of $\mathsf{E}$ and $D_t\mathbf{p} \equiv \partial_t \mathbf{p}+ \mathbf{v} \cdot \nabla \mathbf{p} + \boldsymbol{\omega} \cdot \mathbf{p}$ is the co-moving co-rotating derivative with $\boldsymbol{\omega} = (\nabla\mathbf{v} - (\nabla\mathbf{v})^{\mathsf{T}})/2$ being the vorticity tensor. With this normalization, the flow-alignment terms in the stress and in the polarity dynamics are reactive: their contributions to the entropy production, $\mathsf{\Sigma}_{\rm p}:\mathsf{E}$ and $\mathbf{h}\cdot D_t\mathbf{p}$, cancel, as required by Onsager reciprocity~\cite{julicherHydrodynamicTheoryActive2018}. 

If the relevant order parameter is a nematic tensor $\mathsf{Q}$, then the corresponding free energy is
\begin{align}
\mathcal{F}_{\rm{n}} &= \int d^d\mathbf{x} \big[ \mathcal{A} \; \mathsf{Tr}(\mathsf{Q}^2) + \mathcal{B} \; \mathsf{Tr}(\mathsf{Q}^3) + \; \mathcal{C} \; (\mathsf{Tr}(\mathsf{Q}^2))^2
\nonumber \\
& \qquad \qquad 
+ \overline{\mathcal{K}} \; \nabla \mathsf{Q} : \nabla \mathsf{Q} \big],
\end{align}
where $\mathcal{A}$, $\mathcal{B}$, $\mathcal{C}$ govern the physics of the isotropic to nematic phase-transition, the $\overline{\mathcal{K}}$ is the orientational elasticity modulus, and the corresponding orientational stress is
\begin{align}
\mathsf{\Sigma}_{\rm{n}} = \nu_{\rm n} \mathsf{H} = -\nu_{\rm n} \frac{\delta \mathcal{F}_{\rm n}}{\delta \mathsf{Q}},
\end{align}
where $\mathsf{H}$ is the conjugate molecular field. The order parameter $\mathsf{Q}$ evolves according to $\mathcal{D}_t \mathsf{Q} = \mathsf{H}/\gamma_{\rm n} - \nu_{\rm n} \, \tilde{\mathsf{E}}$ where $\gamma_{\rm n}$ is a kinetic coefficient, $\nu_{\rm n}$ is the nematic flow-alignment parameter and the comoving, corotating derivative for a second rank tensor field is defined as $\mathcal{D}_t\mathsf{Q} \equiv \partial_t \mathsf{Q}+ \mathbf{v} \cdot \nabla \mathsf{Q} + \boldsymbol{\omega} \cdot \mathsf{Q} - \mathsf{Q} \cdot \boldsymbol{\omega}$. 

Within the simplified description used here, the total mechanical stress in an isotropic Newtonian fluid is $\mathsf{\Sigma} = \mathsf{\Sigma}_{\rm f}$, while for anisotropic fluids $\mathsf{\Sigma} = \mathsf{\Sigma}_{\rm f} + \mathsf{\Sigma}_{\rm p}$ or $\mathsf{\Sigma} = \mathsf{\Sigma}_{\rm f} + \mathsf{\Sigma}_{\rm n}$ for polar or nematic fluids, respectively. To complete the hydrodynamic description, we need to supplement these equations with appropriate boundary conditions, and, when the time-evolution is of interest, initial conditions as well.

 Complex anisotropic fluids typically show a viscoelastic response to applied mechanical perturbations. In other words, depending on the time-scales of observation, the material might show an elastic response or flow like a viscous fluid. The hydrodynamic approach can be extended to describe viscoelastic materials. For instance, a simple phenomenological extension of the linear Maxwell viscoelastic model is
$\left(1+\tau\,\mathcal{D}_t\right)
\left(\mathsf{\Sigma}-\mathsf{\Sigma}_{\rm o}+p\mathbb{I}\right)
=\mathsf{\Sigma}_{\rm f}+p\mathbb{I}$,
where $\mathsf{\Sigma}_{\rm o}$ is either $\mathsf{\Sigma}_{\rm p}$ or $\mathsf{\Sigma}_{\rm n}$ as appropriate, and the timescale $\tau$ governs the transition from an elastic behaviour at short times ($t \ll \tau$) to a fluid-like response at long time scales ($t\gg\tau$).

In an ordered medium, such as a group of aligned cytoskeletal filaments, the order parameter measures the local orientation. It is possible for this order parameter to be ill-defined at a finite number of locations. Away from these points, the order parameter must vary smoothly. This implies the existence of localized topological defects in the ordering associated with these points (FIG.~\ref{fig:topo_defects}). A topological charge, decided by the winding number of smooth curves around the defect, characterizes each such singularity. 

 For two dimensional polar ordering, the topological charges are integer-valued whereas for nematic ordering, they are $1/2-$integer valued~\cite{penrose1979topology}. The sum of the topological charges in a domain is governed by the topology and boundary conditions of the domain. Smooth changes of the orientation field or domain geometry that preserve these constraints do not change the total topological charge.  In a system at thermodynamic equilibrium, defect dynamics obeys detailed balance~\cite{chaikin2004principles}. 

In an active system, however, the dynamics of defects is influenced by their topological charges. For instance, a $+1/2$ defect is a polar structure and has a net motility in an active system~\cite{sanchezSpontaneousMotionHierarchically2012}. Such motile defects have been observed in quasi-two-dimensional actin nematics driven by myosin II motors in {\it in vitro} experiments~\cite{kumarTunableStructureDynamics2018}, in developing hydra~\cite{maroudas-sacksTopologicalDefectsNematic2021}, and also in cell collectives~\cite{saw2017}. Recent works have shown that topological defects play an important role in various morphogenetic processes. 

\subsection{Soft-materials driven out of equilibrium}

In the discussion above, our description has focused on systems that are at thermal equilibrium. For cellular processes, whether an equilibrium description is sufficient depends on the process as well as the time-scales of interest. For instance, many chemical reactions that occur on fast timescales can be understood within an equilibrium formalism~\cite{phillipsNapoleonEquilibrium2015}. However, on longer timescales, typically, cellular metabolic processes drive energy and material fluxes, thus necessitating a nonequilibrium description.

For systems with discrete states, breaking detailed balance, i.e., $W_{j\to i} \; P_j \neq W_{i \to j} \, P_i$, can lead to nonequilibrium steady-states (when $dP_i/dt=0$) with non-zero currents. Note that this is a subtle way to be out of equilibrium compared to the case when there are non-zero external forces, possibly time-dependent, acting on the system that would inject energy into the system. Systems that are force-free can nevertheless be out of equilibrium if the distribution across their internal states does not obey the Boltzmann probability.

A typical way to force a system out of equilibrium is to violate the fluctuation-dissipation theorem, i.e., consider sources of dissipation that are not related to the strength of stochastic fluctuations. 
At the hydrodynamic level, systems that are macroscopically force- and torque-free can be driven out of equilibrium in multiple ways. A novel class of nonequilibrium systems with an energy throughput, wherein the energy consumption and dissipation occur at the level of the individual units, are ``active materials'' \cite{finlayson1969convective, julicherActiveBehaviorCytoskeleton2007, prostActiveGelPhysics2015, joannyActiveGelsDescription2009, marchettiHydrodynamicsSoftActive2013, julicherHydrodynamicTheoryActive2018, doostmohammadiActiveNematics2018, ramaswamyActiveFluids2019}. 
For such systems, not requiring microscopic time-reversal invariance generically allows additional terms in the hydrodynamic equations. The active stress contributions for polar and nematic systems are
\begin{align}
\mathsf{\Sigma}^{\rm active}_{\rm p} &= \zeta \; \Delta \mu \; \mathbb{I} + \overline{\zeta} \; \Delta \mu \left[ \mathbf{p} \otimes \mathbf{p} - \frac{(\mathbf{p} \cdot \mathbf{p})}{d} \; \mathbb{I} \right]\label{eq:p_active_stress},
\\
\mathsf{\Sigma}^{\rm active}_{\rm n} &= \zeta \; \Delta \mu \; \mathbb{I} + \overline{\zeta} \; \Delta \mu \; \mathsf{Q} \label{eq:Q_active_stress},
\end{align}
where the coefficients $\zeta \, \Delta \mu$ and $\overline{\zeta} \, \Delta \mu$ result from active processes. The above expressions are obtained through considerations of symmetry for the stress tensor, while $\Delta \mu$ represents the chemical potential difference for ATP hydrolysis, reflecting the non-equilibrium character of the system. The equilibrium equations of evolution mentioned earlier can be appropriately modified to include lowest order terms permitted by symmetry (without assuming time-reversal invariance). This yields
\begin{align}
D_t \mathbf{p} &= \mathbf{h}/\overline{\gamma} - \nu \, \tilde{\mathsf{E}} \cdot \mathbf{p} - \overline{\nu} \, \mathsf{Tr}(\mathsf{E}) \, \mathbf{p} + \lambda_0 \, \mathbf{p} 
\nonumber \\ 
& - \lambda_1 \, (\mathbf{p} \cdot \nabla) \mathbf{p} - \lambda_2 \, (\nabla \cdot \mathbf{p}) \mathbf{p} + \lambda_3 \, \nabla (\mathbf{p} \cdot \mathbf{p}) ,\label{eq:p_evolution}
\\[1em]
\mathcal{D}_t \mathsf{Q} &= \gamma_{\rm n}^{-1} \mathsf{H} - \nu_{\rm n} \, \tilde{\mathsf{E}} + \lambda_{\rm n} \; \mathsf{Q}\label{eq:Q_evolution},
\end{align}
where the terms with the coefficients $\lambda_i$'s and $\lambda_{\rm n}$ are active terms and must generically vanish at equilibrium. However, if $\lambda_2 = 2\lambda_3$ and $\lambda_1=0$, then such terms are permitted even in an equilibrium situation. For instance, in the case of an active polar fluid, equations~\eqref{eq:isotropic_hydrodynamic_equations} and ~\eqref{eq:p_evolution} together with \eqref{eq:isotropic_fluid_stress}, \eqref{eq:polar_passive_stress}, and \eqref{eq:p_active_stress} complete the simplified hydrodynamic description used here when supplemented with appropriate initial and boundary conditions. 

Variants of these equations were initially written down to represent the physics of self-propelled particles. However, they appear to be more generally applicable even for cellular systems, like in the coarse-grained description of the cytoskeleton \cite{prostActiveGelPhysics2015}. Similar considerations apply in the case of an active nematic material as well.

In the previous sections, we have briefly discussed the particulate and continuum descriptions, both at and away from equilibrium, as relevant to the description of soft materials found in living systems. In particular, we have focused on how mechanical forces and stresses can alter the statistical dynamics of particles and fields. In the context of living materials, active stresses typically arise in cellular contexts where the activity of molecular motors plays a dominant role. For instance, coarse-grained descriptions of the cellular cytoskeleton or, at larger scales, entire tissues, can be described in the framework of active materials discussed above. We will revisit these specific topics in section \ref{sec:tissues}. However, it is of paramount importance to be able to quantitatively measure these forces and stresses in living systems. With this in mind, we next discuss how cells infer forces and some of the key techniques developed to infer forces in the laboratory setting.

\section{\label{sec:infer_forces}How do cells infer forces?}

Cells continuously infer the mechanical state of their environment and of their own interior~\cite{katta2015feeling, freund2012fluid}. The molecular basis of this inference is that the conformations and binding affinities of many proteins, and the gating of membrane channels, depend on the forces applied to them~\cite{roca2017quantifying}. In the following we organize cellular force sensing around its underlying physics (force-dependent reaction kinetics) and then follow the mechanical path from the membrane, through adhesions and the cytoskeleton, to the nucleus.

A single picture underlies most of what follows: a mechanosensor is a two-state system whose energy landscape is tilted by force. If the two states are separated by a displacement $\Delta x$ along the loading axis, an applied force $f$ shifts their free-energy difference from its zero-force value $\Delta G_0$ to $\Delta G_0 - f\,\Delta x$, so the switch flips around a characteristic force $f^{*} = \Delta G_0/\Delta x$ with a sensitivity scale $k_BT/\Delta x$; for molecular displacements $\Delta x \sim 1~{\rm nm}$ this scale is a few pN, placing force sensing in the range of FIG.~\ref{fig:force_scales}. Whether the readout is the equilibrium occupancy of the two states, or their force-dependent transition rate when loading is faster than relaxation (the Bell picture below), depends only on the relevant timescales. The mechanosensitive channels, molecular switches, and adhesion bonds discussed in this section are all realizations of this tilted two-state landscape.

\subsection{\label{sec:bond_kinetics}Force-dependent bond kinetics}

The physical basis of essentially every molecular mechanosensor is that the rate of a molecular transition depends on the force applied to it. In the simplest picture, due to Bell, an applied force $f$ tilts the energy landscape along a reaction coordinate so that the unbinding rate of a bond becomes $k_{\rm off}(f) = k_0 \exp(\beta f \, x)$, where $x$ is the distance from the bound state to the transition state~\cite{bell1978models}. This is the Kramers--Bell description of a \emph{slip bond}, whose lifetime decreases with load. The associated force scale, $k_BT/x$, is of order a few pN for $x \sim 1~{\rm nm}$, placing molecular force sensing squarely in the range shown in FIG.~\ref{fig:force_scales}.

Not all bonds slip. \emph{Catch bonds} have lifetimes that initially \emph{increase} with applied force before eventually decreasing, as first demonstrated directly for selectin--ligand complexes~\cite{marshall2003direct}. Catch-bond behaviour requires more than one bound configuration, or a force-dependent switch between binding modes, and provides a generic mechanism by which adhesions can stabilize precisely when they are loaded. The slip/catch distinction is a useful organizing principle: it determines whether a mechanosensor reinforces or releases under load, and single-molecule force spectroscopy with controlled loading rate is the experiment that discriminates between competing molecular models.

\subsection{Mechanosensitive channels and molecular switches}

The fastest mechanosensors are ion channels whose gating is controlled by membrane tension. Bacterial channels such as MscL open above a threshold bilayer tension; in animal cells, the Piezo1 and Piezo2 channels were identified as the molecular components of distinct mechanically activated cation currents~\cite{coste2010piezo}. The energetics of tension gating can be captured by a two-state model in which the open and closed states differ in their in-plane area by $\Delta A$, so that a membrane tension $\sigma_s$ biases gating by an energy $\sigma_s\,\Delta A$ (Section~\ref{sec:soft_matter_theory}). The ions admitted by such channels can in turn feed back on force generation: calcium entry through stretch-activated channels, for example, activates myosin-II and hence cortical contraction~\cite{Salbreux_2007}.

A complementary class of sensors are proteins that act as force-gated conformational switches. The clearest example is talin, which links integrins to the actin cytoskeleton: stretching single talin rods at forces from a few to a few tens of pN unfolds rod domains and exposes cryptic binding sites for vinculin~\cite{delrio2009stretching}. The subsequent binding of vinculin reinforces the linkage, a molecular realization of a load-stabilized switch. Genetically encoded tension sensors based on F\"orster resonance energy transfer (FRET), in which the efficiency of energy transfer between two fluorophores reports their nanometre-scale separation, have made the forces borne by such molecules directly measurable \emph{in vivo}, revealing piconewton-scale tension across vinculin in focal adhesions~\cite{grashoff2010measuring}. More generally, because any protein changes conformation under load, force can alter its binding specificity and enzymatic activity, so that mechanosensitivity is a generic, rather than exceptional, property of cellular proteins.

\subsection{Focal adhesions, the molecular clutch, and rigidity sensing}

Adherent and crawling cells exert and sense forces through focal adhesions, in which integrins couple the extracellular matrix to the actin cytoskeleton via adaptors such as talin and vinculin. The \emph{molecular clutch} model provides a framework for force transmission here~\cite{chanTractionDynamicsFilopodia2008a}: retrograde actin flow, driven by polymerization and myosin contractility, is transmitted to the substrate only when transient adhesion bonds (``clutches'') engage. Coupling the load-dependent unfolding of talin to such a model reproduces the observed biphasic dependence of traction on substrate stiffness, with a threshold above which talin unfolds, vinculin binds, and the transcriptional regulators YAP/TAZ enter the nucleus~\cite{eloseguiartola2016mechanical, dupont2011role}. The predicted tractions are directly comparable with traction-force microscopy (Section~\ref{sec:traction})~\cite{Schwarz2013, schwarz2015traction}, making this a rare case where a non-equilibrium model can be tested against spatially resolved force data.

\subsection{Force transmission to the nucleus}

Mechanical signals are also transmitted physically to the nucleus. The LINC (\emph{linker of nucleoskeleton and cytoskeleton}) complex spans the nuclear envelope, with SUN (Sad1/UNC-84 homology) proteins in the inner membrane binding nesprins in the outer membrane to connect the cytoskeleton to the nuclear lamina~\cite{crisp2006coupling}, whose lamin intermediate filaments set much of the nuclear stiffness. Forces applied at the cell surface can thereby deform the nucleus and stretch chromatin directly, up-regulating transcription on physiological timescales~\cite{tajik2016transcription}; this motivates viewing the nucleus as a mechanosensor in its own right~\cite{kirby2018emerging}. The coupling of chromatin organization to force is developed in Section~\ref{sec:chromatin}.

A theme running through this section is that both force sensing and the representation of information are constrained by geometry. A single bond or switch reports only a scalar, the force component along its own axis; resolving the \emph{tensorial} character of a load (its principal directions, or shear versus isotropic stress) requires extended or oriented sensors such as arrays of adhesion bonds, aligned stress fibres, or the ciliary bundles of hair cells, in which drag displaces the bundle to modulate ion-channel opening~\cite{reichenbachPhysicsHearingFluid2014}. Whether this same geometric constraint also governs how forces carry information is, we suggest, a question worth pursuing. Biochemical signalling acts through scalar concentrations, whereas forces are tensorial; we conjecture that for a force to convey directional information it must act through an extended structure (the cytoskeleton, a transmembrane assembly, or a membrane) capable of converting scalar, binding-driven signals into oriented contractility and deformation.

\section{Measuring forces in the laboratory}
\label{sec:measurement_techniques}

\begin{figure*}
\includegraphics[width = 0.85\textwidth]
{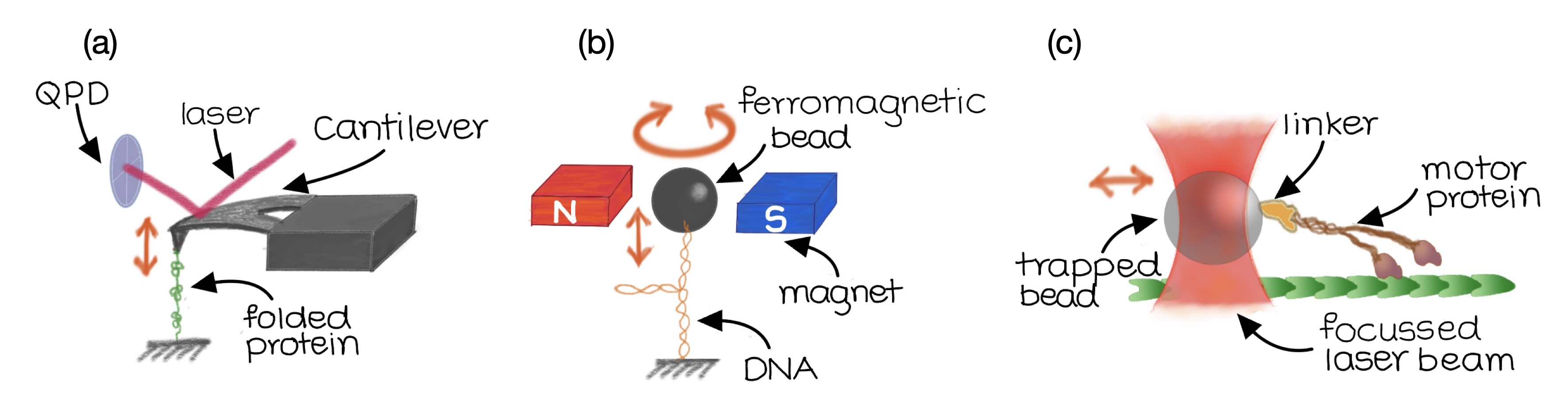}
\caption{Schematic diagrams showing some of the popular techniques used for measuring forces at the molecular scale. (a) An Atomic Force Microscope uses a calibrated triangular cantilever having a pointed tip. The cantilever base is attached to a piezoelectric translator. The deflection of the cantilever is detected using a laser beam which is reflected from the cantilever tip and a quadrant photodiode (QPD) --a device that can provide position information-- as shown. Force-extension curves of proteins like spectrin, for example, can be investigated using this tool. (b) Magnetic Tweezers use a micron-sized ferromagnetic bead to tug or twist molecules like DNA. One end of the DNA strand is attached to a surface and the other is linked to the bead. A permanent magnet can then be used to apply forces by displacing the magnets vertically, or torques by rotating the magnets. The resulting displacements are measured using video microscopy. (c) Optical tweezers use a focused laser beam to trap micron-sized dielectric spheres. The trapped sphere can then be used as a force sensor by calibrating the trapping potential. Several studies of piconewton-level force generation and nanometre-sized stepping of  molecular motors, for example, have been performed using this technique.
}
\label{fig:expt_AFM_tweezers}
\end{figure*}

Experimental information can be derived either from the living cell itself (\emph{in vivo}), from cytoplasmic material extracted from cells (\emph{cell extracts}), or from purified and reconstituted constituents (\emph{in vitro}). A detailed understanding of the mechanisms of force generation and viscoelastic responses in a cellular context requires quantitative measurements of these processes at different scales, ranging from the single molecule level to cells to tissues. 

Mechanical forces at the single molecule level, such as the stepping of single motor protein or elasticity of isolated biopolymer filaments, are intrinsically stochastic in nature. We can only infer a probability distribution of the force from a time-series measurement. At the sub-cellular or single cell level, in addition to stochasticity that can arise from thermal and active processes, spatial heterogeneity can influence mechanical measurements. Several techniques have been developed over the last several decades to overcome these challenges. 

How cells themselves sense such forces (through force-dependent bond kinetics, mechanosensitive channels, and adhesion complexes) was taken up in Section~\ref{sec:infer_forces}; the focus here is on measuring forces in the laboratory.

Techniques for measuring mechanical forces in living matter can be classified based on whether the experimentally measured quantities are macroscopic averages or fluctuations about them. A further classification of the first type is based on responses that are purely mechanical (for example, flows, displacements, strains) versus those that are mechanochemical (for example, reaction rates, chemical currents, conformation changes). Even where such macroscopic responses vanish on average, it is still possible that non-trivial fluctuations can lead to statistical forces which can have observable consequences (for example, enhanced mean-square displacements and depletion forces). Most experimental techniques have an underlying model that relates the measured response with the applied force. 

A full description of these techniques and their many applications is beyond the scope of this article, but details can be found in Refs.~\cite{pullarkat2007rheological, AHMED20153083, wu2018comparison, rodriguez2013review}.  Below we provide the basic operational principles of a few commonly used methods which are summarized pictorially in Figures \ref{fig:expt_AFM_tweezers}--\ref{fig:expt_aspiration}. Table~\ref{tab:techniques} summarizes these methods: the observable each records, the force range it accesses, the model that converts observable into force, and whether it perturbs the sample.

\begin{table*}
\renewcommand{\arraystretch}{1.2}
\begin{tabular}{|p{0.160\linewidth}%
    |p{0.195\linewidth}%
    |p{0.183\linewidth}%
    |p{0.254\linewidth}%
    |p{0.160\linewidth}|}
\hline
\textbf{Technique} & \textbf{Primary observable} & \textbf{Force range} &
\textbf{Signal-to-force model} & \textbf{Invasiveness} \\
\hline
Atomic force microscopy &
Cantilever deflection &
$10$~pN--$10$~nN (to $\sim 100$~nN for tissue indentation) &
Calibrated cantilever (Hooke's law) &
Contact \\ 
\hline
Magnetic tweezers &
Bead displacement / rotation &
$0.01$~pN--$10$~nN ($+$ torque) &
Field-gradient force calibration &
Attached bead \\ 
\hline
Optical tweezers &
Bead displacement in trap &
$0.1$--$100$~pN &
Harmonic trap stiffness &
Non-invasive / attached bead \\ 
\hline
Passive microrheology &
MSD / fluctuation spectrum of embedded probe &
Sub-pN &
Generalized Stokes--Einstein / FDT &
Embedded probe \\ 
\hline
Active microrheology &
Driven response $\chi(\omega)$ &
$\sim$pN--nN &
Response function; FDT comparison &
Probe $+$ forcing \\ 
\hline
Shear rheology &
Bulk stress vs.\ strain; $G'(\omega), G''(\omega)$ &
Moduli $10^{-1}$--$10^{3}$~Pa &
Continuum constitutive relation &
Bulk (\textit{in vitro}), cell monolayer, tissue \\ 
\hline
Traction force microscopy / micropillars &
Substrate-bead displacement / pillar deflection &
$1$--$10$~nN per focal adhesion site &
Elastic inverse problem / pillar bending &
Gel/micropillar substrate \\ 
\hline
Micropipette aspiration &
Aspirated tongue length vs.\ $\Delta P$ &
Tension $10^{-2}$--$10$~mN\,m$^{-1}$ &
Young--Laplace law &
Whole-cell \\ 
\hline
Membrane tether pulling &
Tether force plateau &
$\sim 10$~pN &
$\sigma_{s} = f^{2}/(8\pi^{2}\kappa)$ &
Local, attached bead \\ 
\hline
Fluorescence dyes &
Spectral shift / fluorescence lifetime &
Relative only &
Lipid-packing proxy; requires independent calibration &
Molecular, \textit{in vivo} \\ 
\hline
FRET tension sensors &
FRET efficiency &
$\sim 1$--$10$~pN &
Calibrated elastic-linker probe &
Molecular, \textit{in vivo} \\ 
\hline
Laser ablation &
Recoil velocity of severed structure &
Relative tension &
Viscoelastic recoil model &
Destructive (local) \\ 
\hline
Deformable droplets &
Droplet shape anisotropy &
Stress $0.1$--$3$~kPa &
Laplace / interfacial tension &
Inserted \textit{in vivo} \\ 
\hline
\end{tabular}
\caption{Common laboratory techniques for measuring cellular forces: the primary observable, the order-of-magnitude force range accessed in experiments discussed here, the model that converts the observable into a force, and whether the method perturbs the sample. Ranges depend on implementation.}
\label{tab:techniques}
\end{table*}

\subsection{Atomic force microscopes (AFM)} 
AFM is a popular technique that uses a calibrated cantilever to measure forces~\cite{vinckier1998measuring} (FIG.~\ref{fig:expt_AFM_tweezers}a). The force experienced by the cantilever is measured by reflecting a laser beam from the tip of the cantilever and measuring its deflection using a quadrant photodiode--a device that can provide position information. A feedback loop allows for control of either force or elongation. AFMs have been used to measure ligand-receptor binding~\cite{lee1994sensing}, force-extension behaviour of single protein molecules~\cite{kellermayer1997folding}, mechanics of reconstituted actin networks~\cite{parekh2005loading}, cell and tissue mechanics~\cite{ENGLER2007521, cartagena2016actomyosin}. Typically, in biological systems, AFM can be used to measure forces ranging from $\sim 10~{\rm pN} - 10~{\rm nN}$.

\subsection{Magnetic tweezers} 
Magnetic tweezers use paramagnetic or ferromagnetic beads/particles, which are $\sim 100~{\rm nm} - 5~{\rm \mu m}$ in size. When placed in an external magnetic field gradient, they can be used to apply forces on single molecules or cells~\cite{kollmannsberger2007bahigh} (see FIG.~\ref{fig:expt_AFM_tweezers}b). 
Bead displacements are measured using the methods discussed below (Section~\ref{sec:microrheology}). When a ferromagnetic bead is used, torques can be applied by rotating the external magnetic field. The use of electromagnets allows for precise control of  applied forces and also for their modulation in time. 

Forces in the range of $\sim 0.01~{\rm pN} - 10~{\rm nN}$, a much higher range than provided by optical tweezers, can be achieved by this method. Magnetic tweezers have been used in experiments ranging from measuring the elasticity of DNA strands~\cite{strick1996elasticity, sarkar2016guide}, single protein force-extension relation~\cite{chakraborty2023single, eckels2019mechanical}, and measurements of the local  viscoelasticity of the cell cortex~\cite{berret2016local}.

\subsection{\label{sec:optical_traps}Optical tweezers}
Forces generated by molecular motors, single filament polymerisation forces, ligand-receptor interaction forces, etc., are of the order of a few pN. They are most conveniently measured using optical tweezers (also known as laser tweezers or optical trap)~\cite{bustamante2021optical} (see FIG.~\ref{fig:expt_AFM_tweezers}c). This technique uses a tightly focused laser beam to obtain a 3-dimensional trap for micron-sized dielectric beads~\cite{neuman2004optical}. The trapping is effected by the momentum transferred  to the bead from the light beam as it gets refracted through the bead. The stiffness of the trap is usually measured from the power-spectrum of the trapped bead, which can be approximated as a Brownian particle in a harmonic potential. Any external force acting on the bead is measured by monitoring its displacement from the centre of the trap using a quadrant photodiode. One major advantage of this method is that it is non-invasive and can even be used to trap beads/organelles within a cell without mechanically penetrating the cell~\cite{soppina2009tug}. Forces in the range of $\sim 0.1~{\rm pN} - 100~{\rm pN}$ can be measured using this method.

\subsection{\label{sec:microrheology}Single molecule tracking and microrheology}

\begin{figure}[ht]
\includegraphics[width=0.8\linewidth]
{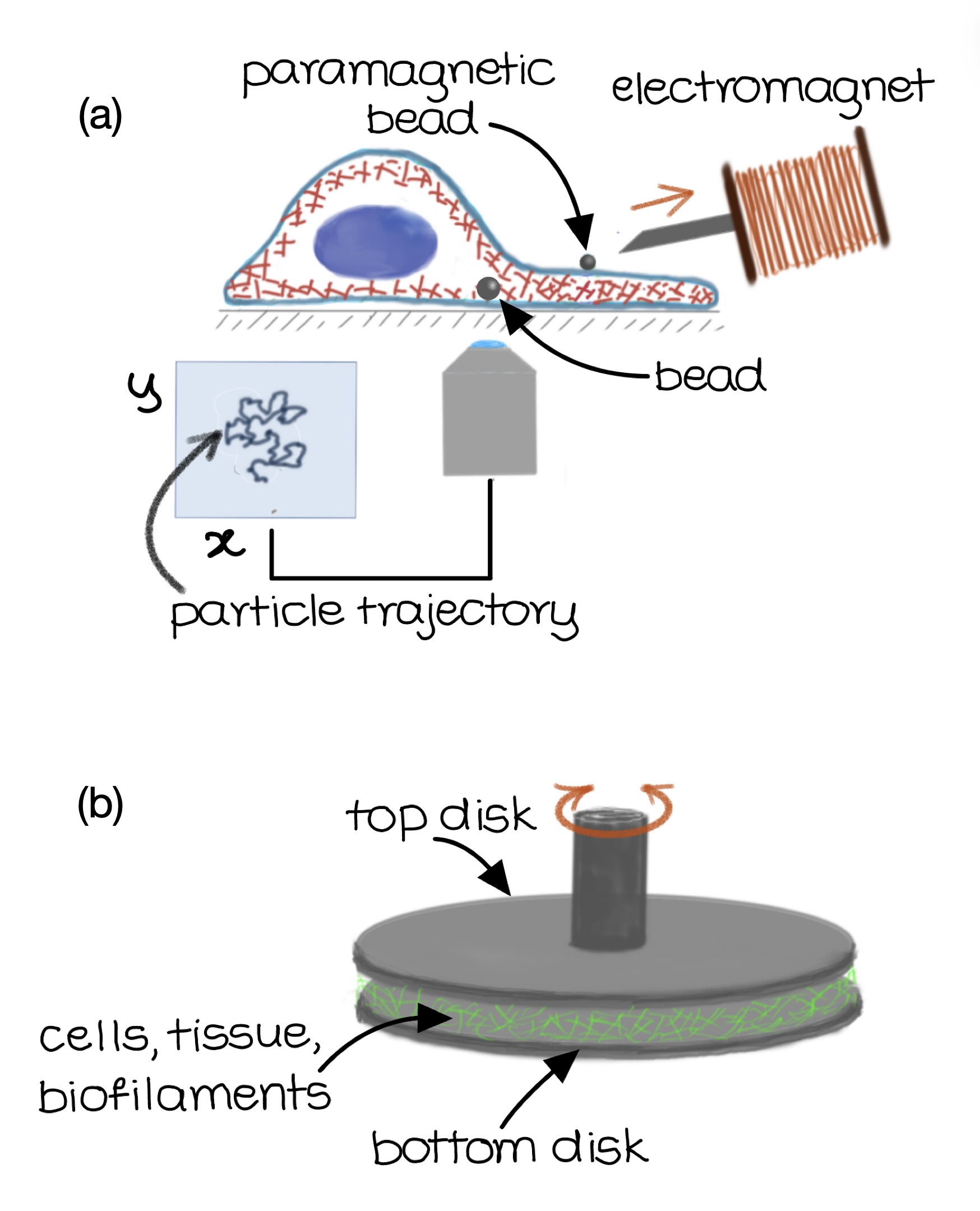}
\caption{Schematic diagrams showing the basic principles behind some techniques used for quantifying mechanical responses of cells or tissues at a mesoscopic scale. (a) Local viscoelastic properties can be explored using either a passive particle embedded in the cell or using a magnetic bead pulled using an electromagnet. In passive microrheology, spontaneous fluctuations are monitored using video-microscopy and the data is analysed to extract material properties. In live cells, this technique is sensitive to the active component of the noise, and the analysis should take this into consideration. On the other hand, in active microrheology, a force is applied to a paramagnetic bead attached to the cell using an electromagnet and the resulting response is analysed. (b) Shear rheology is a popular method used to study viscoelastic properties of {\it in vitro} biopolymer gels.
}
\label{fig:expt_tracking_microrheometer}
\end{figure}

Single molecule tracking is a simple yet powerful technique which relies on the measurement of the trajectory of a microscopic particle (for example, a fluorescent molecule, a microscopic bead or a tiny organelle) (FIG.~\ref{fig:expt_tracking_microrheometer}a). Tracking is performed by video time-lapse microscopy and subsequent image analysis \cite{AHMED20153083}. The spatial resolution for tracking an isolated particle is not limited by the resolving power of a light microscope, which is approximately $0.5\;\mu$m, but rather by the accuracy with which the intensity distribution of the particle can be sampled. Spatial resolutions of about $1~{\rm nm}$ and temporal resolutions of about $100$~Hz can be achieved. For spatially confined particles, like microscopic beads within a cytoskeletal meshwork, tracking at frequencies of the order of a few tens of kHz can be done by scattering a focused laser beam from the particle and imaging the scattered light on to a quadrant photodiode~\cite{yamada2000mechanics}. This allows for the study of frequency dependent responses.

Particle tracking can be used to infer the diffusivity of molecules in membranes by measuring the Mean Square Displacement (MSD) averaged over many realizations~\cite{wilhelm2008out, wirtz2009particle}  (FIG.~\ref{fig:expt_tracking_microrheometer}a). For systems at equilibrium, this technique is also used to measure viscoelastic properties of {\it in vitro} and {\it in vivo} cytoskeletal gels. While this method is fairly straightforward for systems at equilibrium, data from active structures such as the cytoplasm, the actomyosin cortex or the nucleus has to be interpreted with caution~\cite{mizunoNonequilibriumMechanicsActive2007}. Tracking two or more beads simultaneously allows for the measurement of correlation functions, providing information on active stress fluctuations within the gel/cortex~(Section \ref{sec:soft_matter_theory})~\cite{koenderink2006high}. 

The method described in the above paragraph, called passive microrheology, provides information about the spontaneous fluctuations in the medium. In contrast, response properties can also be inferred by subjecting the probe particle to external forcing, e.g., using optical or magnetic tweezers described above~\cite{ayala2016rheological, puig2001measurement}(FIG.~\ref{fig:expt_tracking_microrheometer}a). For a system close to thermodynamic equilibrium, the fluctuation-dissipation relation connects the response function to the measured correlation function (see FIG.~\ref{fig:correlation_response} and section~\ref{sec:linear_response}). However, for an active system, the correlation of the spontaneous fluctuations need not be connected to the response function. In fact, the deviation between these two quantities can serve as a measure of the non-equilibrium nature of the system, provided that instrumental and statistical artefacts have first been excluded~\cite{mizunoNonequilibriumMechanicsActive2007}.

This inference requires care, however, since an apparent breakdown of the fluctuation-dissipation relation is not sufficient evidence of cellular activity. The relation holds for a single, stationary, system, whereas the correlation and response functions are here obtained from two separate measurements (passive tracking and external forcing), often on different probes with independent calibration. Slow drift or ageing (nonstationarity), averaging over a heterogeneous population of local environments, and finite trajectory length or bandwidth can each mimic a violation in the absence of any genuine local activity. Several controls make the inference robust: checking that the two functions coincide at high frequencies, where fast degrees of freedom thermalize; verifying stationarity and, where possible, using the same probe for both measurements; and confirming that depleting ATP or inhibiting motors restores the equilibrium relation~\cite{mizunoNonequilibriumMechanicsActive2007, guo2014probing}.

\subsection{Shear rheology}
Shear rheology using rotating disc-shaped plates is a standard technique to probe the viscoelasticity of soft materials and complex fluids~\cite{macosko1994rheology}. In its simplest configuration, the sample is sandwiched between two circular discs, $\approx 5$ cm in diameter. Shear deformations are applied by the angular displacement (continuous rotation or oscillations) of one of the discs and the resulting shear stress measured~(FIG.~\ref{fig:expt_tracking_microrheometer}b). A feedback mechanism allows for either shear strain or shear stress controlled measurements. When small amplitude, sinusoidal shear stress (or strain) of frequency $\omega$ is applied to a viscoelastic material, the response is captured by a complex shear modulus $\chi_s(\omega) = G'(\omega) + i \, G''(\omega)$, whose real part $G'(\omega)$ is the storage (elastic) modulus and imaginary part $G''(\omega)$ the loss (viscous) modulus (see section~\ref{sec:linear_response}). The stress-to-strain amplitude ratio $|\chi_s|$ and phase $\delta$ with $\tan\delta = G''(\omega)/G'(\omega)$  are related to the material properties, with $\delta = 0$ for a purely elastic material and $\delta = \pi/2$ for a purely viscous material~\cite{fung2013biomechanics}. Frequency dependent responses, and non-linear responses like shear thickening or shear thinning can also be investigated using this method. 

Shear rheology has been used extensively to study reconstituted biopolymer networks~\cite{storm2005nonlinear} with or without associated proteins like crosslinkers or motor proteins~\cite{humphrey2002active, koenderink2009active}. It can also be used to study the rheology of cell monolayers in culture~\cite{fernandez2007shear} and has been used widely to study tissue mechanics~\cite{fung2013biomechanics}.

Both passive and active biopolymer networks are strongly nonlinear, a regime the linear moduli above do not capture. Semiflexible networks stiffen under increasing stress (the differential modulus grows because the force-extension response of an individual filament is itself nonlinear, as noted earlier for the worm-like chain) and they soften, yield and shear-thin at large strains or strain rates as crosslinkers unbind and filaments buckle or rupture~\cite{gardel2004elastic, storm2005nonlinear, broedersz2014modeling}. In active gels the internal stress generated by myosin motors acts much like an externally applied prestress, stiffening the network in an ATP-dependent manner~\cite{koenderink2009active, humphrey2002active}. Separating this active contribution from the intrinsic material response is a recurring experimental challenge: driven (active) microrheology and macroscopic shear measure a response dominated by the passive network, whereas passive microrheology additionally registers the motor-driven fluctuations, so that comparing the two (or, more simply, differencing measurements with and without ATP or motor activity) isolates the active part~\cite{mizunoNonequilibriumMechanicsActive2007}.

\begin{figure}
\includegraphics[width = 0.6\linewidth]
{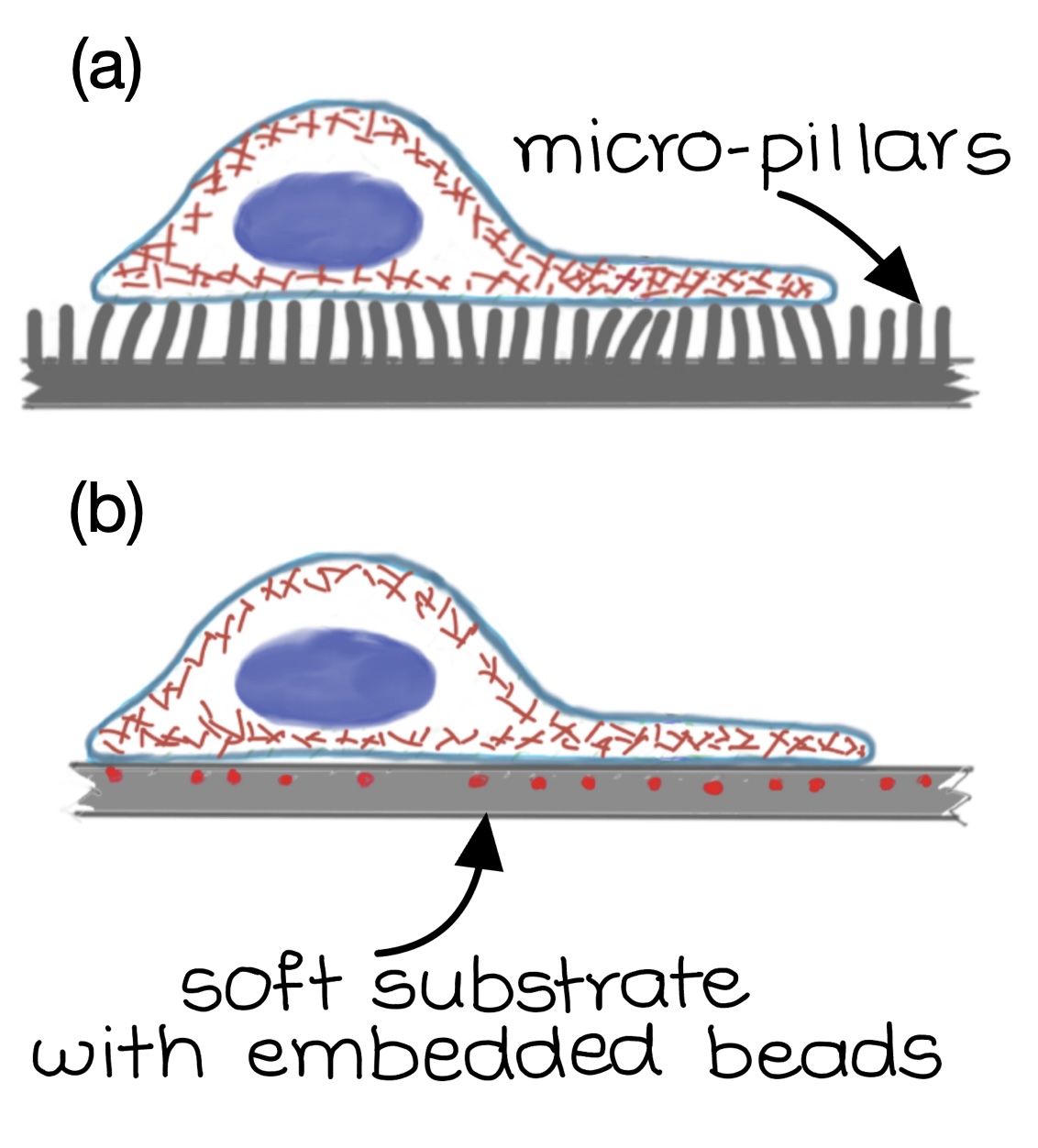}
\caption{Schematic diagrams showing the basic principles behind common experimental techniques used for measuring traction forces exerted by cells on a substrate. (a) The micropillar technique uses an array of micro-fabricated pillars which act as cantilevers. These pillars can be calibrated and the spatial distribution of forces can be inferred. (b) In traction force microscopy, cells are allowed to spread on soft elastic substrates which contain dispersed fluorescent beads (red dots). The displacements of the beads due to traction forces exerted by a cell can be mapped using a microscope and the corresponding stress field calculated from this.
}
\label{fig:expt_micropillars_tractionForce}
\end{figure}
\subsection{\label{sec:traction}Measurement of cell-substrate traction forces}

Traction forces (force per unit area) exerted on a substrate by adherent or crawling single cell, or a group of cells, can be measured using various microscopy techniques~\cite{dembo1999stresses, schwarz2015traction}. In the \textit{micropillar method} (see FIG.~\ref{fig:expt_micropillars_tractionForce}a) cells are grown on a micro-patterned substrate consisting of a two-dimensional lattice of cylindrical pillars with each pillar typically being about a micron in thickness and about $10~{\rm \mu m}$ in height~\cite{tan2003cells}. The traction force exerted by the cell at each position is measured by measuring the deflection of individual pillars, which are calibrated in advance.

In the \textit{traction force microscopy} method (see FIG.~\ref{fig:expt_micropillars_tractionForce}b), cells are grown on a sheet of soft elastomer or gel with a layer of sub-micron fluorescent beads embedded into the top surface. Traction forces exerted by the cells on the gel surface cause the gel to deform, resulting in the displacement of the embedded beads, which is mapped out using image analysis. The surface tractions on the gel are estimated from the measured bead displacements given knowledge of the stress-strain constitutive relationship for the gel. This is an inverse boundary value problem in solid mechanics that is numerically solved using techniques such as Fourier transform and finite element analysis~\cite{schwarz2015traction}.

\subsection{Deformable droplets} Forces in the interior of three-dimensional cells and tissues, where a probe cannot easily be attached from the outside, can be read out by introducing a calibrated deformable inclusion. Biocompatible oil or ferrofluid microdroplets of known interfacial tension are injected between cells. The anisotropic part of the local stress deforms the droplet, and its shape (extracted from microscopy and inverted through the Young--Laplace relation) yields the local anisotropic stress~\cite{mongera2018fluid}. Magnetic droplets can additionally be actuated by an external field to apply a known stress and record the surrounding response~\cite{serwane2017vivo}. The method is minimally invasive and well suited to developing tissues, but it reports only the anisotropic stress, not the isotropic pressure, and requires the droplet stiffness to be matched to that of the tissue.

\subsection{Image-based force inference}
The methods above all apply or attach a calibrated probe. A complementary, non-perturbative strategy infers forces directly from images of the unperturbed system. In a confluent epithelium the geometry of the cell--cell junction network constrains the underlying mechanics. Assuming mechanical equilibrium at each vertex, the junctional tensions and intracellular pressures can be reconstructed, up to an overall scale, from segmented cell shapes: video force microscopy and Bayesian force inference are two such schemes~\cite{brodland2014cellfit, ishihara2012bayesian}, while related approaches read membrane or interface tension from local curvature. Because this inverse problem is underdetermined and rests on an assumed equilibrium and constitutive model, image-based estimates are best cross-validated against a direct measurement such as laser-ablation recoil or an embedded droplet.

\begin{figure}
\includegraphics[width = 0.7\linewidth]
{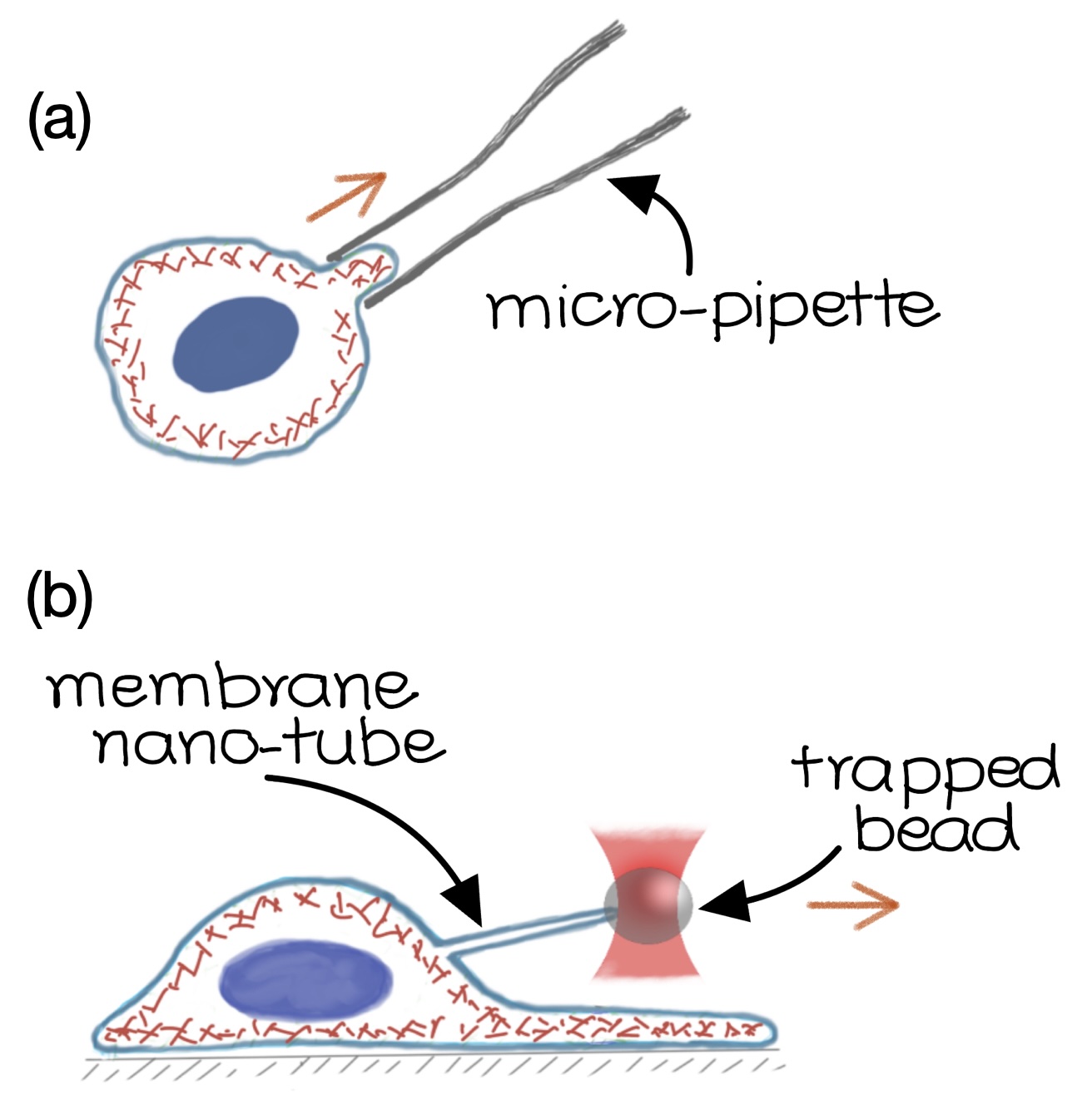}
\caption{Schematic diagrams showing the basic principles behind common experimental techniques used for quantifying mechanical responses of  cell membranes. (a) In micropipette aspiration, a negative pressure is applied to the pipette to suck in a small portion of the cell. Viscoelastic properties of the interface can be calculated using this method. 
(b) In optical tweezers measurements, a micron-sized bead attached to the membrane is pulled using an optical trap to extract a membrane nanotube. Relative changes to cell membrane tension and complex dynamical properties of the membrane can be inferred from such force measurements. 
}
\label{fig:expt_aspiration}
\end{figure}

\subsection{Measuring membrane mechanical responses}\label{sec:membrane_measurements}
Cell membrane tension (see section~\ref{sec:soft_matter_theory}) is a major regulator of processes such as endo/exocytosis, the gating of mechano-sensitive ion channels, the formation of protrusive structures like filopodia and a  number of other examples. A number of techniques have been developed to quantify membrane tension in cells. A popular method used to investigate the cell membrane, usually along with the membrane-associated cortical skeleton, is the \textit{micropipette aspiration technique}~\cite{hochmuth2000micropipette, guevorkian2017micropipette, gonzalez2019advances}. In this method (see FIG.~\ref{fig:expt_aspiration}a), a small part of the cell is aspirated into a glass micropipette whose tip diameter $2R_p$ is small compared to the size of the cell. At steady state, the effective tension $\sigma_s$ is calculated from the pressure difference $\Delta p$ required to aspirate, the radius of curvature of the end cap of the aspirated portion $R_p$, and the radius of curvature of the cell outside the pipette $R_c$, using the Laplace formula (section~\ref{sec:soft_matter_theory}): $\sigma_s = (\Delta p/2) \, (1/R_p - 1/R_c)^{-1}$. Several other applications like measurement of cortical viscoelasticity, mechanics of cell aggregates, cell-cell adhesion have also been developed based on this technique ~\cite{guevorkian2017micropipette, gonzalez2019advances}. 

Another popular method used is the \textit{tether extraction method} (see FIG.~\ref{fig:expt_aspiration}b) where a membrane nano-tube is pulled out of the cell surface, usually using optical tweezers~\cite{evans1973new}. This is done by first attaching a micron-size bead to the lipid bilayer without any mechanical linkages to the cytoskeleton. 
Beyond a threshold force, a cylindrical membrane nano-tube, usually devoid of cytoskeletal elements, $\approx 100$ nm in diameter, extends from the cell to the bead. The effective membrane tension $\sigma_s$ is related to the measured force as $\sigma_s = f^2/(8\pi^2 \kappa)$, where $f \sim 10$ pN is the force on the bead and $\kappa \sim 10^{-19}$ Nm is the bending modulus of the membrane~\cite{derenyi2002formation} (see section~\ref{sec:soft_matter_theory}). In cells, the value of $\sigma_s$ may also be controlled by adhesion to the cortex and compositional variations in the membrane ~\cite{hochmuth1996deformation}. If a step-like extension is applied to an existing tether, the force exhibits a multi-timescale relaxation response. This relaxation occurs through various dissipation mechanisms, such as intra-bilayer friction, membrane-cytoskeleton friction, and fluid flow inside the tube~\cite{datar2015dynamics, shi2018cell}.

\subsection{Fluorescence techniques}

Fluorescent dyes can also be used to gauge in-plane membrane tension or relative changes in tension, an approach that has grown popular because the measurements are relatively easy, compatible with live cells. Two mechanisms are exploited. In the first, environment-sensitive dyes report lipid packing: as the membrane is stretched and the lipids spread apart, water penetrates between the head groups and shifts the emission spectrum~\cite{zhang2006laurdan}. In the second, mechanosensitive ``flipper'' probes respond to lipid packing through their fluorescence lifetime, which increases with membrane tension~\cite{colom2018fluorescent}. Because both readouts sense lipid order as a proxy rather than mechanical tension directly, they report relative changes most reliably and require independent calibration to recover absolute tensions. The signal can also be confounded by variations in membrane composition, temperature, and probe partitioning \cite{roffay2024tutorial}.

Genetically encoded molecular tension sensors, introduced in Section~\ref{sec:infer_forces}, turn F\"orster resonance energy transfer (FRET) into a quantitative force readout~\cite{lacroix2015construction}. In the tension-sensor module, a donor and acceptor flank an elastic peptide (often derived from spider silk) that lengthens under load, so that increasing tension lowers the energy-transfer efficiency; calibrating the module against force converts the measured efficiency into a tension. Beyond the vinculin sensor noted earlier~\cite{grashoff2010measuring}, analogous constructs report the tension borne by E-cadherin at cell--cell junctions~\cite{borghi2012ecadherin} and by talin during substrate-rigidity sensing~\cite{austen2015extracellular}.

\subsection{Laser ablation}

Laser ablation, a method by which a laser spot is briefly directed at a localized zone, can be used to perturb the mechanical integrity of a cell or tissue. In combination with theoretical models, this method can be used to obtain quantitative estimates of local tensions in bio-filaments, cells, and tissues~\cite{mayer2010anisotropies, wu2012investigation,shivakumar2016laser}.

In the previous two sections, we have discussed theoretical and experimental methods used to understand mechanical forces. In cells and tissues, mechanical forces and biochemical signals are tightly interconnected. It is, however, unclear what general physical principles underlie such mechanochemically organized structures, a point we turn to in the next section.

\section{Principles of mechanochemical transduction}
\label{sec:mechanochemical_transduction}

The linear irreversible thermodynamics introduced in Section~\ref{sec:soft_matter_theory} gives the natural language for force generation. Close to equilibrium, a generalized force $F_j$ drives conjugate currents $J_i$ through the Onsager coefficients $L_{ij}$~\cite{onsagerReciprocalRelationsIrreversible1931, onsagerReciprocalRelationsIrreversible1931a},
\begin{align}
J_{i} = \sum_{j} L_{ij}F_j \label{eq:onsager_flux}.
\end{align}
The diagonal terms are ordinary transport coefficients, but the off-diagonal $L_{ij}$ ($i\neq j$) couple otherwise distinct degrees of freedom: an electric current driving a heat flux in the Peltier effect~\cite{onsagerReciprocalRelationsIrreversible1931}, or, as below, a chemical reaction driving mechanical motion. Left to itself the system relaxes to equilibrium, where every $J_i = 0$; but when metabolism holds the chemical flux from vanishing, non-zero fluxes are generically sustained in the coupled mechanical degrees of freedom.

A molecular motor is an example of such a system where transitions between different chemical conformations are maintained out of detailed balance by an external supply of a fuel. This biases the chemical reaction in a particular direction~\cite{julicherModelingMolecularMotors1997} (this is true of all metabolic processes). The chemical flux, as a result of the coupling between chemical and mechanical degrees of freedom, can drive material currents. This is the essence of \emph{mechanochemical} coupling.

\begin{figure}
\centering
\includegraphics[width=0.65\linewidth]{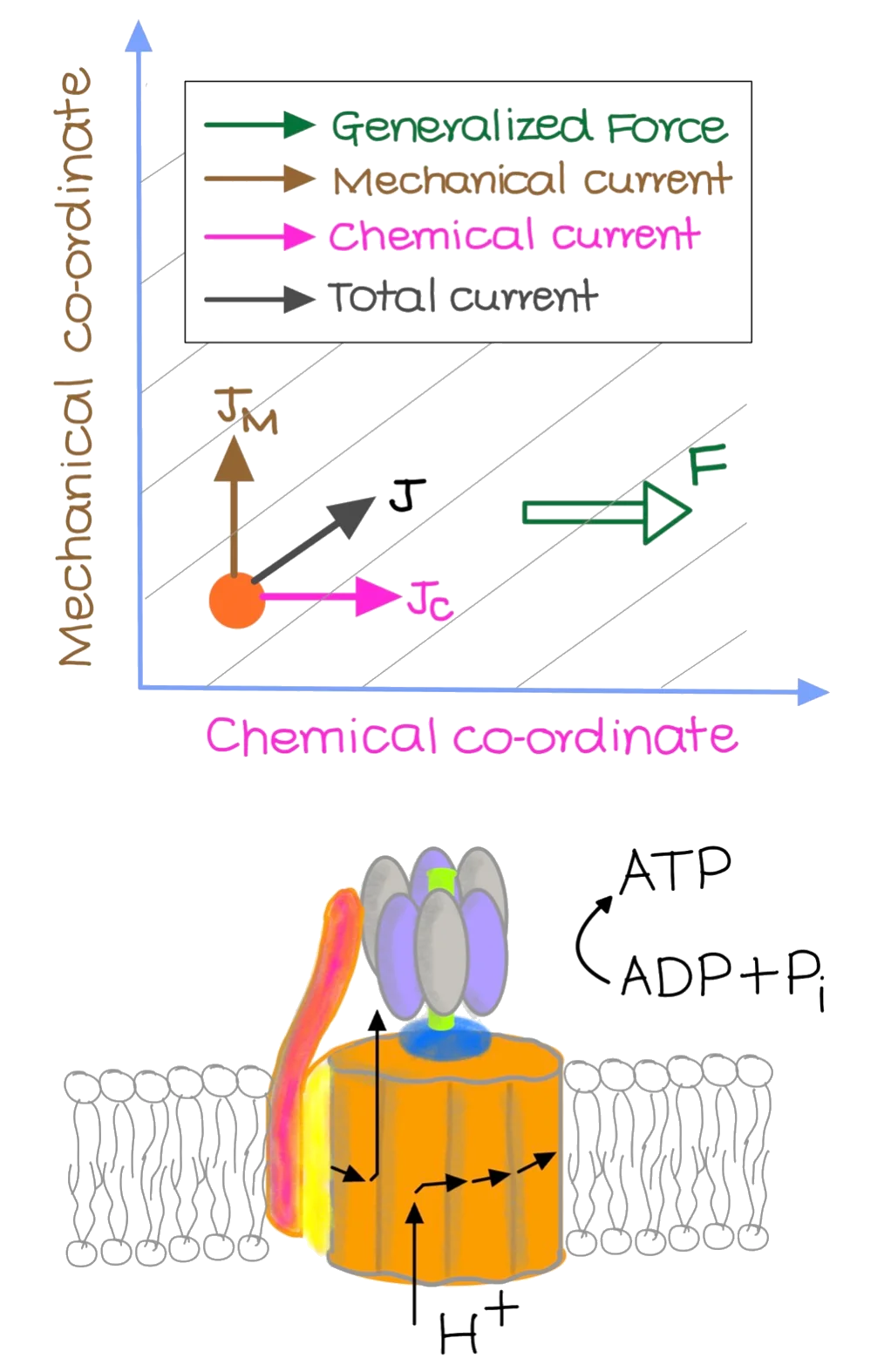}
\caption{Top: A schematic representation of the coupling between the chemical (horizontal axis) and the mechanical (vertical axis) degrees of freedom of a mechanochemical entity. In the case of a molecular motor, the horizontal axis represents the chemical state of ATP hydrolysis of the motor while the vertical axis represents its position along a filament. The oblique lines represent non-zero off-diagonal Onsager coefficients $L_{MC}$ and $L_{CM}$ in Eq.~\eqref{eq:onsager_flux}, which couple the mechanical current $J_M$ to the chemical force and the chemical current $J_C$ to the mechanical force. A generalized force $F$ applied in the chemical direction therefore drives not only a chemical current $J_C$ but also a mechanical current $J_M$, so that the total current $J$ is oblique; conversely, a force applied in the mechanical direction generates a chemical current. The tight coupling between these two degrees of freedom implies that the system is constrained to move parallel to the oblique lines. Bottom: The $F_0F_1$-ATP synthase runs this coupling in reverse. Protons (H$^+$) flowing down their electrochemical gradient across the membrane drive the rotation of the membrane-embedded $F_0$ ring and the central stalk; this mechanical rotation drives conformational changes in the catalytic $F_1$ head that synthesize ATP from ADP and P$_i$.}
\label{fig:chemo_mechanical}
\end{figure}

An intuitive way to visualize such a mechanochemical coupling is shown in FIG.~\ref{fig:chemo_mechanical}. Here, the coordinates shown provide an abstract representation of the chemical and mechanical degrees of freedom of a motor-like molecule. The coupling between these degrees of freedom is depicted by the diagonal lines which represent a generalized non-diagonal mobility tensor. In other words, the diagonal ``grooves'' in this mechanochemical space are such that a net force in one direction, say, the chemical direction leads to a current in both the directions. This would be the normal operation of motors~\cite{julicherModelingMolecularMotors1997}. The opposite action, wherein chemical currents are generated from mechanical forces, is seen in nanomachines such as the $F_{0}F_1$-ATP synthase present in the membranes of mitochondria~\cite{boyerATPSYNTHASESPLENDID1997, hayashiFluctuationTheoremApplied2010}.

Two caveats temper this picture. First, the linear Onsager relations hold only near equilibrium, whereas molecular motors operate far from it: a single ATP hydrolysis releases $\approx 20~k_BT$, so the chemical affinity driving the coupling is large, not infinitesimal. The mechanochemical cartoon therefore motivates the coupling but does not by itself describe motor operation, which is genuinely nonlinear and is treated as such in Section~\ref{sec:molecular_motors}. Second, the figure depicts \emph{tight} coupling, in which each chemical cycle is rigidly slaved to one mechanical step; real motors frequently operate with loose or variable coupling (the ratchet limit discussed there), so tight coupling is best read as an idealization that maximizes thermodynamic efficiency rather than as the generic case. The coupling is, moreover, not automatic: in an isotropic medium a scalar chemical affinity cannot by itself drive a vectorial mechanical current, so directed force generation requires a broken symmetry (here the structural polarity of the filament track), in accordance with the Curie principle invoked in Section~\ref{sec:molecular_motors}. The efficiency, defined as the ratio of mechanical output power to chemical input power, is bounded by unity and is revisited for the ratchet models of Section~\ref{sec:molecular_motors}.

The principles of mechanochemical transduction discussed above are generically applicable in a wide range of cellular processes. In the next few sections, we illustrate these principles by discussing physical models for chromatin organization, force generation by polymerization of biopolymers, stochastic motion of molecular motors on asymmetric tracks, and, at a larger scale, the organization of the cellular cytoskeleton to orchestrate cell polarity, motility, and division. 

\section{\label{sec:poly_forces}Polymerization forces}

\begin{figure}[ht]
\centering
\includegraphics[width=0.7\linewidth]
{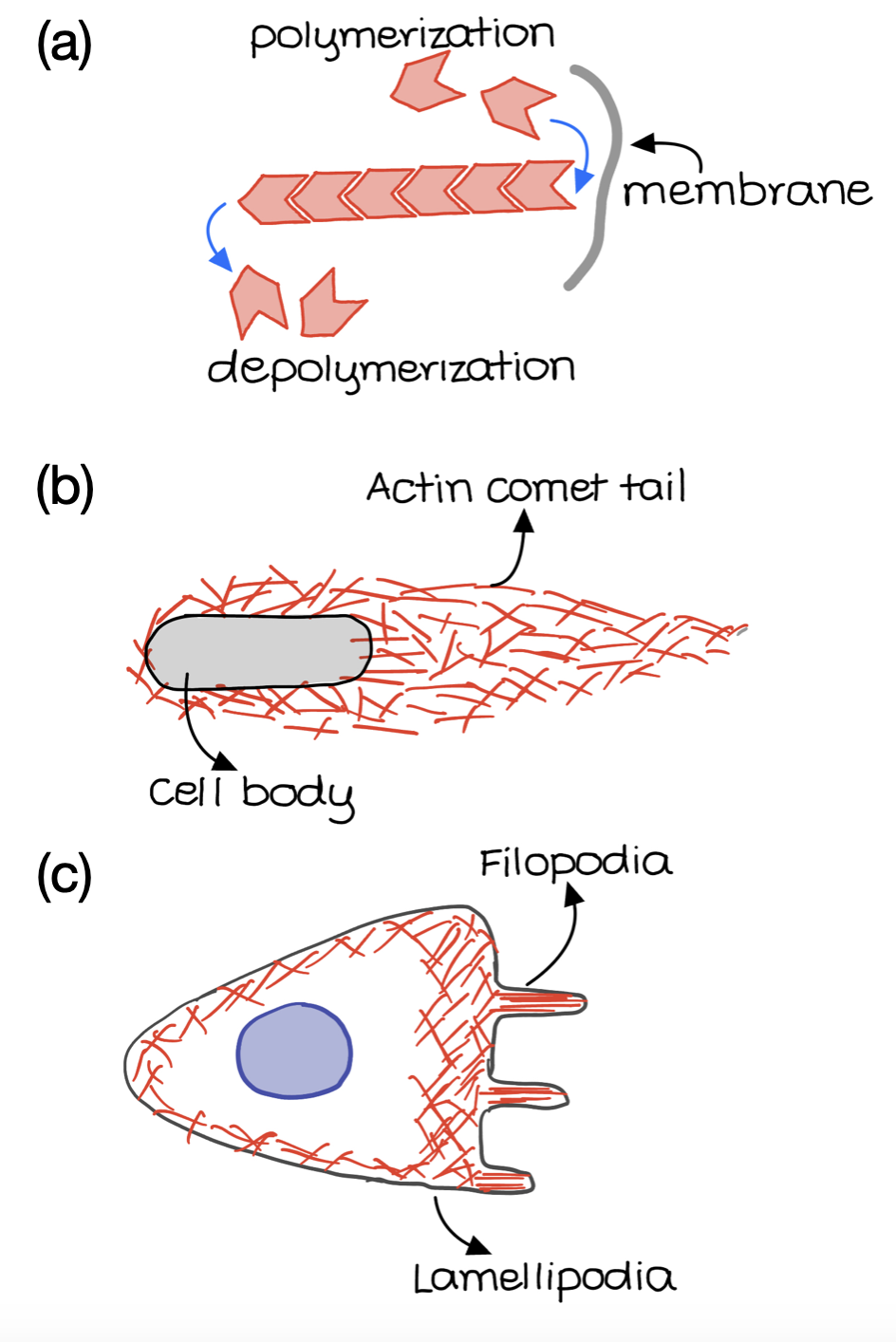}
\caption{Some cellular phenomena wherein forces are generated by polymerization of filaments. (a) Force-generation by the polymerization/depolymerization processes at the level of a single filament pushing against a fluctuating barrier, (b) Listeria propulsion using actin polymerization, and (c) Lamellipodial and filopodial protrusions in a cell.}
\label{fig:polymerization}
\end{figure}

Microtubules and actin filaments are biopolymers that polymerize and depolymerize in the presence of NTP~\cite{alberts2014molecular}. Both are polar filaments: the two ends (designated plus and minus) of the filament are chemically distinct. Microtubules are composed of protofilaments forming a hollow cylinder approximately $25~{\rm nm}$ in outer diameter, with each individual protofilament formed from dimeric units of tubulin monomers. 


Microtubules remodel primarily at their plus-ends via dynamic instability, characterized by cycles of slow growth and rapid shrinkage (catastrophe) \cite{mitchison1984dynamic}. In contrast, actin filaments in steady state exhibit ``treadmilling'', in which the rate of addition of monomers at the plus end of actin is compensated by the loss of monomers at its minus end. This can be described in terms of an effective advection of the centre of mass of the filament. 

Composite networks of these two types of biofilaments along with molecular motors are the primary active constituents of the cellular cytoskeleton. Below, we discuss how the dynamics of these filaments contribute to emergent phenomena such as positioning cellular organelles, bacterial propulsion, and growth of cellular protrusions. 

\subsection{Force-generation due to polymerization}

Polymerization and depolymerization of biopolymers are important sources of force generation in cells. Polymer growth/shrinkage can generate pushing/pulling forces on  mechanical obstacles. These forces are used in processes such as cell migration, cell division, etc. It has been demonstrated {\em in vitro} that polymerizing actin filaments can generate protrusions in membrane vesicles~\cite{miyata1999protrusive}, or buckle when they are made to grow against a barrier~\cite{kovar2004insertional}, with typical forces about 1 pN per filament~\cite{footerDirectMeasurementForce2007}.

The simplest model for force generation due to polymerisation is one in which a sub-unit binds to the filament with a rate $k_b$ and unbinds at a different rate $k_u$. Considering a filament with one dynamic end and the other end anchored to a stable seed, the master equation governing the probability $P_n(t)$ of finding a polymer with $n > 0 $ monomers at time $t$ is 
\begin{align}
\frac{dP_n}{dt} = k_{b} P_{n-1} - k_{b} P_n - k_{u} P_n + k_{u} P_{n+1}.
\end{align}
We assume that a reservoir of actin monomers is available for polymerization, that a stable nucleating centre for polymers is always present, and that we can neglect shape changes of the filament. The calculation is carried out using the reflecting boundary condition $dP_0/dt = -k_b P_0 + k_u P_1$, corresponding to a stable $n = 0$ configuration. With the assumption $k_b < k_u$, the steady-state distribution and the average filament length, respectively, are~\cite{bressloff2021stochastic}
\begin{align}
    P_n = \left(1 - {k_b \over k_u}\right)\left({k_b \over k_u} \right)^n \quad \mbox{and} \quad \langle n \rangle = {k_b/k_u \over 1 - k_b/k_u}.
\label{eq:Pn_ave_n}
\end{align}

Let the nucleating centre be anchored while the freely polymerizing end encounters a barrier (such as the membrane; FIG.~\ref{fig:polymerization}(a)). This process can generate a pushing force when the force-independent rates are such that $k_b^0 > k_u^0$. The reaction force $f$ from the membrane can change the binding-unbinding rates to $k_{b} = k^0_{b} e^{-\beta \, f \alpha a}$ and $k_{u} = k^0_{u} e^{ \beta \, f (1-\alpha) a}$, obtained via a Kramers approximation, where $a$ is the size of the subunits, and $0<\alpha<1$ is a load distribution factor which indicates the differential force dependence of the binding and unbinding rates~\cite{KolomeiskyBook}. The corresponding rate ratio is $k_b/k_u = (k^0_b/k^0_u) e^{-\beta f a}$.  For $k_b/k_u > 1$, the filament grows with mean velocity
\begin{eqnarray}
v_g = [k^0_{b}e^{-\beta f\alpha a} - k^0_u e^{\beta f (1-\alpha) a}]a. 
\label{eq:growth_force_change}
\end{eqnarray}
At a critical force $f_s$, called the stall force, this growth velocity vanishes, i.e., $v_g = 0$.  The stall force $f_s$ is
\begin{align}
f_{s} = \frac{\Delta \mu}{a} \label{eq:f_stall},   
\end{align}
where $\Delta \mu = k_BT \ln(k^0_b/k^0_u)$ is the free energy of binding of a single monomer -- this can also be written in terms of the ratio $c/c_{\rm crit} = k_b^0/k_u^0$, where $c$ and $c_{\rm crit}$ are the actual and critical concentrations of actin monomers at the polymerizing end~\cite{footerDirectMeasurementForce2007}. 

In actin polymerization, the free energy gain $\Delta \mu$ per added monomer depends on the monomer concentration and nucleotide state and is typically several $k_BT$. Taking the representative value $\Delta \mu = 4 k_BT$, with monomer axial spacing $a \approx 2.7~{\rm nm}$ gives 
\begin{align}
f_\text{s} = \frac{\Delta \mu}{a}
\;=\;
\frac{4\,k_BT}{2.7\;\mathrm{nm}}
\;\approx\;
\frac{4\times 4.1\;\mathrm{pN\,nm}}{2.7\;\mathrm{nm}}
\;\approx\;
6\;\mathrm{pN}.
\end{align}
This is the ideal stall-force estimate for the chosen monomer
chemical potential. Experiments report stall forces of order
$1~{\rm pN}$, consistent with the predicted single-filament
stall force under their experimental conditions~\cite{footerDirectMeasurementForce2007}.
We adopt $\sim 1~{\rm pN}$ as a characteristic per-filament
force scale in the estimates that follow. This model can further be extended to incorporate, for example, differential binding/unbinding rates at the two ends of the filaments and explicitly distinguish the ATP/ADP bound states of the monomers~\cite{ranjith2009nonequilibrium}. 


The above calculation is for the force generated by a single polymerizing filament with one end fixed and pushing against a soft barrier at the other end. A more general situation is a bundle of $N$ filaments growing from a common nucleating centre towards a barrier at distance $d$. For simplicity, we consider a $1$-D geometry and neglect interactions between the filaments, so that they are statistically equivalent. Only filaments in contact with the barrier push on it, and the mean force on the barrier is
\begin{align}
f_N(d) = N \, p_c(d) \, \bar{f}_c(d), \label{eq:repulsive_polymerization}
\end{align}
where $p_c(d)$ is the probability that a given filament is in contact with the barrier and $\bar{f}_c(d)$ is the mean force it exerts while in contact. Both depend on how the barrier moves and on how the load is shared among the contacting filaments, which in turn feeds back on their growth~\cite{wang2014load}. Since each contacting filament carries at most its stall force, $f_N \lesssim N f_s$. Note that the stationary distribution~\eqref{eq:Pn_ave_n} cannot be used for $p_c$: it exists only for $k_b < k_u$, i.e.\ above the stall force, where an unconstrained filament shrinks rather than pushes. A simple closure, appropriate when filament lengths are limited by turnover, is an exponentially decaying force--distance law,
\begin{align}
f_N(d) \simeq F_0 \, e^{-d/\ell_f}, \label{eq:bundle_force_exp}
\end{align}
where $\ell_f$ is a characteristic filament length and $F_0 \lesssim N f_s$. We stress that~\eqref{eq:bundle_force_exp} is an assumed form, not a consequence of the single-filament model above. This is the force that $N$ filaments polymerizing from a fixed nucleating centre exert on a soft barrier. Conversely, if the barrier is fixed, the same repulsive force can lead to translocation of the nucleating centre. This expression will be used in Section~\ref{sec:spindle} to estimate the positioning forces generated by the mitotic spindle during cell division. Pushing forces generated by multiple actin filaments are harnessed by bacteria (such as \emph{Listeria}) and also by eukaryotic cells to generate tubular extensions such as filopodia.

\subsection{Listeria propulsion}

A well-investigated example of polymerization forces generated by actin filaments is the propulsion machinery of {\it Listeria monocytogenes} — a micron-long, cylindrical, pathogenic bacterium that infects mammalian cells, including humans (FIG.~\ref{fig:polymerization}(b))~\cite{bray2000cell}. Once inside a host cell, \textit{Listeria} hijacks the actin supply of the eukaryotic cell to become motile and uses this motility to  penetrate neighbouring cells~\cite{tilney1989actin, theriot1992rate}. Typical propulsion speeds are of the order of $0.1-0.2~\mu{\rm m}/s$. 

\textit{Listeria} does this by expressing actin nucleating proteins on its surface.  This leads to the formation of a filamentous actin tail which can be tens of microns long~\cite{theriot1992rate}. Although polymerization from the bacterial surface requires a loose coupling between the surface and the filament, the bacterium is always attached to the tail due to the large number of filaments. Moreover, crosslinking proteins bind these filaments together as they grow, forming an elastic gel with a finite shear modulus. This tail, made of sub-micron length actin filaments, gradually depolymerizes from the rear, thus giving a motile \textit{Listeria} its characteristic comet-like appearance. Remarkably, motility outside the living cell has been demonstrated using a minimal system consisting of micron sized beads coated with an actin nucleating protein ~\cite{noireaux2000growing}. 

Two complementary frameworks for actin-based propulsion are worth contrasting. The microscopic, single-filament \emph{polymerization ratchet}, in the elastic and tethered forms developed by Mogilner and Oster~\cite{mogilnerCellMotilityDriven1996, mogilnerForceGenerationActin2003}, builds the propulsive force from the thermal bending fluctuations of individual semiflexible filaments; treating the filaments as independent, the total mean force is the sum of the individual contributions and scales with the number of contacting filaments, Eq.~\eqref{eq:repulsive_polymerization}. This picture captures the per-filament stall force and load--velocity relation, but because it does not enforce global force balance with the surrounding network, it becomes questionable once the filaments are densely crosslinked and elastically coupled. The continuum \emph{elastic-gel} picture derived below~\cite{gerbal1999listeria, gerbalElasticAnalysisListeria2000} instead treats the comet as a crosslinked elastic medium whose build-up and release of stress drives motion; it naturally explains the spontaneous symmetry breaking of the actin shell around a bead, but coarse-grains away the individual filaments. The two are discriminated experimentally by whether the instantaneous speed depends only on the current load (non-interacting filaments) or also on the loading history (a collectively coupled gel): micromanipulation of actin-propelled beads reveals a pronounced history dependence~\cite{parekh2005loading}, favouring a collective, elastically coupled description when the network is dense. More recent treatments fold filament recycling and turnover into this collective picture~\cite{colin2023recycling}.

To arrive at an expression for the propulsion speed, we consider a spherical bead on which the actin filaments are polymerizing. An approximate value for the typical thickness of the polymerizing gel can be obtained from the balance of forces resulting from polymerization/depolymerization and the mechanical stresses generated in the elastic medium. Large tangential strains developed in the polymerizing gel lead to a spontaneous symmetry breaking, and hence to the motility of the bead \cite{noireaux2000growing}. 

Consider a spherical bead of radius $r$ covered with an actin nucleating protein, with average spacing $\xi$, on which actin monomers, of size $a$, polymerize. The total free energy gained from adding the first layer of actin mesh on the bead surface is $U_\xi = \Delta \mu \,(4\pi r^2/\xi^2)$, where $\Delta \mu$ is the free energy associated with addition of a single actin monomer to the gel. However, polymerization of subsequent actin layers generates both radial and tangential  stresses in the crosslinked actin gel with elastic modulus $E$. The polymerization is energetically favorable only when the magnitude of free energy reduction $U_\xi$ is greater than the added work required to polymerize against the radial stress $\mathsf{\Sigma}_{rr}$. If the thickness of the gel $h \ll r$, then the strain in the gel is $\sim h/r$ and hence, the elastic energy cost is  $U_e \sim E a\;(h/r)^2 \; 4\pi r^2$. The balance of elastic energy and the free energy gained from polymerization provides an estimate for the thickness of the actin gel $h \approx r \sqrt{\Delta \mu/E a\xi^2}$~\cite{noireaux2000growing}. This estimate assumes that the gel  is radially symmetric and is in a steady state at mechanical equilibrium. 

For directional motility, the initially uniform actin gel must break symmetry. A simple approximation for this symmetry breaking threshold is to consider tangential strains $h/r \sim 1$. An estimate for the elastic modulus is $E \approx k_BT\;\ell_T/\xi_c^4$, where $\ell_T$ is the persistence length of a single filament and $\xi_c$ is the mean spacing between the crosslinked actin filaments in the gel~\cite{boal2012mechanics}. Therefore, taking the nucleator spacing to be comparable to the network mesh size ($\xi \approx \xi_c$), an estimate for a possible symmetry breaking transition is $(\xi_c/\sqrt{\ell_T a}) \sqrt{\Delta \mu/k_BT} \sim 1$~\cite{noireaux2000growing, gerbal1999listeria, gerbalElasticAnalysisListeria2000}. These expressions assume that polymerization timescales are much faster compared to the viscoelastic relaxation time of the actin gel. 

Once symmetry breaking occurs, a simple way to estimate the motility speed $v$ of the bacterium is to consider the actin tail as an incompressible elastic gel and estimate the resulting flow speed arising from force-balance and volume conservation. This yields the expression \cite{gerbal1999listeria}
\begin{align}
\frac{v}{v_p} = 1 - \frac{F_{\rm ext}}{E \; S_b},
\end{align}
where $v_p$ is the polymerization rate, $F_{\rm ext}$ is the external force exerted by the medium on the bacterium, and $S_b$ is the cross-sectional area of the bacterium. Incorporating the change in the polymerization rate due to the applied forces, as in \eqref{eq:growth_force_change}, does not lead to a qualitative change in the above expression.  The external load may include drag or friction exerted by the surrounding medium arising from detachment between listeria and the surrounding actin network, but not the internally generated polymerisation force.  The value of $E \, S_b$ is  $\approx 1 ~{\rm nN}$~\cite{gerbal1999listeria}. Hence,  $100-1000$ simultaneously polymerizing actin filaments with stall force of $1~{\rm pN}$ each would generate a stall force $F_{\rm ext} \approx 0.1 - 1~{\rm nN}$. In summary, we see that force-generation due to actin polymerization induced stresses leads to a spontaneous motility of the \textit{Listeria}.

\subsection{Filopodial extension}

Another well known example of actin-based force generation is the growth process of filopodia. These are thin, tubular protrusions, usually seen in motile cells (FIG.~\ref{fig:polymerization}(c)). Filopodia have a typical diameter of about $100-300$~nm and consist of $\sim$ 10 to 20 actin filaments organized in a polar bundle. Filopodial growth is driven by actin polymerisation, which generates a pushing force on the membrane as discussed above. 

To extend a cylindrical membrane tube of diameter $d_m \sim \sqrt{\kappa/\sigma_s}$, the pushing force must exceed the passive restoring force $F_m \sim \sqrt{\sigma_s\kappa}$ exerted by the membrane, where $\sigma_s$ is the in-plane tension and $\kappa$ is the bending rigidity of the membrane -- typically, $d_m \sim 100~{\rm nm}$ and $F_m \sim 10~{\rm pN}$. However, the force exerted by a single polymerizing actin filament is $\sim 1~{\rm pN}$. The large active forces required for pushing the membrane in a filopodium result from the coordinated polymerization of multiple actin filaments driven by bundling proteins. 

Filopodia are highly dynamic structures and play an important role in sensing chemical and mechanical cues during cell locomotion or axonal growth cone guidance~\cite{bray2000cell}. While the growth of filopodia is reasonably well explained, their mechanosensitivity and filopodial retraction are still not fully explained and multiple mechanisms~may be involved~\cite{chanTractionDynamicsFilopodia2008a, leijnse2022filopodia, bornschlogl2013filopodial}.

This section focused on force-generation due to filament polymerization and neglected the effect of molecular motors that translocate on these filaments using chemical energy. Such motor activity is another route to force-generation in cells and tissues.

\section{Molecular motors}\label{sec:molecular_motors}

How is directed transport of cargo achieved within the crowded and stochastic environment of a cell? Brownian diffusion is effective for transport in small micron-sized cells (e.g., bacteria). However, for larger cells (e.g., neurons), diffusive transport is highly ineffective. Transport mechanisms that involve specialized protein molecules called molecular motors moving unidirectionally on asymmetric polymeric tracks provide a significantly more efficient and fast way of transporting material within cells.

This distinction between diffusive and motor-driven transport has an evolutionary dimension. Prokaryotes possess homologues of the cytoskeletal filaments actin and tubulin~\cite{busiek2015bacterial}, but none of the processive motors (myosin, kinesin, and dynein) that walk along them. These motor families are essentially a eukaryotic innovation, and molecular-phylogenetic reconstructions indicate that a diverse toolkit of them, comprising multiple classes each of kinesin and dynein together with several myosins, was already present in the last eukaryotic common ancestor~\cite{wickstead2011evolution, koumandou2013molecular}. Coupling these motors to a dynamic, treadmilling cytoskeleton furnished the machinery for directed transport, phagocytosis, mitotic chromosome segregation, and ciliary motility~\cite{vale2003molecular}. In this sense the advent of active force generation lifted the diffusion-limited constraint on cell size and internal organization noted above, and can be read as a threshold in cellular complexity, though such motor innovations arose alongside other eukaryotic novelties, so isolating their causal role is difficult.

For an object to move unidirectionally in the absence of external forces in a viscous environment, it must be able to transduce energy internally (be active) as well as derive its sense of direction through the absence of a fore-aft symmetry. This is the essence of the Curie principle~\cite{curie1894symetrie}. Molecular motors are examples of the operation of the Curie principle at the cellular level.

The molecular motor is a nano-scale entity, fruitfully thought of as having multiple internal states. The motor moves on a one-dimensional and periodic biopolymer track (e.g., microtubules with a periodicity of $\sim 8 ~{\rm nm}$). Due to their small size, these motors function in a highly noisy environment. For a motor to move unidirectionally at a steady mean velocity, work must be performed against a drag force from the ambient medium. This requires the transduction of energy at a local scale. This drives conformational changes in the molecular motor. 

\begin{figure}[ht]
\centering
\includegraphics[width=0.9\linewidth]
{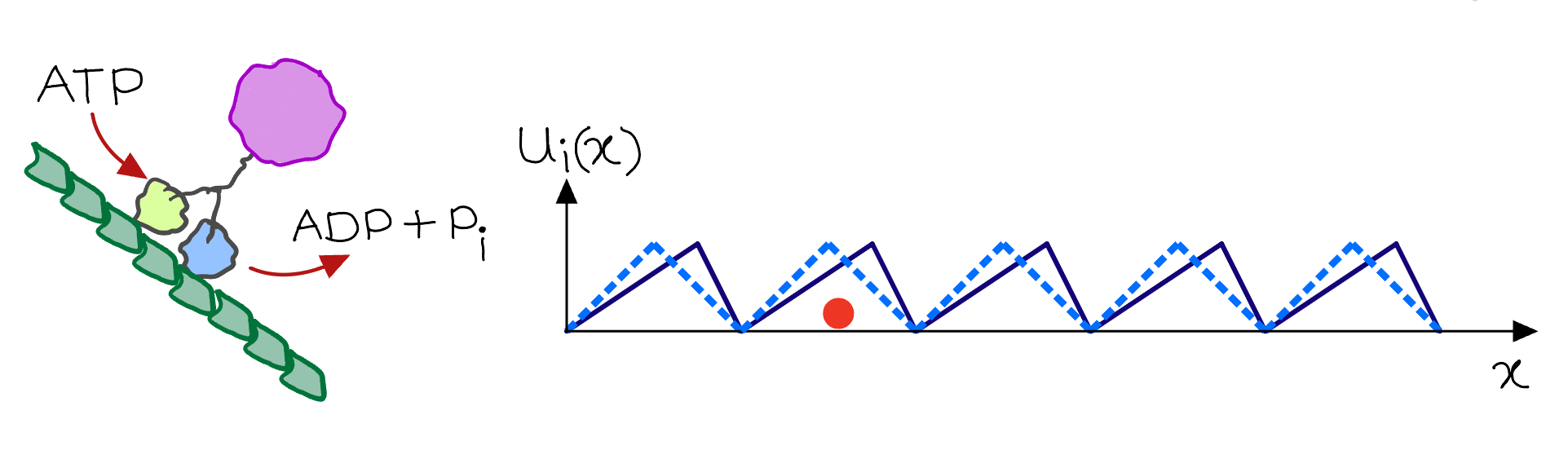}
\caption{Directional movement of a typical molecular motor along an asymmetric track. An effective Brownian ratchet model to capture the statistical dynamics of the motor can be visualized as a point particle moving in a spatially periodic but asymmetric potential shown as $U_i(x)$. Depending on the configurational state of the motor, the potential switches between the profiles represented by the continuous and dashed lines. The transitions between the internal states of the motor are biased by ATP hydrolysis and thus do not satisfy detailed balance. This non-equilibrium energy input, combined with the spatial asymmetry of the track leads to directional transport. }
\label{fig:molecular_motors}
\end{figure}

The interaction of the molecular motor with the track can be represented by a fore-aft asymmetric and periodic potential representing the polar nature and the structural periodicity of the filament, respectively (FIG.~\ref{fig:molecular_motors}). The conformational changes of the molecular motor can be interpreted as switching between different local interaction potentials with the track.  If the transitions between these different conformations (interaction potentials) do not satisfy detailed balance (due to energy flux from ATP hydrolysis), then coupled with the polar nature of the track, the Curie principle implies that we can expect a non-zero mean velocity of the motor. 

 The original Feynman-Smoluchowski pawl-and-ratchet model illustrates how irreversible motion might potentially happen in a thermal environment for macroscopic systems \cite{feynmann1963feynmann}. The pawl-and-ratchet model describes a situation where a pawl can move along a ratchet, which is an asymmetric object constructed so that it moves more easily in one direction as opposed to the other. However, thermal fluctuations allow the ratchet to ``slip back'' so that, on average, there is no net unidirectional motion that can be imparted. 

To circumvent this, the advancing and the falling back of the pawl must be induced by the switching of the ratchet potential at rates that do not obey detailed balance. This can be done through the injection of energy, such that the entire system is out of equilibrium. The introduction of a non-equilibrium element destroys the time-reversal symmetry of the original problem, allowing for directed motion. 

The force-velocity relation for molecular motors can be obtained by using optical tweezers to measure forces exerted by the motor (see Section~\ref{sec:optical_traps}). The motion of the motor in an ATP saturating environment can be studied statistically, leading to the result that the fundamental periodicity of the filament, either microtubules or actin, sets the fundamental step size of the motor. Applying increasing loads finally leads to motor stalling. By mapping this onto a particle moving in an asymmetric potential, the stall force can be related to fundamental aspects of the motor~\cite{julicherModelingMolecularMotors1997, reimannBrownianMotorsNoisy2002}.

A word on what does \emph{not} drive these motors. The directed motion is not powered by spatial temperature gradients: thermal equilibration is essentially instantaneous on nanometre scales, so a motor and its surroundings remain isothermal~\cite{julicherModelingMolecularMotors1997, mugnaiTheoreticalPerspectivesBiological2020}. This is why a purely thermal Feynman--Smoluchowski ratchet cannot serve as a biological motor: rectifying equilibrium fluctuations at a single temperature would violate the second law, and directionality must instead come from breaking detailed balance through the chemical energy of ATP hydrolysis. A related question is the level of description. All-atom simulations resolve the individual conformational transitions, but coarse-graining over them recovers the effective few-state ratchet dynamics used above, which is why these simple models capture the observed stepping despite ignoring atomic detail.

\subsection{Brownian ratchet models}

The thermodynamic ideas underlying the theory of motors, invokes multiple states, in each of which the system rapidly achieves thermal equilibrium. Multiple conformational states of the translocating motor protein have been identified in experiments (up to five or six different states have been  identified)~\cite{KolomeiskyBook}. 

A simple model is the ``particle fluctuating between states'' model, where one imagines at least two states $i=1,2$ with each described phenomenologically by a periodic potential $U_i(x)$. This potential must, of necessity, be fore-aft asymmetric \cite{magnascoForcedThermalRatchets1993}. The non-equilibrium component is added by accounting for transitions between states that occur independently of the energy of that state. This breaks detailed balance. 

Let $P_i(x,t)$ be the probability of finding the particle at position $x$ in the state $i$. The evolution of these probabilities is governed by the Fokker-Planck equations:
\begin{align}
\partial_t P_1 &= -\partial_x J_1 - W_{1\to2}(x) \, P_1 + W_{2\to1}(x) \, P_2 ,
\\
\partial_t P_2 &= -\partial_x J_2 + W_{1\to2}(x) \, P_1 - W_{2\to1}(x) \, P_2,
\end{align}
where $W_{i\to j}(x)$ is the space-dependent transition rate from state $i$ to the state $j$. We assume that the switching rates share the spatial symmetries of the potentials. The spatial current in each state is of the form
\begin{equation}
J_i = \mu_i \left (-k_BT \partial_x P_i -P_i \partial_x U_i + P_i \, F_{\rm ext} \right ) ,\
\end{equation}
where $\mu_i P_i \, F_{\rm ext}$ is the probability current resulting from an applied external force $F_{\rm ext}$. The total current $J = J_1 + J_2$ can be written in terms of the total probability $P = P_1 + P_2$ as
\begin{equation}
J = J_1 + J_2 = \mu_{\rm eff} \left (-k_B T \partial_x P - P \partial_x U_{\rm eff} (x) + P \, F_{\rm ext} \right).
\end{equation}
In the stationary state, with constant state mobilities $\mu_i$,  $\lambda(x) = P_1(x)/P(x)$, $\mu_{\rm eff}(x) = \mu_1 \, \lambda + (1 - \lambda) \, \mu_2$ is the effective mobility, and the effective potential is
\begin{align}
U_{\rm eff}(x) - U_{\rm eff}(0) &= \int_0^x dy \; \bigg[ \frac{\mu_1 \lambda \partial_y U_1 + \mu_2 (1-\lambda) \partial_y U_2}{\mu_{\rm eff}(y)}\bigg]
\nonumber \\
& \qquad
+ k_BT \big[ \ln \mu_{\rm eff}(y) \big]_{0}^x.
\end{align}
In the absence of an applied force, to get directed motion, the effective potential $U_{\rm eff}(x)$ needs to have a net tilt over the periodicity of the underlying potential. If both the potentials $U_i(x)$ are fore-aft symmetric or if detailed balance is satisfied, then $U_{\rm eff}(x)$ cannot have a net tilt~\cite{julicherModelingMolecularMotors1997}.  The net force generated by the ratchet mechanism described above can be used to tug cargo (e.g., vesicles) effecting intracellular transport. The efficiency of this process, i.e., the work done per ATP molecule divided by the free energy released per hydrolysis, can be high compared with that of many macroscopic engines. With appropriate modifications, models of the sort that we discussed above can also be used to describe rotary motors as well.

The isothermal ratchet is only one of two limiting descriptions of how chemical energy is converted into directed motion, and it is worth stating the alternative explicitly. In the rectified-diffusion (ratchet) picture developed above, the motor is \emph{loosely coupled}: it explores the track by thermal diffusion, and the non-equilibrium switching between potentials merely biases fluctuations that are already present, so there is no designated mechanical working stroke. In the competing \emph{power-stroke}, or lever-arm, picture a specific, tightly coupled conformational change (the swing of the myosin lever arm, or the docking of the kinesin neck linker) drives a largely deterministic step for each ATP hydrolysed, with diffusion relegated to a subsidiary search~\cite{HowardBook}. Real motors combine both elements. Muscle myosin-II makes a genuine lever-arm power stroke, but its low duty ratio keeps it detached for most of the cycle, so a sarcomere's output is a statistical average over many transiently attached heads; highly processive kinesin and myosin-V instead take many tightly coupled steps without releasing, each step a power stroke followed by a diffusive search of the free head for its next binding site. The two limits make different, testable predictions, and single-molecule optical trapping is the experiment that separates them: the step and substep sizes, the load- and ATP-dependence of the velocity, and the randomness parameter (the variance of the displacement divided by the product of the step size and the mean displacement, whose inverse gives a lower bound on the number of states in a sequential kinetic model) together fix where a given motor lies. Tight coupling, with one step per ATP and a step size locked to the track periodicity (kinesin advances $8~{\rm nm}$ per hydrolysis), points to the power-stroke limit, whereas large load-tunable diffusive excursions and a mechanochemical coupling ratio departing from unity point to the ratchet limit. Discrete-state mechanochemical models interpolate between the two and are the natural framework for fitting such force- and ATP-resolved stepping data.

\subsection{Collections of molecular motors}

Many biophysical situations involve a collective action of groups of molecular motors. Coupling many processive motors can lead to interesting phenomena, such as collective oscillations and metachronal waves~\cite{guerin2010coordination, gilpin2020multiscale}. These responses are associated with important biological processes such as flagellar and ciliary motion, muscle contraction, cardiomyocyte beating, and the beating of insect wings. Below, we discuss a few cellular processes which are driven by motor generated forces discussed above. 

\subsubsection{Cilia and flagella}

The beating of cilia and flagella provides another example of emergent phenomena from interacting collections of motors and filaments. These are both filamentous cell protrusions which actively beat repeatedly by consuming ATP. Eukaryotic flagella (not to be confused with bacterial flagella -- see Section~\ref{sec:neq_active_cells}) can be several tens of microns long. They are seen in small numbers (usually < 5) towards the anterior end of swimming organisms like Chlamydomonas. 

Cilia, on the other hand, are shorter ($<10~{\rm \mu m}$), usually occur as a dense carpet (often $\gg 100$ per cell) and are ubiquitous. In protozoans like \emph{paramecium}, they beat synchronously to generate wave like patterns that propel the cell forward. In filter feeding microorganisms like rotifers, cilia generate fluid flow that drive food particles into the feeding apparatus. In mammals, ciliary carpets line the surface of epithelial tissues, the layers of cells that cover the surface of organs, driving fluid flow. In the respiratory tract, cilia drive mucus towards the nostrils,  in the brain cavity, they drive fluid circulation, and within fallopian tubes, they help push the egg cell forward. 

Cilia and flagella are structurally almost identical, each being membrane protrusions dense with microtubule bundles (FIG.~\ref{fig:cilia_sarcomere}(a))~\cite{linck2016axoneme}. Typically, they contain a ring of nine longitudinally aligned microtubule doublets arranged with their plus end away from the cell~\cite{euteneuer1981polarity}. Each doublet consists of one complete tubule with about $13$ protofilaments attached along its length to a partial tubule with fewer protofilaments. All doublets are anchored at one end. Additionally, a pair of central microtubules is also common. The entire microtubule structure is interlinked by elastic linker proteins. The circularly arranged doublets are coupled to their neighbours by an array of dynein motors such that one doublet acts as a cargo while the heads bind to and push against the adjacent doublet (FIG.~\ref{fig:cilia_sarcomere}(a)). Thus this entire structure forms an active elastic bundle. It is known as an axoneme.

It is known that active systems consisting of collection of molecular motors coupled to a viscoelastic solid like medium can generate sustained spontaneous oscillations~\cite{kruse2005oscillations}. The beating of the axoneme may be such an example. It is known that the dynein motors generate longitudinal forces between adjacent microtubule doublets~\cite{ishibashi2020force}. In the axoneme, these longitudinal forces are exerted with a phase lag across the circumferential ring. The differential stress then translates the longitudinal force into bending deformations~\cite{ishibashi2020force}. A study of chiral effects in cilia beating can be found in~\cite{hilfingerChiralityCiliaryBeats2008}.

How the axoneme selects its beat is itself an open question, and competing pictures differ in what feedback turns steady dynein activity into an oscillatory waveform. The best-developed picture is one of spontaneous dynamic instability: the collective load-dependence of dynein detachment destabilizes the straight state, so oscillation emerges once activity exceeds a threshold, with no externally imposed regulation~\cite{kruse2005oscillations, julicherModelingMolecularMotors1997}. Alternative proposals hold that dynein activity is instead regulated by feedback from the axoneme's own deformation: either from the local curvature~\cite{machin1958wave, brokaw1971bend} or from the transverse forces that set the spacing between doublets and hence dynein engagement, the geometric-clutch hypothesis~\cite{lindemann1994geometric}. Because these mechanisms can reproduce similar waveforms, they are best discriminated by perturbation: measuring the phase relation between curvature and sliding, imposing external curvature, or varying the viscous load and asking whether beat frequency and amplitude respond as each predicts~\cite{ishibashi2020force, hilfingerChiralityCiliaryBeats2008}.

Another prominent example of the collective dynamics of filaments and motors occurs in the auditory hair-cells of the inner ear in animals \cite{reichenbachPhysicsHearingFluid2014, hudspethIntegratingActiveProcess2014}. Here, rows of hair-cells with a staircase-like pattern of actin filled stereocilia projecting out of each cell are arranged in an ordered fashion. Each generates active forces that are used to open and close ion channels. Similar mechanisms operate as pressure sensors in the fish lateral line.

\subsubsection{Muscle contraction}

\begin{figure}
\centering
\includegraphics[width=0.9\linewidth]{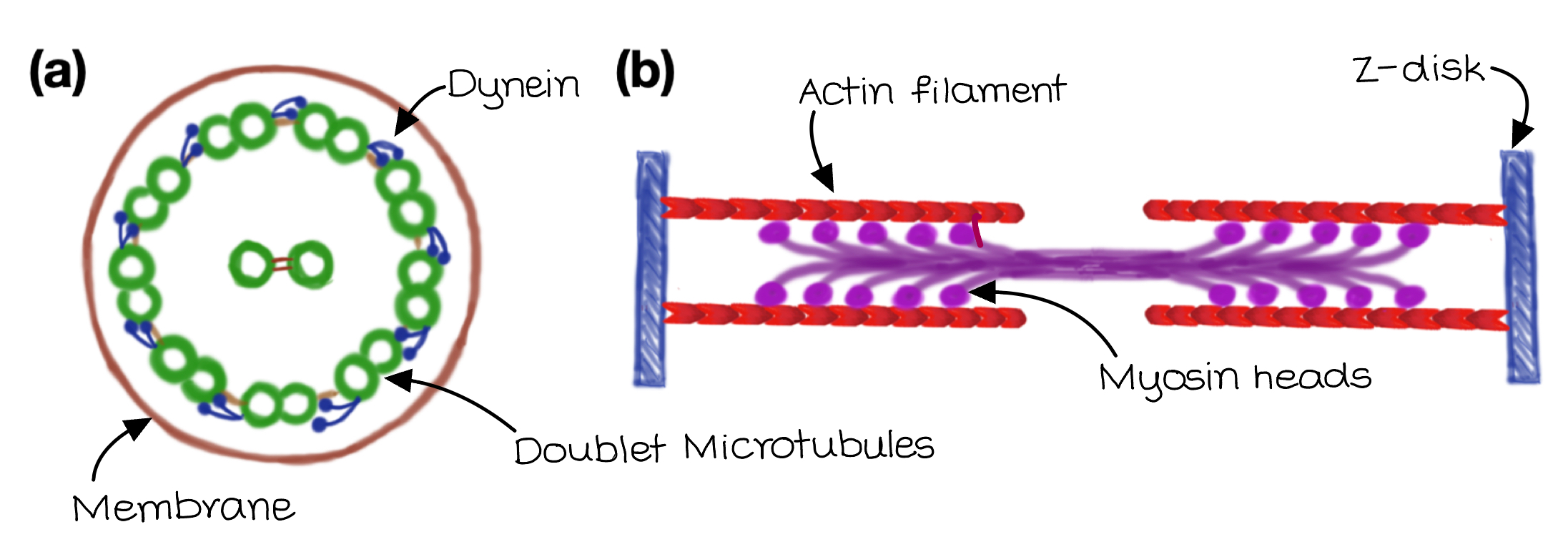}
\caption{Molecular motors often operate in a collective manner to generate large forces. Two prominent examples are the beating of cilia and the contraction of muscle cells. (a) Cross section through an axoneme (a $9+2$ arrangement of microtubules of overall diameter $\sim 100$ nm) showing the arrangement of microtubules and the interspersed dynein motors. The microtubules are arranged in doublets and pairs of these filaments can slide past each other via the action of molecular motors. Synchronous forces generated in this manner lead to the large-scale beating of a cilium. (b) The basic units for contractile force generation in muscle cells are sarcomeres (each unit a few $\mu$m long). These consist of bundles of antiparallel actin filaments interdigitated with bipolar myosin thick filaments. The ATP consuming activity of this collection of motors generates a net contractile force.}
\label{fig:cilia_sarcomere}
\end{figure}

A prominent example of the collective action of filament-motor systems is muscle contraction.  Sarcomeres are the minimum contractile unit of muscle cells. Here actin filaments bundle into parallel arrays with opposing polarity. In a sarcomere, one end of each bundle is anchored onto a protein rich structure called as Z-disk (see FIG.~\ref{fig:cilia_sarcomere}(b)). Individual muscle myosin motors are non-processive (a low duty cycle), with only $\approx 4-10\%$ attached to actin filaments at any instant. However, in sarcomeres, myosin motors form long chain structures called myosin thick filaments where only a fraction of motor heads are attached at any given time. The coordinated action of such myosin thick filaments pulling on parallel actin bundles generates a net macroscopic contractile force, of about 100 nN per sarcomere \cite{pavlov2009mechanical, leonard2010force, caruelPhysicsMuscleContraction2018}. Cooperative contraction of large numbers of such sarcomere units operate muscles. The regulation of muscle contraction is controlled by external signals, such as from neurons. Muscle cells can also show spontaneous oscillations of the contractile force they generate \cite{ishiwata1991spontaneous, ishiwata2010molecular}. A recent review can be found in~\cite{caruelPhysicsMuscleContraction2018}.

Muscle contraction is the classic setting in which competing frameworks can be placed side by side. The phenomenological description, due to Hill, represents the muscle as a contractile element in series and parallel with passive elastic elements, and summarizes the data through a hyperbolic force--velocity relation~\cite{fung2013biomechanics}. 

Hill's model is compact and predictive at the tissue scale but silent about mechanism. The competing, mechanistic framework is the Huxley cross-bridge model, in which each myosin head attaches, executes a force-generating power stroke, and detaches with strain-dependent rates~\cite{huxley1957muscle, HowardBook}, with the swinging-lever refinement of Huxley and Simmons~\cite{huxleySimmons1971proposed} tying the tension to a discrete conformational change of the bound head. The two are distinguished experimentally by rapid length-step (tension-transient) measurements: the fast elastic recovery exposes the sub-millisecond power stroke of individual cross-bridges, which the lumped Hill element cannot resolve. Modern treatments connect the pictures, deriving the force--velocity behaviour from the thermodynamics of the cross-bridge cycle~\cite{goupil2019thermodynamics} and, at coarse-grained scales, modelling muscle as a soft active hydraulic material~\cite{shankarActiveHydraulicsOdd2024}.

\section{Actomyosin cortex}
\label{sec:actomyosin_cortex}

Most eukaryotic cells are known to have a membrane associated cortical actin mesh consisting of actin filaments, cross-linkers and myosin motors \cite{alberts2014molecular, chughActinCortexGlance2018}. This submicron thick meshwork~\cite{clarkMonitoringActinCortex2013} is responsible for orchestrating important cellular processes such as division, motility, and polarity. Compared to the scales of a typical eukaryotic cell ($20~{\rm \mu m}$), the actomyosin cortex ($\approx 0.5~{\rm \mu m}$) can be treated as an effective two-dimensional thin film. Early studies noticed large-scale flows in the layers underneath the cell-membrane that were prominent during cytokinesis \cite{brayCorticalFlowAnimal1988}. More recent studies have revealed a highly branched and cross-linked meshwork of actin filaments in the actomyosin cortices of moving fibroblast cells \cite{svitkinaAnalysisActinMyosin1997}.

The actomyosin cortex is an active material (see  FIG.~\ref{fig:cortex_patterns}(a-b)). As discussed in Section~\ref{sec:molecular_motors}, ATP-fed myosin motors tug on actin filaments. Moreover, the entire actomyosin meshwork is remodelled through the continuous polymerization and de-polymerization of actin filaments with a typical turnover time of several tens of seconds~\cite{saha2016determining}. 
An appropriate coarse-grained description of the actomyosin cortex would be that of an active viscoelastic gel. The crossover timescale from the short-time elastic to the long-time fluid behaviour is typically controlled by the turnover rate of the components of the cortex. 

Experimental observations do not reveal significant long-range anisotropic spatial ordering of actin filaments such as found in sarcomeric structures. In an idealized network with equivalent contractile and extensile
configurations, their stress contributions would cancel on average. Recent work has shown that the nonlinear elasticity of actin filaments (which can lead to buckling), coupled with the nonequilibrium activity of motors, can break the symmetry between contractile and extensile behaviour. This  generates a net contraction of the actomyosin networks \cite{lenzContractileUnitsDisordered2012, lenzRequirementsContractilityDisordered2012,murrellFactinBucklingCoordinates2012}. Further, it appears that the breakdown of detailed balance in the binding-unbinding kinetics  of cross-linking proteins is sufficient to generate contractile stresses in polymeric gels without explicit force-generation from the activity of molecular motors \cite{chenMotorFreeContractilityActive2020}. Which of these mechanisms dominates (and, more basically, why actomyosin networks contract rather than extend) remains an open question: the sign of the active stress is an emergent property of filament nonlinearity, motor duty ratio, and crosslinker turnover rather than something fixed by symmetry alone.

Another mechanism that has been demonstrated experimentally for both microtubules and actin filaments is an entropy-driven contractility that arises when the filaments are crosslinked by diffusible crosslinkers~\mbox{\cite{Lansky2015,Kucera2021}}. This is akin to the expansion of an ideal gas confined in a cylinder closed by a movable piston. Actin filaments can also exert contractile forces when they are linked to a substrate via proteins of the formin class, such as mDia1. In this case, force generation originates in the fact that these proteins remain coupled to the barbed end even as the filament depolymerises, so that the retreating end drags on the anchor and places the filament under tension~\mbox{\cite{Jegou2013}}. This is similar to the force-generation mechanism proposed for chromosome separation by depolymerising microtubules, where a persistent coupling is maintained between the tubule and the kinetochore~\mbox{\cite{Koshland1988,Grishchuk2005}}. 

More recently, it has been shown that actin filaments can generate contractile stresses through allosteric effects caused by the binding of actin-severing proteins of the ADF/cofilin class~\mbox{\cite{Ashraf2025}}: binding of ADF/cofilin deforms the actin monomers, and this deformation manifests itself as a pulling force when the filament is attached to a substrate. Finally, an entropy-driven contraction generated by polymer assembly has been proposed for the ultrafast contraction of the stalk of the unicellular organism \textit{Vorticella}~\mbox{\cite{Mahadevan2000}} (see FIG.~\ref{fig:force_scales}). \textit{Vorticella} has a cell body anchored to the substrate by a long (\mbox{$\sim\!100~\mu$m}) stalk that is filled with charged polymers aligned along the long axis of the stalk. Influx of calcium ions screens these charges, making the polymers floppy and thereby causing the stalk to shrink.

Contractile stresses in the cortex have been measured by compressing rounded fibroblast cells with a flat AFM cantilever or using micropipette aspiration technique (see Section \ref{sec:measurement_techniques}), giving a tension of the order of 400 pN/$\mu$m. Deactivation of myosin-II motors results in a significant drop in this tension~\cite{tinevez2009role}.

Active hydrodynamic descriptions have been successful in explaining patterning in actomyosin cortex. We will discuss a couple of them in this section.  A simplified one-dimensional model \citep{boisPatternFormationActive2011,kumarPulsatoryPatternsActive2014} for pattern formation in the actomyosin cortex considers a diffusible molecule with a concentration field $c(x,t)$ which regulates active stresses. In the fluid limit, the total stress $\mathsf{\Sigma}$ generated in the cortex is
\begin{align}
\mathsf{\Sigma} = \eta \partial_x v + \zeta \Delta \mu \, f(c(x))
\label{eq:bois_consitutuve_eqn}
\end{align}
where $v$ is the hydrodynamic velocity field in the cortex with a viscosity $\eta$. Here stress is integrated across the cortical thickness and has
units of force per length; $\eta$ is the corresponding surface
viscosity. The active stresses are regulated by the function $f[c(x)]$, $\zeta$ is an activity coefficient, and $\Delta \mu$ is the chemical potential difference associated with ATP hydrolysis. The single coefficient $\zeta\Delta\mu$ carries the activity, and its sign fixes the character of the stress: $\zeta\Delta\mu > 0$ is contractile (the biologically prevalent case, whose microscopic origin is the filament buckling and detailed-balance-breaking crosslinker kinetics noted above), while $\zeta\Delta\mu < 0$ is extensile. The measured cortical tension, approximately
$400~{\rm pN}/{\rm \mu m}$ ($0.4~{\rm mN/m}$) in the example
above, provides an experimental scale for the active stress
$\zeta\Delta\mu f(c)$. This scalar active stress is the isotropic reduction of the general polar and nematic active stresses of Section~\ref{sec:soft_matter_theory} (Eqs.~\eqref{eq:p_active_stress},~\eqref{eq:Q_active_stress}): with filament orientation disordered only the isotropic part $\zeta\Delta\mu\,\mathbb{I}$ survives, whereas the tensorial $\overline{\zeta}\Delta\mu\,\mathsf{Q}$ contribution re-enters once nematic order develops, as in the cytokinetic ring discussed below. Low-Reynolds number physics dictates that any fluid element has a force-balance condition imposed on it at every instant of time. With a simple frictional approximation of the form $F_{\rm ext} = -\gamma \, v$ for the traction forces arising from the underlying cytoplasm and the plasma membrane above, the force-balance condition reads
\begin{align}
\partial_x \mathsf{\Sigma} = -F_{\rm ext} = \gamma \, v.
\label{eq:bois_force_balance}
\end{align}
Using \eqref{eq:bois_consitutuve_eqn} in \eqref{eq:bois_force_balance} leads to a flow equation
\begin{align}
\eta \partial_x^2 v - \gamma v = - \zeta \Delta\mu \, \partial_x f(c(x)).
\label{eq:flow_equation}
\end{align}
This equation predicts that gradients in the cortical tension lead to cortical flows. This picture has been experimentally verified by measuring large-scale cortical flows in the zygotes of $C.~elegans$, monitored using the spatial profile of myosin as an input~\cite{mayer2010anisotropies}.

\subsection{Cortical pattern formation}

The flows resulting from contractility gradients, via \eqref{eq:flow_equation}, will advect any molecules present in the cortex. In particular, the elements that are responsible for the generation of active stresses, such as myosin motors, will also be subjected to advective fluxes. The dynamics of the concentration $c$ of the diffusible molecule  can be modelled by
\begin{align}
\partial_t c = -\partial_x j + R(c), \qquad j = v \, c - D \partial_x c,
\label{eq:bois_conc_eqn}
\end{align}
where $D$ is a diffusion coefficient and $R(c)$ represents the changes to the concentration occurring from local chemical reactions, for instance the ever-present turnover of molecules between the cortical surface to the cytoplasmic bulk. 

Equations \eqref{eq:flow_equation} and \eqref{eq:bois_conc_eqn} provide a closed set of autonomous equations for the evolution of the concentration field $c$ and the hydrodynamic velocity field $v$. A trivial solution is a spatially uniform concentration field $c_0$ which is a solution of $R(c_0)=0$ and zero velocity $v=0$. However, this uniform state can become unstable upon increasing the strength of the active stresses, i.e., the coefficient $\zeta\Delta\mu$. 

\begin{figure}
\centering
\includegraphics[width=0.9\linewidth]{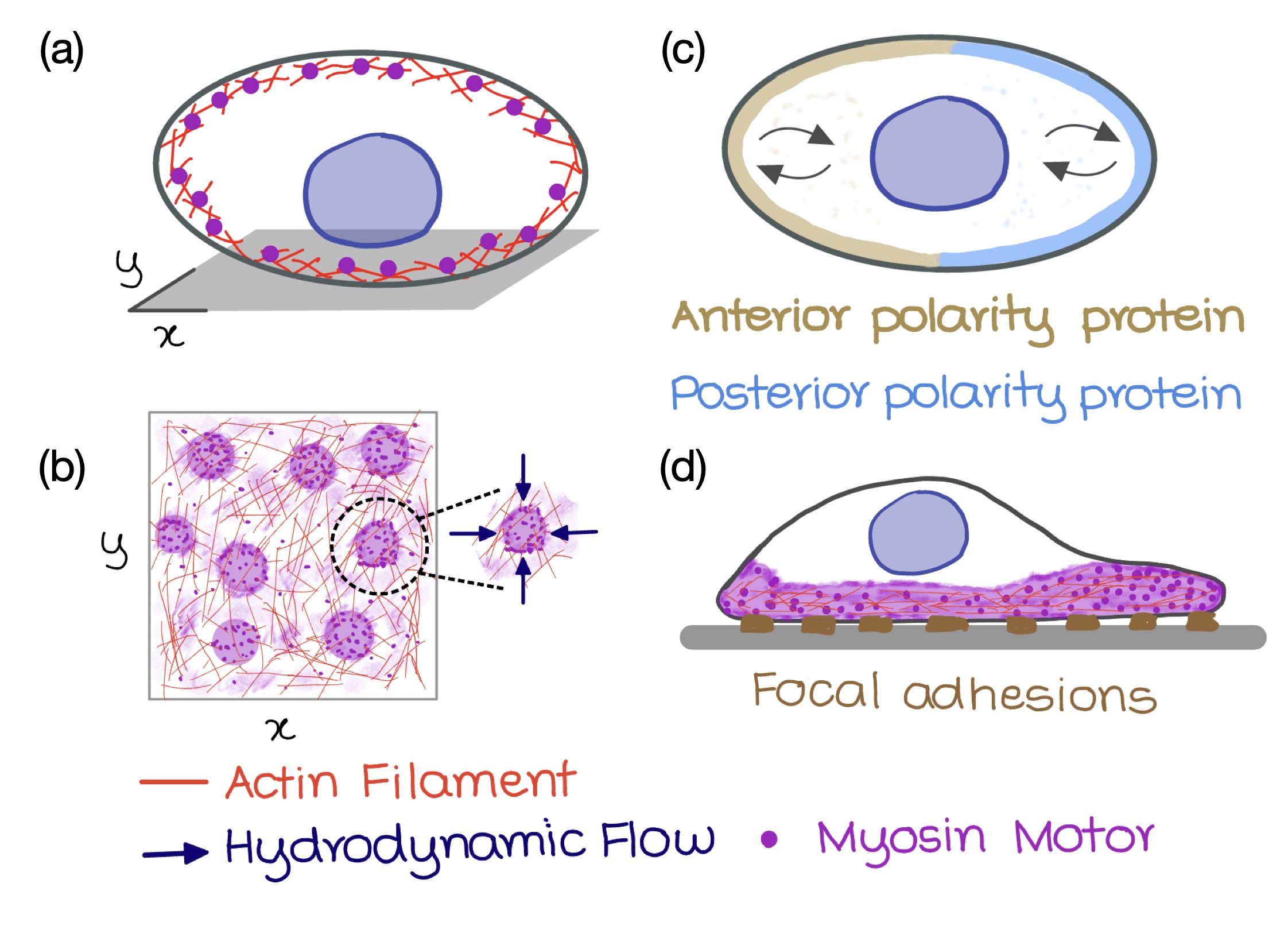}
\caption{(a) The actomyosin cortex is a thin meshwork of actin filaments, myosin motors and crosslinkers. Pattern formation in the actomyosin cortex typically results from the competition between surface diffusion, turnover kinetics between the cytoplasm and the cortex, and hydrodynamic flows arising from actomyosin contractility gradients. (b) At high contractility, spontaneous patterns can emerge in both the concentration of active stress regulators and hydrodynamic flows. (c) Such flows can also couple to surface concentrations of cell polarity proteins. The macroscopic segregation of such cell-surface proteins and signalling molecules leads to the establishment of an antero-posterior axis in {\em C. elegans}. (d) In crawling cells, gradients of acto-myosin contractility between the front and rear end of the cell generate asymmetric traction forces transmitted to  the substrate via focal adhesions, thereby contributing to motility.}
\label{fig:cortex_patterns}
\end{figure}

An intuitive way of understanding this instability is as follows: while diffusive fluxes tend to homogenize concentration fluctuations, advective fluxes generate contractile flows that tend to enhance them.  High activity, i.e., large P\'eclet numbers $\mathrm{Pe}=\zeta\Delta\mu/\gamma D \gg 1$, can destabilize the homogeneous state and lead to a patterned state with an inhomogeneous concentration and flow profile \cite{boisPatternFormationActive2011}. Specifically, with a choice $R(c) = -k \, (c-c_0)$, a linear-stability analysis of the above equations shows that concentration fluctuations $\delta c_q$ at a wavenumber $q$ around the homogeneous state $c=c_0$ have a growth rate $\lambda_q$ with
\begin{align}
\lambda_q = -k - D \, q^2 \left(1 - \frac{\mathrm{Pe} \, c_0 \, \partial_c f(c_0)}{1 + q^2\ell_h^2} \right)
\end{align}
where $\ell_h = \sqrt{\eta/\gamma}$ and $\mathrm{Pe}$ is the P\'eclet number defined above~\cite{boisPatternFormationActive2011}. For small $\mathrm{Pe} \ll 1$, we see that these concentration fluctuations are suppressed while for large activities $\mathrm{Pe} \gg 1$, these concentration fluctuations grow exponentially until the nonlinear terms suppress their growth. For $k\neq0$, the growth rate $\lambda_q$ has a maximum for a non-zero wave number $q$, which defines the length-scale of the emergent concentration patterns. Figure \ref{fig:cortex_patterns}(b) shows such an active mechanochemical pattern in the concentration field of myosin motors in the actomyosin cortex of cells.

It is straightforward to generalize \eqref{eq:bois_conc_eqn} to $n$ species each of which could, in principle, regulate the active stress $\zeta\Delta\mu \; f(\{c_1, c_2, \ldots, c_n\})$. Such extensions lead to active pulsatory patterns \cite{kumarPulsatoryPatternsActive2014} and also extend classical Turing patterns for reaction-diffusion systems \cite{turingChemicalBasisMorphogenesis1952} to mechanochemical patterns. 

Generalizing the equations \eqref{eq:bois_force_balance}-\eqref{eq:bois_conc_eqn} to a curved shape and coupling them to bulk cytoplasmic flows leads to spontaneous patterns that can either resemble cell-polarity or those corresponding to cell division, depending on the relative values of the cytoplasmic and cortical viscosities \cite{mietkeMinimalModelCellular2019}. The formation of the cytokinetic ring with enhanced myosin levels in the middle of the cell, the associated orientation patterns of actin filaments and the ensuing contractile flows can thus be accounted for within simplified hydrodynamic models. Furthermore, studies with dynamical shapes can also explain the temporal progression of the cell shape driven by the constriction of the contractile ring towards cytokinesis \cite{turlierFurrowConstrictionAnimal2014}.

Systems with anisotropic order parameters, such as a polarity field $\mathbf{p}$ or a nematic tensor field $\mathsf{Q}$, have a richer set of pattern forming instabilities \cite{julicherActiveBehaviorCytoskeleton2007, doostmohammadiActiveNematics2018}.  For instance, recognizing that cortical dynamics can influence the dynamics of membrane associated proteins, it has been demonstrated that polar orientational patterns in the actomyosin cytoskeleton regulate the formation of nano-clusters of GPI-anchored proteins. These are involved in various signalling mechanisms. The emerging view is that of an active composite cell surface wherein the composition and dynamics of the cellular periphery is actively regulated by its internal components \cite{gowrishankarActiveRemodelingCortical2012}.

Active anisotropic patterns also explain the orientation of actin filaments in the cytokinetic ring during cell division. The nematic orientation $\mathsf{Q}$ is driven by hydrodynamic flows which are themselves guided by gradients of active stress. Within a simplified two-dimensional theory that is appropriate to a thin-shell description of the cortex, an imposed active stress profile that corresponds to enhanced myosin levels at the cell centre generates contractile flows and a corresponding alignment of actin filaments similar to those seen in the cytokinetic ring \cite{zumdieckContinuumDescriptionCytoskeleton2005, salbreuxHydrodynamicsCellularCortical2009}. Such flow driven orientation patterns during cytokinesis have been experimentally measured. They agree quantitatively with hydrodynamic descriptions \cite{reymannCorticalFlowAligns2016}.

\subsection{Cell polarity\label{sec:cell_polarity}}

Most cell divisions are symmetric: the two daughter cells are identical in volume, shape, and genetic content. However, some cell divisions, particularly those that occur during cell-fate specification or early embryonic phases, are asymmetric, with the two daughter cells differing in their physical sizes or in the segregation of the cytoplasmic components during cytokinesis. Such asymmetric cell division implies the establishment of a cell polarity which distinguishes the two halves of the dividing cell.

An archetypal example of asymmetric cell division driven by cell-polarity is the first division of the \emph{C. elegans} zygote. Incidentally, this very first division also sets the organismal polarity by defining the antero-posterior axis of the organism. Cell polarity here is established by the asymmetric segregation of PAR (partitioning defective) proteins on the surface of the cell (see  FIG.~\ref{fig:cortex_patterns}(c)). Recent studies have clearly demonstrated that PAR polarity in \emph{C. elegans} zygotes is driven by large-scale hydrodynamical flows generated in the cortex in response to gradients in myosin contractility \cite{goehringPolarizationPARProteins2011}. The spatiotemporal profiles of the protein concentrations and cortical flows can be quantitatively accounted for using an active gel theory of the actomyosin cortex coupled to reaction-diffusion-advection equations for the concentrations \cite{grossGuidingSelforganizedPattern2019}. 

A typical hydrodynamic model for the establishment of cell polarity takes into account (i) the transport of polarity marker proteins by both passive diffusion and active advection, (ii) biochemical interactions between the components such that spatially homogeneous and patterned profiles coexist within the same parameter regime, and (iii) spatiotemporal cues that trigger the transition of the system from the unpolarized state to the polarized state. Experimental evidence suggests that weak stochastic fluctuations cannot lead to a spontaneous transition of the system from an unpolarized state to a polarized state. This transition is effected when a sufficiently strong upstream signal triggers the system.

 Establishing cell polarity is the first step towards selecting a direction for cell movement. In the next subsection, we use the simple model for polarity patterns in the actomyosin cortex developed here to address the motility of crawling cells.

\subsection{Cell motility}

Cell migration necessarily requires the development of a front-rear polarity (see  FIG.~\ref{fig:cortex_patterns}(d)). The coupling of this polarity to mechanochemical processes can lead to the emergence of cell motility~\cite{bray2000cell, carlssonMechanismsCellPropulsion2011, rechoMechanicsMotilityInitiation2015}. As discussed in Section~\ref{sec:cell_polarity} on cell polarity, the emergence of a front-rear asymmetry is typically achieved by internal chemical reactions with antagonistic dynamics. Such reactions can lead to bistability between a homogeneous and polarized distribution in the cell~\cite{mori2008wave}. Such polar patterns in chemical concentrations, when coupled to active stresses in viscous/viscoelastic acto-myosin cortex, can lead to a net velocity for the cell. We now discuss a simple model for cell migration based on this picture~\cite{aransonPhysicalModelsCell2016}.

Consider the cell as a one-dimensional, moving viscous active fluid element of length $L(t) = l_+(t) - l_{-}(t)$, where $l_\pm$ are the locations of the front and rear ends. Note that $d l_\pm/dt = v(l_\pm)$, where $v$ is the velocity field. The total stress is of the form $\mathsf{\Sigma} = \eta \partial_x v + \mathsf{\Sigma}_a$, where we assume an active stress of the form $\mathsf{\Sigma}_a = (c/c^*)\zeta \Delta \mu$, with $c$ and $c^*$, respectively, the actual and reference myosin concentrations. Using the force balance equation ~\eqref{eq:bois_force_balance} to eliminate the velocity field $v$, the total stress follows 
\begin{align}
\label{eq:mot_sigma}
\ell_h^2\partial_{xx} \mathsf{\Sigma}  - \mathsf{\Sigma} = -(c/c^*)\zeta \Delta  \mu,
\end{align}
where $\ell_h = \sqrt{\eta/\gamma}$. Following \eqref{eq:bois_conc_eqn}, the evolution of $c$ is given by
\begin{align}
\label{eq:mot_c}
\partial_t c + \gamma^{-1}\partial_x (c \partial_x \mathsf{\Sigma}) = D \partial^2_{xx}c.
\end{align}
At the moving boundaries, in addition to the kinematic relations 
\begin{equation}
d l_\pm/dt = \gamma^{-1}\partial_x \mathsf{\Sigma}\vert_{x = l_\pm},
\end{equation}
we assume that $c$ and $\mathsf{\Sigma}$ satisfy
\begin{align}
\partial_x c(l_{\pm}, t) = 0, 
\qquad
\mathsf{\Sigma} (l_\pm)  = -K\; (L(t) - L_0)/L_0,
\end{align}
where the stress boundary conditions represent an effective
elastic resistance to changes in cell length, with $K$ an
effective compression modulus and $L_0$ the preferred cell length.

At small values of activity $\zeta \Delta \mu$, the only solution is a uniform profile of $c(x)$ and a spatially uniform stress, yielding $v(x) = 0$. On the other hand, at high activity, the instability of the homogeneous state can lead to appearance of spontaneous patterns in $c$ and hence in $\mathsf{\Sigma}$. Assuming a steady state patterned solution $c(x)$ and $\mathsf{\Sigma}(x)$ satisfying \eqref{eq:mot_sigma} and \eqref{eq:mot_c} with $dl_{+}/dt = dl_{-}/dt$, i.e., $dL/dt = 0$, the centre of mass velocity of the cell is given by
\begin{equation}
    V_{\rm cell} = -\frac{\zeta \Delta \mu}{\eta c^*}\int_{-L_s/2}^{L_s/2} \frac{\sinh(z/\ell_h)}{2 \sinh(L_s/2\ell_h)} c(z)dz,
\end{equation}
where $z$ is the spatial coordinate in the centre of mass ($z = 0$) frame, and $L_s$ is the steady-state length~\cite{carlssonMechanismsCellPropulsion2011}. Note that if either $\zeta \Delta \mu = 0$ or $c(z)$ is an even function, then $V_{\rm cell} = 0$. Clearly, it is the odd part of $c(z)$ which breaks the front-rear symmetry, i.e., makes the cell polar. Therefore, for cell motility, we need both non-equilibrium energy input ($\zeta \Delta \mu \neq 0$) and a mechanism to break the fore-aft symmetry ($c(z) \neq c(-z)$) -- another illustration of the Curie principle discussed in Section~\ref{sec:mechanochemical_transduction}. 

The mechanism for cell migration discussed above involved the generation of active stresses by the action of molecular motors. However, other mechanisms that can break detailed balance such as treadmilling coupled with spatial symmetry breaking can also lead to cell migration. For instance, in the case of migratory cells known as keratocytes, cell movement is mostly driven by actin treadmilling where polymerization occurs predominantly at the leading edge and depolymerization occurs away from the leading edge~\cite{mogilner2020experiment}. 

More broadly, cell migration is conventionally divided into two limiting modes. In the protrusion- or treadmilling-dominated mode, motion is powered by actin polymerization at the leading edge (the polymerization-ratchet mechanism of Section~\ref{sec:poly_forces}), with the resulting retrograde flow transmitted to the substrate through the molecular clutch at focal adhesions; fast-gliding fish keratocytes are the canonical example~\cite{mogilner2020experiment, chanTractionDynamicsFilopodia2008a}. In the contractility-dominated mode, illustrated by the model above, myosin-generated cortical stresses and flows propel the cell, a mechanism that becomes dominant for amoeboid and bleb-based migration under weak adhesion or strong confinement. Which mode prevails is set largely by adhesion strength and confinement, and the two can be dissected experimentally: inhibiting Arp2/3 or formins suppresses the protrusive mode, whereas inhibiting myosin~II suppresses the contractile one~\cite{sengupta2021principles}.

\subsection{Coupling to membrane organization}

The integrity of the actomyosin cortex can sometimes be compromised, leading to a rupture of the cortical meshwork from the membrane. The interfacial tension generated in the membrane-cortex composite is balanced by the pressure difference across the cell surface in accordance with Laplace's law. After separation of the cortex from the membrane, the turgor pressure in the cytosol leads to a localized ballooning of the cell membrane -- a bleb. The restoration of the cortex allows the bleb to be reabsorbed~\cite{charras2008blebs}.

Coupling between the actomyosin cortex and the membrane plays an important role in regulating traffic of cargo-laden vesicles into and out of the cell. For instance, actin polymerization can influence the protrusion of the membrane to form vesicles~\cite{zhangModelingEndocytosisYeast2015}.

Over the past few decades, the role of the coupling between the cell cortex and proteins embedded in the plasma membrane has attracted renewed attention. An early hypothesis was that of membrane ``rafts'', local fluid-like micro-phase-separated regions on the cell surface which are enriched in specific proteins, signalling molecules and cholesterol. Such rafts were initially thought to be equilibrium objects, formed through phase separation, consistent with a ``fluid mosaic'' model for the cell membrane. 

Recent work suggests an intriguing alternative closer to available data. The enrichment of nano-domains on living cell membranes is induced by the motion of myosin motors on the cell cortex just below the membrane surface, connected to the relevant integral membrane proteins through appropriate linkers, and modulated by the presence of cholesterol. Motor motion on an actively remodelling cytoskeleton can lead to the formation of star-like structures (asters), with a definite polarity. Motors linked to membrane proteins can drive their cargo to accumulate at points defined by the centres of asters, leading to an enrichment of specific components. Passive lipid--lipid interactions are thus not the sole driver of lateral organization; they are supplemented, and often dominated, by the active behaviour of the actomyosin cortex~\cite{gowrishankarActiveRemodelingCortical2012}. The complex spatio-temporal organization of myosin motors on reconfiguring actin tracks leads to the formation of spontaneous nano-scale domains that live as long as they are stabilized through protein attachments, after which they dissolve.

Active forces generated in the actomyosin cortex can also lead to large scale fluid flows within cells. Such cytoplasmic flows are implicated in positioning various organelles and even cell motility \cite{mogilnerIntracellularFluidMechanics2018}.

These two pictures rest on different assumptions and are distinguished by experiment. In the equilibrium view, descending from the fluid-mosaic model~\cite{singer1972fluid}, lateral heterogeneity arises from passive phase separation set by lipid composition and cholesterol, with domain size fixed by the balance of line tension against mixing entropy; model membranes indeed show micron-scale liquid-ordered/liquid-disordered coexistence, and giant plasma-membrane vesicles sit close to a demixing critical point~\cite{simons1997functional, veatch2008critical}. This picture, however, does not account for the domains inferred in living cells, which are nanometric and short-lived rather than macroscopic and static. The active view instead attributes these nanodomains to ATP-driven myosin activity on cortical actin, which clusters linked membrane components and sets their size and lifetime~\cite{gowrishankarActiveRemodelingCortical2012}. The frameworks are most cleanly discriminated by perturbing activity rather than composition: equilibrium domains should depend on temperature and lipid composition but be insensitive to ATP, whereas active nanodomains dissolve, and lose their characteristic concentration-independent clustering, upon ATP depletion or actin/myosin disruption.

\section{Chromatin, the nucleus, and cell division}

\begin{figure}[ht]
\centering
\includegraphics[width=0.7\linewidth]
{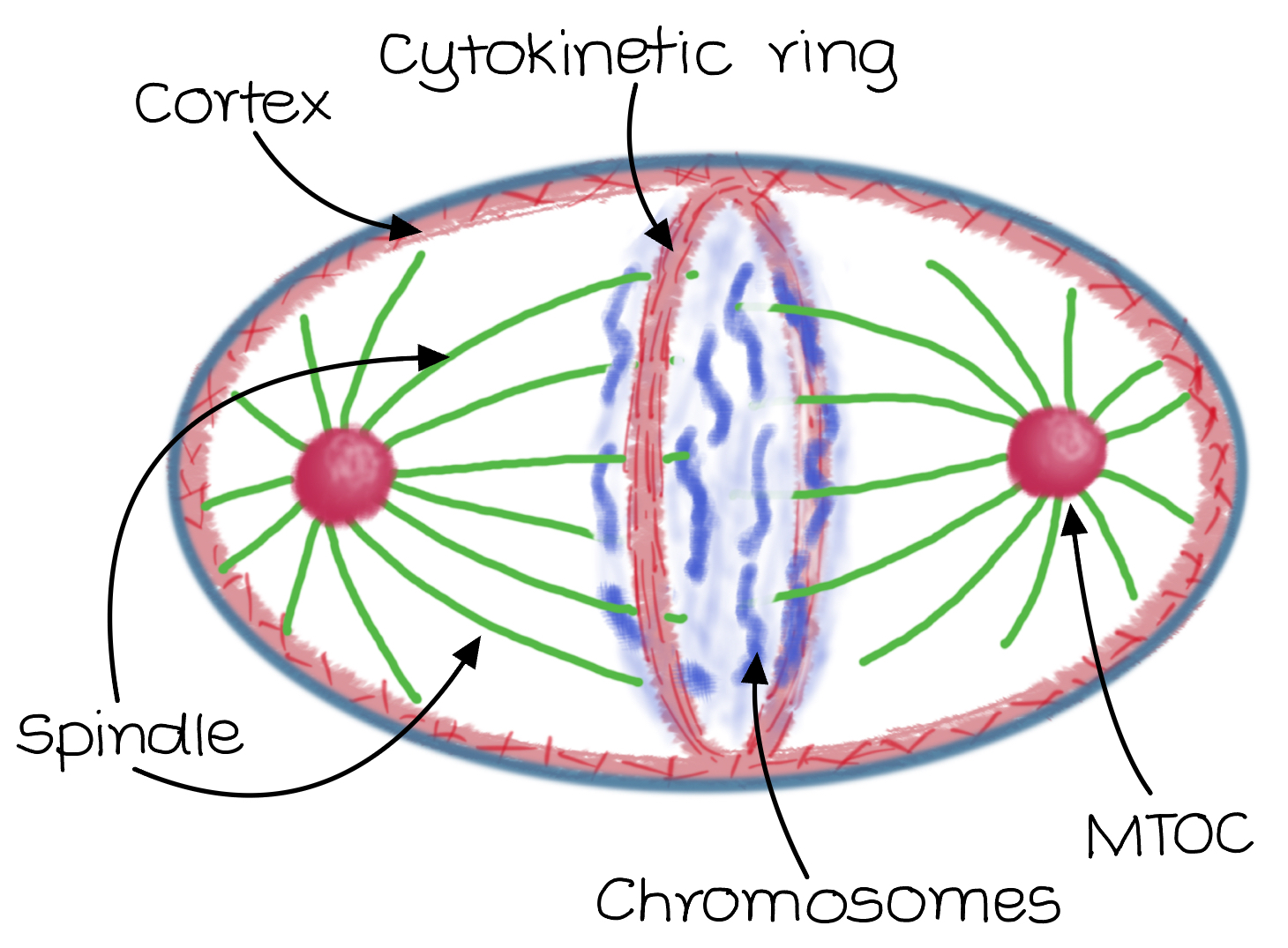}
\caption{A dividing cell assembles a highly coordinated complex machinery of the actomyosin and microtubule cytoskeleton to drive segregation of the chromosomes. The mitotic spindle as shown is a structure that facilitates the segregation of the duplicated chromosomes. The spindle involves microtubules that emerge from the two microtubule organizing centres (MTOC). One set of microtubules contact the chromosomes, while the other set (astral microtubules) interacts with the cortex. Large-scale cortical flows can contribute to the assembly of the cytokinetic ring on the cellular surface which ``pinches'' the mother-cell into two daughter cells. }
\label{fig:cell_division}
\end{figure}

A fundamental property of living systems is their ability to replicate. Cell division is achieved in two stages. In the first stage, genetic material in the form of DNA is duplicated, followed by a doubling in cytoplasmic components. In the second stage, the two DNA copies are segregated into opposite sides of the mother cell. Concomitantly, the process of cytokinesis cleaves the mother cell into two daughter cells. Beyond microscopic processes involved during DNA replication, the segregation of chromosomes and cytokinesis must involve cell-scale forces (FIG.~\ref{fig:cell_division}).

The duplication of DNA is achieved by the translocation of helicase motor proteins which unzip the double-stranded DNA, followed by the duplication of the genetic sequence aided by DNA polymerases. Both the translocation of the helicase motor and the action of the DNA polymerase involves mechanical forces. The movement of the helicase can be modelled using the Brownian ratchet formalism discussed in Section~\ref{sec:molecular_motors}~\cite{burnham2019mechanism}. More broadly, the replication fork is a force-generating and force-sensitive machine: the helicase advances against the base-pairing and torsional resistance of the template, and single-molecule experiments that vary the tension on the template show that the replication rate of a single DNA polymerase depends strongly on the applied force, exposing the mechanical coupling between synthesis and template deformation~\cite{maier2000replication}. As the fork proceeds it overwinds the DNA ahead into positive supercoils, and the torque so generated must be relieved by topoisomerases for replication to continue, the same torsional bookkeeping met above for transcription.

In eukaryotes, the second stage of cell division involves the formation and positioning of the mitotic spindle and the subsequent scission of the mother cell into two daughter cells. Both these processes involve mechanical forces in an essential manner. Below, we discuss these processes only in the context of eukaryotic cells.

\subsection{\label{sec:chromatin}Chromatin and chromosome organization}

Chromatin is the term for the organized complex of DNA and proteins that is found in eukaryotic cells.  In mammalian cells, DNA is wound around a structure made by dimerizing a complex of four histone proteins to form an octamer. DNA winds around such histone octamers. The combination of DNA and histones is a nucleosome core particle~\cite{cremer2010chromosome}. 

During various physiological processes, such as packaging into chromosomes, replication, repair, protein binding and transcription, the DNA is subjected to stretching, twisting, and bending. mRNA production, called transcription, is carried out by RNA polymerase (RNAP), a molecular machine that moves on DNA. An estimate for the force that can be generated by the motor can be obtained from $F_{\rm max} = \Delta G/\delta$, where $\Delta G$ is the free energy available for mechanical work per nucleotide addition (a few $k_BT$) and $\delta = 0.34$~nm is the size of a typical nucleotide. Because the step-size is small, the force that can be generated for RNAP is large, up to $\sim 25$--$30$~pN, consistent with direct single-molecule measurements~\cite{wangForceVelocityMeasured1998}. This is much larger than the forces generated by MT or actin associated motors. The forces generated by RNAP are predominantly spent on untwisting DNA, separating its strands and then moving along the separated strands. This process involves the formation of higher order coiled structures of DNA (supercoils) and requires application of internal torques. 

\begin{figure}
\centering
\includegraphics[width=0.9\linewidth]{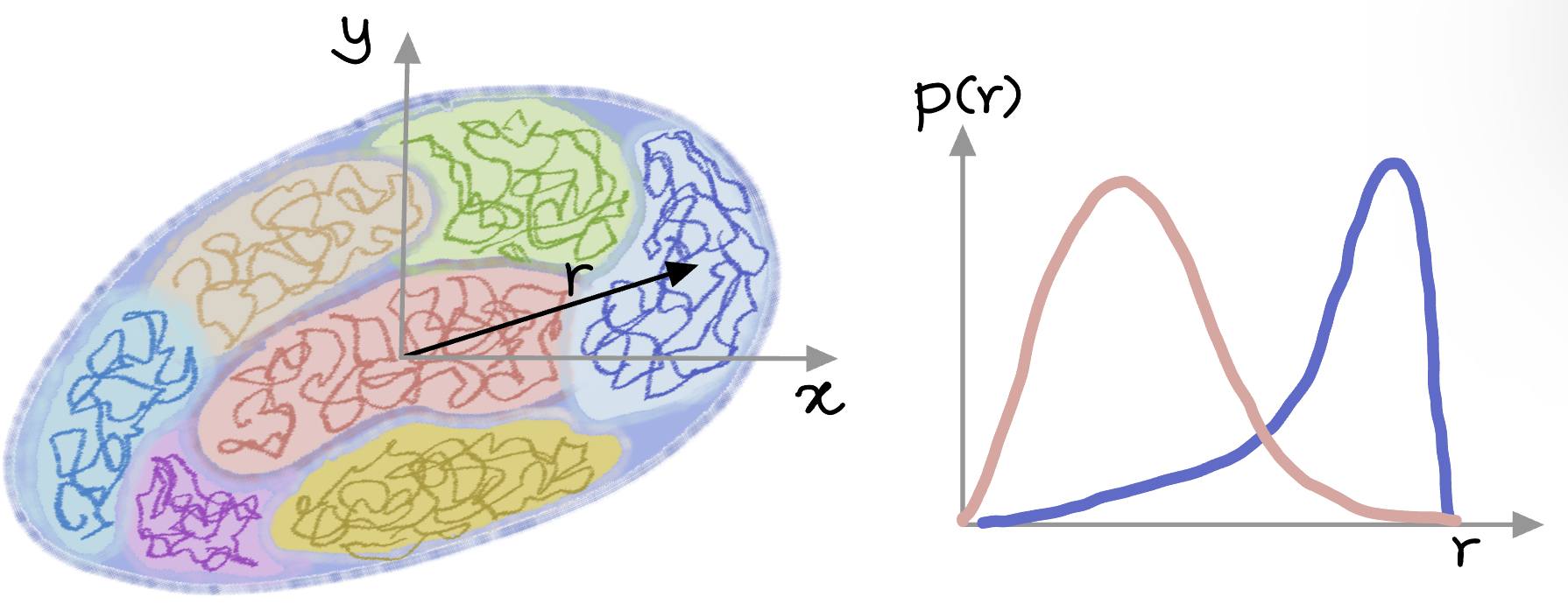}
\caption{Chromosomes in higher eukaryotic cells are organized spatially into largely non-overlapping chromosome territories. The spatial arrangement of individual chromosomes is also non-trivial, with more gene-dense chromosomes tending to be found at more interior locations than gene-poor ones. Such non-trivial positioning is apparent in fluorescence in-situ hybridization experiments in which individual chromosomes can be labelled and their distributions across an ensemble of cells quantified.}
\label{fig:chromosome_territory}
\end{figure}

Within the genome, highly compact {\em heterochromatin} regions, where nucleosome core particles are tightly packed, are typically gene poor and transcribed at low levels, while more loosely bound {\em euchromatin} regions are gene rich and transcribed more.  Fluorescent labelling experiments show that they occupy largely non-overlapping ``chromosome territories'' within the nuclei of higher eukaryotes (FIG.~\ref{fig:chromosome_territory}). The location of these territories is not random. Much evidence points to a favoured arrangement of chromosomes in which the regions close to the nuclear centre tend to be gene-rich, while gene-poor and more dense regions of chromatin appear more peripherally, towards the nuclear envelope. The creation of such non-trivial spatial patterns requires a coherent movement of large molecules across micron sized regions, arguing for the crucial importance of forces at the nuclear scale.

A distinct and now-central source of force in chromatin organization is \emph{loop extrusion} by SMC-family motor complexes. Condensin and cohesin are ATP-driven machines that capture DNA and reel it in to grow progressively larger loops; single-molecule imaging shows a single condensin extruding tens of kilobases at up to $\sim 1.5~{\rm kb/s}$ against loads of order a piconewton~\cite{ganji2018realtime}. Such loop-extruding motors organize interphase chromatin into topologically associating domains and compact chromosomes for mitosis, and polymer models in which processive extrusion competes with diffusive loop growth reproduce the domain structure seen in chromosome-conformation-capture data~\cite{fudenberg2016formation}. Loop extrusion thus supplies an explicit, motor-level force-generation mechanism complementing the coarse-grained active descriptions below.

Current questions regarding force generation at the level of chromatin begin by considering chromosomes as polymers. Some approaches emphasize single polymer properties, such as the statistics of contacts in comparison to the chromosome capture data. Experiments suggest the presence of coherent motions of chromatin across micron size regions over tens of seconds within the nucleus~\cite{zidovska2013micron}. This dynamics is ATP dependent, suggesting that activity should be a core component of any description of nuclear architecture~\cite{weber2012nonthermal}. 

Two independent, contemporaneous approaches tackle this question, and they prove complementary. The first is hydrodynamic: active effects are incorporated into a coarse-grained stress tensor for chromatin~\cite{bruinsmaChromatinHydrodynamics2014}, and a multiphase continuum model of an active two-phase liquid adds crosslinking proteins within heterochromatin that generate contractile forces~\cite{Rautu2025}. A peripheral heterochromatin layer then emerges from a competition between three mechanical effects: heterochromatin contractility, active euchromatin stresses, and a confining boundary. These calculations reproduce the coherent, micron-scale motions of chromatin seen in experiments, but do not resolve individual chromosomes, their positioning, or the commonalities of chromatin organization across cell types. Related active-polymer descriptions along similar lines have also been developed more recently~\cite{Mahajan2022}.

The second approach represents the same activity differently, mapping the differential transcriptional activity across roughly one-megabase segments of the genome onto a local effective temperature~\cite{ganai2014chromosome}. Brownian-dynamics simulations of human chromosomes under this mapping reproduce the radial organization of chromosomes by gene density, the separation of euchromatin and heterochromatin, the differential positioning of the two X chromosomes in female cells, and chromosome-specific density and centre-of-mass distributions~\cite{agrawal2020nonequilibrium}. That differential activity within and across chromosomes might provide a general and comprehensive explanation for many different properties of large-scale nuclear architecture was first suggested in Refs~\cite{ganai2014chromosome, agrawal2020nonequilibrium}. The limitations of this approach mirror that of the hydrodynamic model: the effective-temperature mapping is phenomenological, replacing the vectorial, correlated non-equilibrium forcing by a single scalar calibrated to transcription data rather than derived from force generation, so it captures steady-state organization rather than the coherent currents and non-equilibrium dynamics themselves.

The two frameworks are thus complementary rather than competing: the hydrodynamic description captures the coherent dynamics but not single-chromosome positioning, whereas the effective-temperature description captures positioning and compartmentalization but not the active currents. Unifying the two, by deriving the effective activity parameters from the underlying force-generating processes and predicting both the static architecture and its dynamics within one framework, remains an open problem.

\subsection{The nucleus}

The nucleus is a dynamical organelle that exhibits mechanoresponsive behaviour, modulating cellular functions in response to mechanical stimulation. Some part of this primarily viscoelastic response involves the ability to deform the nucleus, changing its volume and shape.  The field of nuclear mechanotransduction studies how local forces might influence transcription programs. There appear to be three mechanisms with which microenvironment regulates gene expression. These are via the control of the nuclear import of different transcription factors, changes in nuclear organization, and spatiotemporal mechanoregulation of co-transcribed genes~\cite{uhlerRegulationGenomeOrganization2017}. All of these can be affected through the dynamics of the cytoskeleton outside the nucleus, itself influenced by the extracellular environment. 

The nuclear volume is determined by a balance of pressures between inside and outside. Osmotic pressure differences originating in preferentially localized soluble molecules are believed to largely determine this volume. The outward pressure includes contributions from chromatin, the fluid component of the nucleoplasm, as well as from proteins and small molecules. The pressure acting inwards originates in the cytoplasm and has a component reflecting forces arising from the cytoskeleton. 

Physical connections between chromatin and the nuclear envelope determine the mechanics of the nucleus. The nuclear envelope is composed of outer and inner nuclear membranes, forming a double lipid bilayer. The nuclear lamina, a mesh-like protein network below the inner membrane, is comprised primarily of lamin proteins. These anchor chromatin to the nuclear envelope, aided by a variety of other proteins.  The transmission of forces between the cytoskeleton and the nucleus regulates nuclear movement and positioning, as required, for example, in cell migration. 

The outer nuclear membrane connects to the endoplasmic reticulum. This provides a reservoir of lipids that fuel nuclear surface area changes. Nuclear Pore Complexes (NPCs) embedded in the nuclear envelope regulate the active nuclear transport of large macromolecules, while similarly embedded mechanosensitive ion channels respond to changes in nuclear membrane tension, allowing for the flux of ions and small molecules.  NPCs respond to nuclear membrane tension by increasing their diameter, so that mechanoresponsive transcription factors (such as YAP) can be imported.  Other stretch-sensitive ion channels respond to tension by allowing the release of calcium from the ER, enhancing cell contractility. 

Forces acting on the nucleus can lead to changes in the lamin network. Other responses to an applied force include nuclear stiffening, arising from the recruitment of lamins to the nuclear envelope. The liquid-liquid phase separation of intrinsically disordered proteins within the nucleus could also participate in nuclear mechanosensing.

Stem cells can also exhibit unusual mechanical properties, such as  a negative Poisson's ratio in a transitional regime~\cite{pagliara2014transition}. While conventional materials expand in the transverse direction while subjected to a longitudinal compressive force, auxetic materials contract. A theory for such mechanics that incorporates the coupling of nuclear dimensions to chromatin states that arise as a consequence of activity  has been proposed in~\cite{tripathiChromatinCompactionAuxeticity2019}. 

\subsection{The mitotic spindle}
\label{sec:spindle}

\begin{figure}[ht]
\centering
\includegraphics[width=0.9\linewidth]{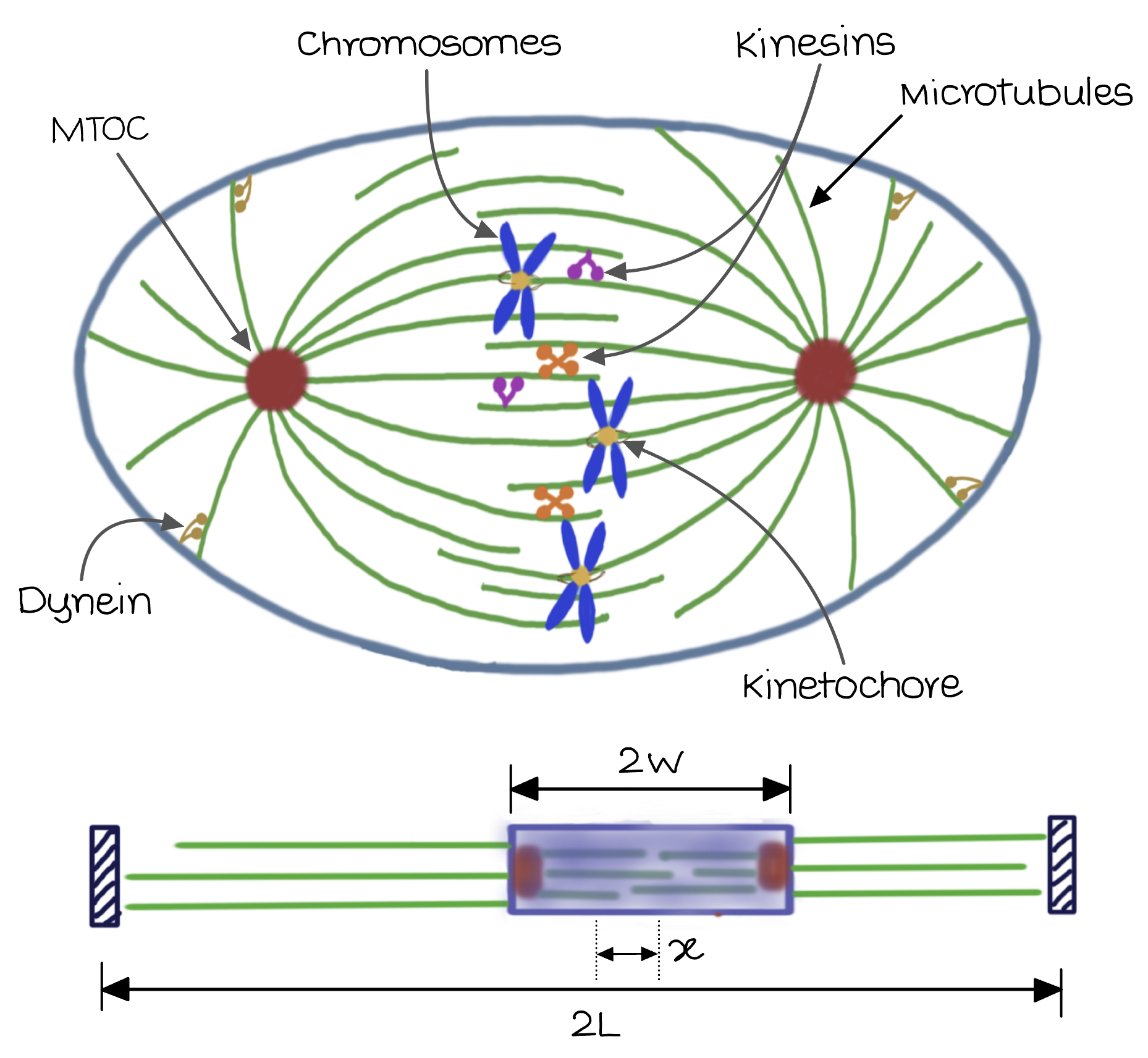}
\caption{Top: Schematic showing the typical microtubule arrangement in the mitotic spindle. These microtubules position the condensed chromosome pairs (sister chromatids) along the mid plane of the cell and subsequently help in partitioning them as cell division progresses. Polymerization and depolymerization of microtubules and motor proteins from the kinesin and dynein superfamilies generate forces within this system and at the confining boundary of the cell. Bottom: A simplified one-dimensional model  for the positioning of the spindle in which the astral microtubules are a linear bundle of polymerizing and depolymerizing filaments. Here $2w$ is the length of the spindle, $x$ the displacement of its centre from the cell midpoint (dashed lines), and $2L$ the cell length.}
\label{fig:spindle}
\end{figure}

One critical process in eukaryotic cell division is the segregation of chromosomes. The mitotic spindle consists of two microtubule organizing centres (MTOC) from each of which emanate a radial collection of microtubules, forming a bipolar structure (FIG.~\ref{fig:spindle}). Through a combination of microtubule polymerization and interaction with the actomyosin cortex, the spindle centres symmetrically within the cell~\cite{grillDistributionActiveForce2003}. This process also positions the duplicated chromosomes near the equatorial region of the cell. 

Some microtubules emanating from the MTOC attach to condensed chromosome pairs (sister chromatids) via motor proteins and specialized protein complexes known as kinetochores. Subsequent force generation in the spindle microtubules by the action of motor proteins such as kinesins, and depolymerization forces (of the order of $10~{\rm pN}$ per chromatid pair), lead to the segregation of sister chromatids to the opposite ends of the cell~\cite{nicklas1983measurements, joglekar2010mechanisms, akiyoshi2010tension}. 

In the metaphase stage of cell division, the microtubules emanating from each of the MTOCs interact with the cell cortex and also with the microtubules emerging from the partner MTOC. These interactions, which include active and passive components, determine the effective mechanical properties and stability of the spindle structure (FIG.~\ref{fig:spindle}). A highly simplified picture, in which only polymerization forces acting between the two MTOCs and between MTOCs and the boundary are considered, can provide insight into the mechanisms involved in centring the spindle. 

Consider a $1$-D toy model of the mitotic spindle with two MTOCs, shown in FIG.~\ref{fig:spindle}. The spindle has total length $2w$, the cell boundaries are at $\pm L$, and $x$ denotes the displacement of the spindle centre from the midpoint of the cell, positive to the right. Assuming that the microtubules growing towards the chromosomes are rigidly coupled, the net force on the spindle is the difference between the pushing forces exerted by the astral microtubules of the two MTOCs against the nearer boundaries, at distances $d_{\rm left} = L - w + x$ and $d_{\rm right} = L - w - x$:
\begin{align}
f_{\rm sp}(x) = f_N^{\rm left}(d_{\rm left}) - f_N^{\rm right}(d_{\rm right}),
\label{eq:spindle_force}
\end{align}
where $f_N^{\rm left, right}$ are of the form~\eqref{eq:repulsive_polymerization}, with possibly different filament numbers and dynamics at the two poles. For astral microtubules, the exponential law~\eqref{eq:bundle_force_exp} has a natural origin: in the bounded regime of dynamic instability, microtubule lengths are exponentially distributed with mean $\ell_f = v_g v_s/(v_s f_{\rm cat} - v_g f_{\rm res})$, set by the growth and shrinkage speeds $v_{g,s}$ and the catastrophe and rescue frequencies $f_{\rm cat,res}$~\cite{dogterom1993physical}. The fraction of microtubules reaching the cortex then decays as $e^{-d/\ell_f}$, while contacting microtubules can exert pushing forces of a few
pN~\cite{dogterom1997measurement}. For identical poles,
\begin{align}
f_{\rm sp}(x) = -2F_0 \, e^{-(L-w)/\ell_f} \sinh(x/\ell_f) \simeq -k_{\rm sp}\, x,
\end{align}
where $k_{\rm sp} = (2F_0/\ell_f)\, e^{-(L-w)/\ell_f}$ is the effective centring stiffness of the spindle. More generally, for identical force laws $f_{\rm sp}(x) \simeq 2x f_N'(L-w)$, so any pushing force whose magnitude decreases with the gap is restoring, and the centred position $x = 0$ is a stable stationary state; asymmetries between the poles shift this position. Experiments using microneedles and magnetic tweezers show that transverse perturbations to the spindle alignment decay within timescales $\sim 15 \, \textrm{s}$. This relaxation demonstrates a restoring response together with
dissipation, and supports the stability of spindle positioning
and orientation. The pushing picture used here is one of two competing classes of positioning mechanism. In the pushing scenario, astral microtubules polymerize against the cortex and buckle, well documented for centring the nucleus in fission yeast~\cite{tran2001mechanism}; in the competing pulling scenario, cortically anchored dynein motors pull on astral microtubules, force imbalances between the poles setting the position, as inferred from cortical force-generator counts in the \emph{C.~elegans} embryo~\cite{grillDistributionActiveForce2003}. The two are discriminated by whether severing or de-anchoring cortical motors relaxes or reverses the centring; pulling dominates in most animal cells, pushing in small cells and yeast.

Forces resulting from the filamentous structures within the spindle and their interaction with the boundary of the cell lead to its positioning. An understanding of the geometrical shape of the spindle should necessarily take into account the cohesive forces between the microtubules, mediated by crosslinkers and motors. The question of the large-scale shape of the spindle is conveniently described within a continuum approach. In this case, the mitotic spindle can be considered as a droplet of active matter \cite{oriola2020active}.  The action of molecular motors and ATP dependent microtubule dynamics are the sources of activity. 

In cell-extracts, giant-number fluctuations observed in the orientational ordering of the spindle microtubules justify an active continuum description~\cite{bruguesPhysicalBasisSpindle2014}. Further this approach allows one to estimate the material parameters of the spindle such as the strength of active stress ($\approx70~{\rm Pa}$),  effective spindle surface tension ($\approx 140~{\rm pN/\mu m}$), and the elastic constant of the orientation elasticity ($\approx 400~{\rm pN}$). 

Using this continuum approach and the experimentally determined parameters, it is shown that stable spindle shapes can emerge from a competition between bulk active stress due to activity, nematic elasticity of microtubules and surface tension on the boundary. This model can account for the observed shape transitions seen in cell extracts as a function of dynein activity~\cite{oriola2020active}. While active matter theory accounts for the spindle shapes seen in cell extracts, the situation within the cell is, of course, far more complex ~\cite{Schaeffer2024.06.11.598451}.

\subsection{Cytokinesis}
\begin{figure}[ht]
\centering
\includegraphics[width=0.9\linewidth]
{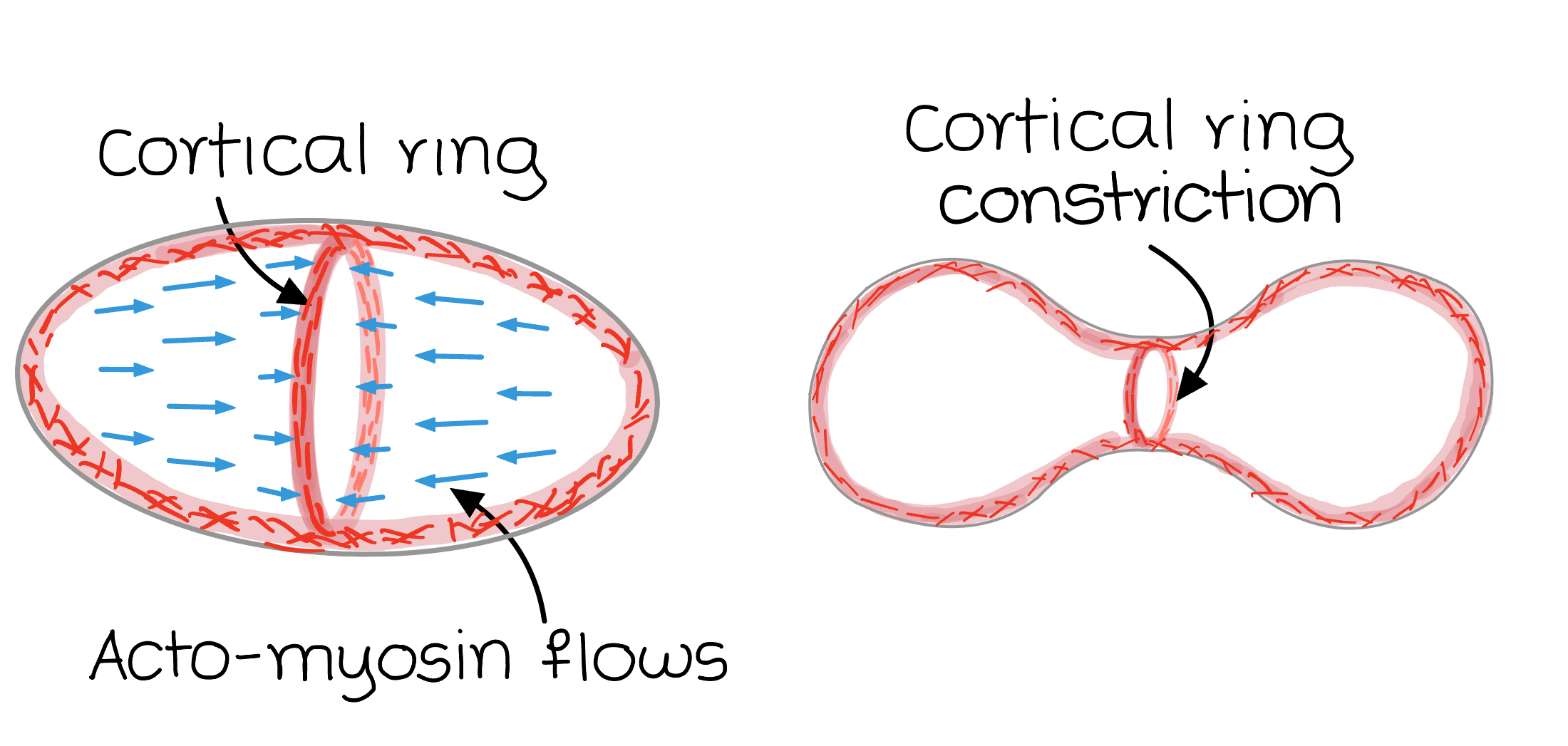}
\caption{Flow-induced alignment of actin filaments due to large-scale
cortical flows can contribute to cytokinetic-ring formation. This enhances actomyosin contractile forces at the equatorial region of the cell and leads to constriction eventually culminating in cytokinesis.}
\label{fig:cytokinesis}
\end{figure}

Eukaryotic cells typically divide by the formation and subsequent constriction of actomyosin rings~\cite{cheffings2016actomyosin} (see Fig~\ref{fig:cytokinesis}). In most cell types, polar actin filaments within the ring (of width $\approx 100~{\rm nm}$) organize in a nematic order and also undergo continuous turnover. Constriction of such rings by contractile forces appears to be a general mechanism to ``cleave'' a volume into two parts, with cytokinetic rings composed of FtsZ filaments achieving a similar effect even in bacterial cells.
 
All cells maintain an internal clock via cyclic variations in the concentration of cell-cycle associated molecules~\cite{morgan2007cell}. At suitable time-points, signals regulated by the cell-cycle enhance the contractility of the actomyosin cortex. Signals from upstream processes can trigger gradients in these active stresses which, in turn, lead to large scale cortical flows. These flows, with speeds $5-10~{\rm \mu m/min}$, lead to the emergence of spatially non-uniform actomyosin patterns. These eventually form a ring-like structure.

Theoretical studies show that such ring-like structures can emerge spontaneously from an instability of the uniform state of the actomyosin cortex~\cite{mietkeMinimalModelCellular2019}. The convergent nature of these flows compresses actin filaments and aligns them along the equator of the cell (FIG.~\ref{fig:cytokinesis}). This further enhances the contractile forces within the ring~\cite{whiteMechanismsCytokinesisAnimal1983, reymannCorticalFlowAligns2016} (Sections~\ref{sec:soft_matter_theory}~and~\ref{sec:actomyosin_cortex}).  The patterns in the actomyosin cortex and the shape of the cell can thus form a coupled dynamical system for concentration, flow, and stress fields, and geometrical descriptors of the surface. The analyses of such a dynamical system shows that cytokinesis-like patterns in both myosin concentrations and cell shape can arise spontaneously~\cite{RajputActiveGeometrodynamics}.

The nematic order parameter $\mathsf{Q}$ for the alignment of actin filaments is non-zero at the equator of the cell (FIG.~\ref{fig:cytokinesis}). Specifically, taking $y$ along the ring, the diagonal component
$Q=Q_{yy}=-Q_{xx}$ of $\mathsf{Q}$ evolves according to~\cite{salbreuxHydrodynamicsCellularCortical2009}
\begin{align}
    \partial_t Q = -v \partial_x Q - \frac{\overline{\beta}}{2} \partial_x v - \frac{1}{\tau} Q  + \frac{\ell^2}{\tau} \partial^2_x Q,\label{eq:Q_alignment}
\end{align}
where $v$ is the $x-$component of the cortical velocity (i.e., perpendicular to the actomyosin ring; see FIG.~\ref{fig:cytokinesis}), $\overline{\beta}$, $\tau$, and $\ell$ are material parameters respectively characterizing the flow alignment, the relaxation time of $Q$, and the orientational correlation length. With an experimentally measured profile $v(x)$ of the active flows, Eq.~\eqref{eq:Q_alignment} quantitatively accounts for the steady-state spatial profile of $Q$, thereby supporting the idea of flow-induced alignment of actin filaments along the cytokinetic ring, as demonstrated in {\it C. elegans} embryos~\cite{reymannCorticalFlowAligns2016}. In other kinds of cells, such as bacteria and yeast, the emergence of the cytokinetic ring is driven by different mechanisms~\cite{ghosh2008origin, cheffings2016actomyosin}.

Enhanced actomyosin levels generate a pattern of active contractile stresses. While the tangential components of these stress gradients drive in-plane cortical flows, the normal components lead to forces that can deform the cell surface forming a cleavage furrow~\cite{salbreux2012actin}.  These active forces have to overcome the forces arising from the bending elasticity of the cell membrane and external forces (from the cytoplasm and the cell exterior). In yeast cells, for instance, the circumferential tension generated within the contractile ring has been measured by micropipette aspiration to be $\approx 400~{\rm pN}$~\cite{stachowiak2014mechanism} which is sufficient to deform and constrict the cell surface. With a given spatiotemporal pattern of these contractile forces,  the cellular surface buds in a manner akin to cytokinesis~\cite{turlierFurrowConstrictionAnimal2014}. Constriction seems to proceed at a uniform rate for many cells up to the point where the membranes of the two constricting regions meet~\cite{mabuchiCleavageFurrowTiming1994, khaliullinPositivefeedbackbasedMechanismConstriction2018}. A merger of these membrane surfaces and subsequent scission event completes cytokinesis.

Although there is clear evidence for the presence of myosin in the contractile ring from yeast to mammalian cells, myosin-free contraction of rings have also been observed in some  cell types~\cite{gerisch2000cytokinesis}. Additionally, it has theoretically been argued that myosin-free contractile forces can be generated in the ring if the actin filaments undergoing turnover are connected via end-tracking crosslinkers~\cite{zumdieck2007stress}.

\section{Tissue mechanics}
\label{sec:tissues}

\begin{figure}
\centering
\includegraphics[width=0.9\linewidth]{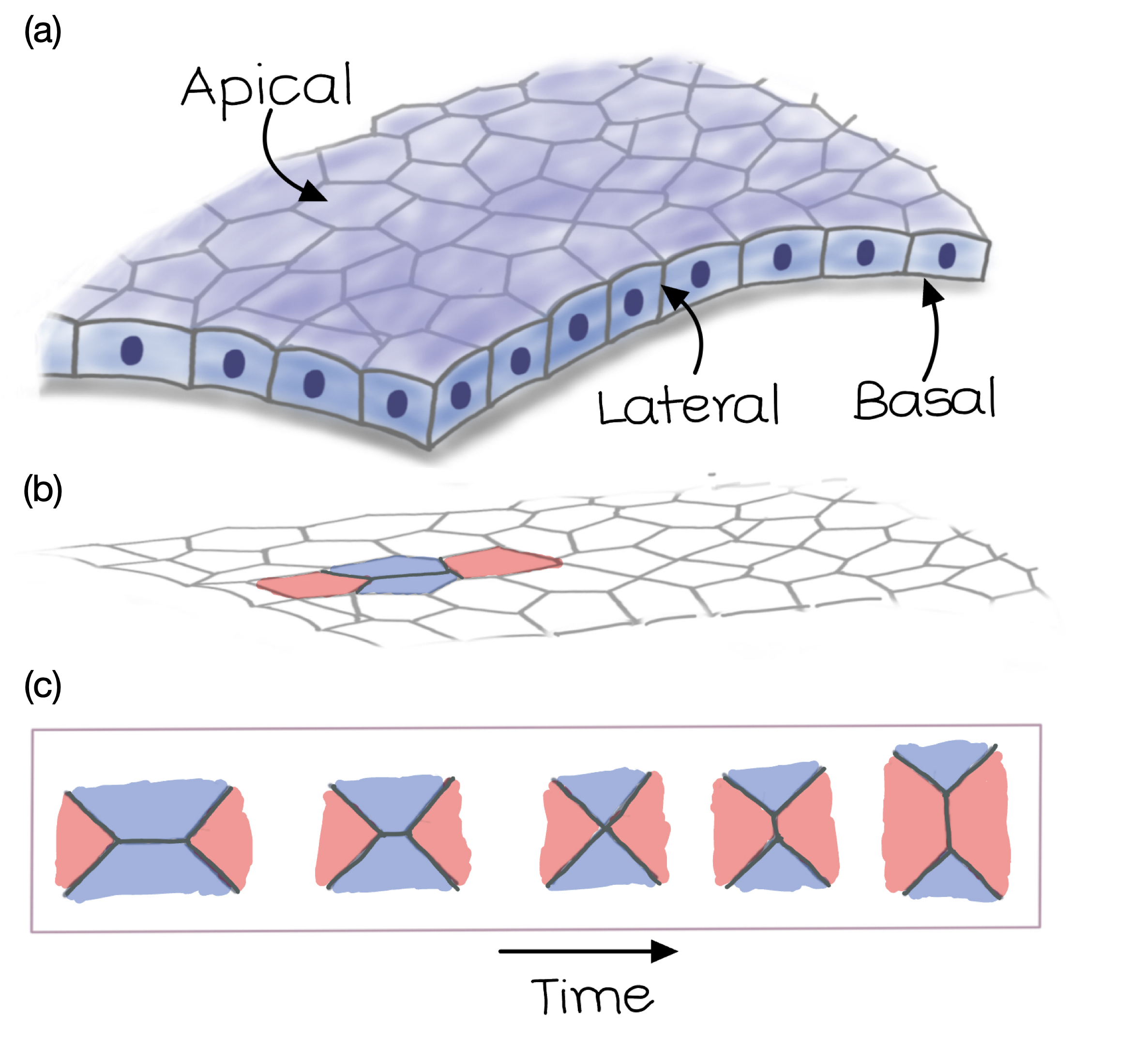}
\caption{At scales of many cells (tissues), the interactions between neighbouring cells through junctional elasticity lead to a coarse-grained description of the entire tissue. For epithelial cell layers as shown, which are quasi-two-dimensional structures, deformation of the tissues with time can lead to morphogenetic changes, such as folds and invaginations. (a) A nearly planar quasi-two-dimensional tissue. (b) Apical projection of the tissue in (a) leads to the planar vertex model. In addition to junctional elasticity, the vertex model can also incorporate local anisotropy via polarized bond tensions. (c) Active fluctuations in cell edge contractility can result in tissue fluidization  via neighbour exchanges.}
\label{fig:tissue}
\end{figure}

Tissues are dense, cohesive collections of cells together with the extracellular matrix (ECM) they secrete, organized to serve specific physiological functions. Their cells interact mechanically (through cadherin-based adhesion, junctional actomyosin, and the shared matrix) and chemically, through signalling molecules that diffuse between neighbours and modulate local force generation. A tissue is therefore best viewed not as a simple sum of its cells but as an active, strongly-correlated many-body collective, in which forces generated at the molecular scale are transmitted through cell--cell junctions and the matrix to reshape the tissue as a whole. Two features set such a collective apart from ordinary condensed matter. First, cells in a confluent tissue tile space without gaps, mechanically coupling their shapes and rearrangements. Second, cell number is not conserved because division and death act as local sources and sinks of material; cells also continually exchange neighbours, allowing the tissue to rearrange and flow while remaining structurally intact. The central problem of tissue mechanics is to bridge scales in the presence of these features: to connect molecular force generation to cell-level rearrangements, and those in turn to tissue-scale changes of form during morphogenesis, homeostasis, and repair. In much of what follows we concentrate on the junction-dominated mechanics of epithelial monolayers, returning to the extracellular matrix and to fully three-dimensional deformations later in the section.

\subsection{\label{sec:vertex_model}The vertex model and its parameters}

The most studied tissues, both experimentally and theoretically, are epithelial monolayers: confluent, single-cell-thick sheets in which cells tile space and adhere to their neighbours through cadherin-based junctions (FIG.~\ref{fig:tissue}(a)). Such cells are polarized along their apico-basal axis, and frequently acquire an in-plane, or planar cell polarity (PCP), as well. Their in-plane mechanics is governed by the interplay of the actomyosin cortex, the contractile cell--cell junctions, and, where present, adhesion to a substrate, and ranges from nearly elastic to fluid-like depending on the length and time scales probed~\cite{perez2019active, xi2019material}. A minimal but remarkably successful description of this mechanics is the \emph{vertex model}, which represents the sheet as a tiling of polygons and tracks the positions $\{\mathbf{r}_i\}$ of the cell vertices.

A configuration is assigned an energy-like function
\begin{align}
\mathcal{E} = \sum_{\alpha} \frac{K_\alpha}{2} (A_\alpha - A_{0\alpha})^2 + \sum_{\alpha} \frac{\Gamma_\alpha}{2} P_\alpha^2 + \sum_{\alpha \beta} \Lambda_{\alpha \beta} l_{\alpha \beta},\label{eq:VM_W}
\end{align}
where, for each cell $\alpha$, $A_\alpha$ and $A_{0\alpha}$ are the actual and preferred areas, $P_\alpha$ is the perimeter, and $l_{\alpha\beta}$ is the length of the junction shared with a neighbour $\beta$; the last sum runs once over each distinct junction. Each term has a direct biological meaning. The first, with area stiffness $K_\alpha$, penalizes deviations from a preferred cell area and encodes cell volume incompressibility, osmotic pressure, and the out-of-plane resistance of the cortex. The remaining two terms describe the mechanics of the cell boundary, and their physical origins are distinct. The quadratic perimeter term is an \emph{elastic} response: with modulus $\Gamma_\alpha$, it represents the resistance of the cortical and junctional actomyosin to changes in its length. The linear edge term is an \emph{effective line tension} $\Lambda_{\alpha\beta}$ on each junction, itself the net result of two competing effects: cadherin-mediated adhesion, which favours longer contacts and lowers the tension, and junctional actomyosin contractility, which favours shorter contacts and raises it. It is the combination of the two, not either alone, that fixes a cell's preferred perimeter: completing the square (below) converts the modulus $\Gamma$ and the tension $\Lambda$ into a contractile target perimeter $P_0$. Keeping them separate matters because they have different molecular origins and respond differently to perturbations of adhesion versus contractility. The mechanically stable states of the tissue are the vertex configurations that locally minimize $\mathcal{E}$.

A common way to relax towards such a state is overdamped gradient descent,
\begin{align}
\gamma \frac{\partial {\mathbf r}_i}{\partial t} = -\frac{\partial \mathcal{E}}{\partial {\mathbf r}_i},\label{eq:VM_dynamics}
\end{align}
where $\gamma$ is an effective friction between the vertices and their surroundings. Two modelling choices built into Eqs.~\eqref{eq:VM_W}--\eqref{eq:VM_dynamics} are worth making explicit, since both bear on the theme of this review. First, $\mathcal{E}$ is an energy-like function and \eqref{eq:VM_dynamics} is relaxational gradient descent towards its minima: the minimal vertex model is thus a \emph{passive}, equilibrium-like description, and active force generation is not intrinsic to it. Activity is absent from \eqref{eq:VM_W} and must be added by hand (through motile forces on cells, time-dependent or fluctuating junctional tensions, or stochastic remodelling), as we do for the active vertex model later in this section. Second, the friction $\gamma$ is with an external medium or substrate, so momentum is not conserved; \eqref{eq:VM_dynamics} describes a tissue coupled to its surroundings rather than an isolated, momentum-conserving one.

Unlike an ordinary elastic solid, a confluent tissue continually remodels its connectivity: cells exchange neighbours through T1 transitions (FIG.~\ref{fig:tissue}(c)), are lost by extrusion or death (T2), and are added by division. These rearrangements relax stress and allow the sheet to flow~\cite{fletcher2014}. Whether they can occur freely is itself set by the tissue's mechanics: the vertex model predicts a rigidity transition in which the energy barrier to a T1 rearrangement vanishes as the cells' preferred geometry changes. When adhesion dominates junctional contractility ($\Lambda<0$), the bulk energy of a homogeneous tissue ($K_\alpha=K$, $\Gamma_\alpha=\Gamma$, $\Lambda_{\alpha\beta}=\Lambda$) can be rewritten as
\begin{equation}
    \mathcal{E} = \sum_\alpha {K \over 2} (A_\alpha - A_{0})^2 + \sum_\alpha{\Gamma \over 2} (P_\alpha - P_0)^2,
\end{equation}
with a preferred perimeter $P_0=-\Lambda/(2\Gamma)$. The control parameter is the dimensionless target shape index $s_0 = P_0/\sqrt{A_0}$, set by the balance of adhesion and contractility. For $s_0 \lesssim s_0^\ast \simeq 3.81$ (approximately the shape index of a regular pentagon), disordered confluent vertex models predict a rigid solid with finite energy barriers to T1 transitions; above this threshold, the barriers vanish and the tissue fluidizes through neighbour exchanges under applied shear~\cite{staple2010mechanics, bi2015}.
Because the minimal confluent vertex model is considered at fixed mean density, density cannot serve as the control parameter as it does in particulate jamming; that role is played by cell shape through $s_0$. The resulting shape-driven unjamming is distinct in origin from density-driven jamming. Two caveats temper this clean picture. The threshold $s_0 \approx 3.81$ is a zero-temperature, ground-state result: thermal and active fluctuations round the transition and shift the effective critical shape index, so that fluidization is set jointly by cell shape and by the amplitude of active tension fluctuations~\cite{bi2016motility}. Its precise value is also model-dependent, differing between the vertex and Voronoi realizations of the same idea.

The vertex model's appeal is that it links cell shape directly to tissue mechanics through parameters with clear biological interpretations. Its limitations are equally clear. The parameters $K$, $\Gamma$ and $\Lambda$ are effective and hard to measure independently, and the mapping from molecular actomyosin and adhesion onto them is not unique: different molecular scenarios can produce the same $(\Gamma,\Lambda)$, an identifiability problem we return to below. The description assumes a confluent sheet with well-defined polygonal junctions; it omits the extracellular matrix and basement membrane, whose mechanics is folded into the single friction $\gamma$; and it is intrinsically two-dimensional. It is readily extended (with cell motility, active junctional stresses, morphogen-dependent or anisotropic bond tensions, and stochastic remodelling~\cite{noll2017active, krajnc2018, sato2015cell, okuda2018combining}), and we return to such active variants, and to fully three-dimensional descriptions, below.

\subsection{Continuum and hydrodynamic descriptions}

While the cell-level description within the vertex model is useful in providing insights into various aspects of epithelial morphogenesis, the phase space of parameters is large and difficult to constrain. In the spirit of hydrodynamic descriptions discussed in Section~\ref{sec:soft_matter_theory}, it is natural to ask for coarse-grained continuum descriptions of tissue dynamics. The kinematics of planar tissue deformation during morphogenesis separates naturally into an isotropic (trace) part and an anisotropic (traceless) part, associated with size and shape changes of the tissue respectively. The isotropic part collects changes in individual cell area together with cell division and apoptosis, all of which alter the local density. The anisotropic part collects changes in the elongation of individual cells together with the traceless contribution of oriented topological events: neighbour exchanges (T1), oriented divisions, and extrusions (T2). A given process can contribute to both: an oriented division, for instance, both adds cell area and elongates the local texture. Such contributions can be coarse-grained over space and time and reformulated in the language of continuum mechanics~\cite{ranft2010fluidization, etournay2015, guirao2015}. For example, taking $\rho$ to be the number of cells per unit area, the cell-number balance takes the following form
\begin{equation}
    {\partial \rho \over \partial t} + \nabla \cdot (\rho {\mathbf v}) = k_d(\rho) - k_a(\rho),\label{eq:mass_balance}
\end{equation}
where $k_d$ and $k_a$ are the local rates per unit area at which cells are added by division and removed by apoptosis, respectively, and these rates may depend nonlinearly on $\rho$. Their imbalance is the source term that renders cell number non-conserved, as anticipated at the start of this section. The tissue velocity ${\mathbf v}$ arises from the force balance condition, 
\begin{align}
\nabla \cdot \mathsf{\Sigma} = \gamma \mathbf{v},
\label{eq:tissue_force_balance}
\end{align}
where $\gamma$ is the friction coefficient and the stress tensor $\mathsf{\Sigma} = \tilde{\mathsf{\Sigma}} - p{\mathbb I}$. Here, the tissue pressure $p$ and the deviatoric stress $\tilde{\mathsf{\Sigma}}$ are obtained from 
\begin{align}
p &= \mu_a \ln {\rho \over \rho_0}, 
\label{eq:iso_stress}\\
\tilde{\mathsf{\Sigma}} &= 2\mu_s \mathsf{Q} + \zeta \Delta \mu \;  \mathsf{q} \label{eq:aniso_stress},
\end{align}
where $\mu_a$ and $\mu_s$ are the area compression and shear
moduli, respectively, and $\rho_0$ is the reference cell density
at which $p=0$. The anisotropies in cell shapes and those arising from chemical polarities are captured by the tensors $\mathsf{Q}$ and $\mathsf{q}$, respectively. The cell shape anisotropy evolves according to 
\begin{equation}
    \mathcal{D}_t \mathsf{Q} =  \left(\mathsf{E} - {1\over 2} \mathsf{Tr}(\mathsf{E})\: \mathbb{I}\right) - \mathsf{R}, \label{eq:DtQ}
\end{equation}
where the strain rate tensor $\mathsf{E}$ is given in \eqref{eq:shear_strain_rate} and $\mathsf{R}$ represents the contribution from the topological rearrangements of cells. The dynamics of $\mathsf{q}$ is controlled by the manner in which chemical polarities of cells interact with each other as well as with signalling cascades and tissue mechanics. Typically, $\mathsf{q}$ reaches a steady state profile relatively quickly. In such a scenario, a static spatial profile of $\mathsf{q}$ can be assumed. The shear strain rate resulting from topological transitions evolves according to
\begin{align}
(1 + \tau_R {\mathcal D}_t)\mathsf{R} = {1 \over \tau_Q}\mathsf{Q} + \lambda \mathsf{q}.\label{eq:R}
\end{align}
Here, $\tau_Q^{-1}$ sets the rate at which cell-shape anisotropy $\mathsf{Q}$ is converted into topological rearrangements (which in turn relax the shape through Eq.~\eqref{eq:DtQ}), while $\tau_R$ is the relaxation time of $\mathsf{R}$ itself. The second term, proportional to $\lambda$, captures the driving of oriented rearrangements by the active nematic polarity $\mathsf{q}$.

To summarize, \eqref{eq:mass_balance}, \eqref{eq:tissue_force_balance}, \eqref{eq:DtQ}, and \eqref{eq:R} govern the evolution of density $\rho$, cell shape $\mathsf{Q}$, and topological rearrangements $\mathsf{R}$, together with the constitutive relations \eqref{eq:iso_stress} and \eqref{eq:aniso_stress}. The model discussed above largely follows \cite{popovic2017active}; alternative descriptions~\cite{ishihara2017, grossman2022instabilities, guirao2015} for epithelial morphogenesis also exist. We will illustrate the formalism by considering a simple case. 

In the simplest scenario, the edge tension $\Lambda_{\alpha \beta}$ is homogeneous across all bonds. However, the presence of morphogen signalling fields can induce both anisotropy and heterogeneity in bond tensions and lead to tissue deformation and flows. Planar cell polarity (PCP) can itself emerge from the interplay of local cell--cell interactions and global tissue-scale gradients~\cite{singh2024emergence}. In the \textit{Drosophila} male genital epithelium, a polarized myosin-II distribution is associated with anisotropic junctional remodelling during directional tissue rotation~\cite{sato2015left}.

One way to construct an active vertex model (FIG.~\ref{fig:tissue}(b)) is to modify the bond
tensions according to~\cite{sato2015cell} 
\begin{equation}
   \Lambda_{\alpha \beta} = \Lambda_0 + \Lambda_1 \cos2(\theta_{\alpha \beta} - \theta_0) + \xi_{\alpha \beta}(t),\label{eq:aniso_bond}
\end{equation}
where $\Lambda_0$ is the base-level cell-cell bond tension which is amplified or reduced, respectively, by an amount $\Lambda_1$ depending on its orientation with respect to a preferred orientation $\theta_0$, which, in turn, is dictated by PCP. The active noise $\xi_{\alpha \beta}(t)$ is modelled as an Ornstein-Uhlenbeck process, whose amplitude and correlation time $\tau$ set the magnitude and persistence of junctional-tension fluctuations, respectively. Sufficiently large fluctuations can shrink junctions and trigger T1 transitions,   thereby fluidizing the tissue. At a coarse-grained level, the effects of the modified bond tensions in  \eqref{eq:aniso_bond} can be represented phenomenologically by the active terms proportional to $\zeta$ in \eqref{eq:aniso_stress} and $\lambda$ in \eqref{eq:R}. Thus, we effectively have an active, fluidized tissue, with anisotropic bond tensions that result in T1 transitions as well as anisotropic active stresses. 

For $\gamma=0$ in \eqref{eq:tissue_force_balance} and a spatially homogeneous PCP tensor $\mathsf{q}$ with $q_{xy} \ne 0$, a tissue confined to a rectangular geometry of size $L_x\times L_y$, with periodic boundary conditions along the $x$ direction, no-slip ($\mathbf v=\mathbf 0$) at the bottom wall ($y=0$), and zero shear stress ($\mathsf{\Sigma}_{xy}=0$) at the top wall ($y=L_y$), develops, in the linear regime and for $R_{xy}(0) = 0$, the shear rate 
\begin{equation}
    \partial_y v_x = 2\left(\lambda - {\zeta \Delta \mu\over 2 \mu_s\tau_Q} \right)q_{xy}(1 - e^{-t/\tau_R}).
\end{equation}
 This leads to a Couette-flow-like steady-state velocity profile
\begin{equation}
v_x(y) = 2y \left(\lambda - {\zeta \Delta \mu \over 2 \mu_s \tau_Q} \right)q_{xy}.
\end{equation}
Note that this velocity profile, arising spontaneously from the activity, produces a sliding (simple-shear) deformation of the tissue, accompanied by a steady-state cell elongation $Q_{xy} = -\zeta \Delta \mu/(2\mu_s)q_{xy}$ set by the balance between the active and elastic shear stresses.

\begin{figure}
\centering
\includegraphics[width=0.8\linewidth]{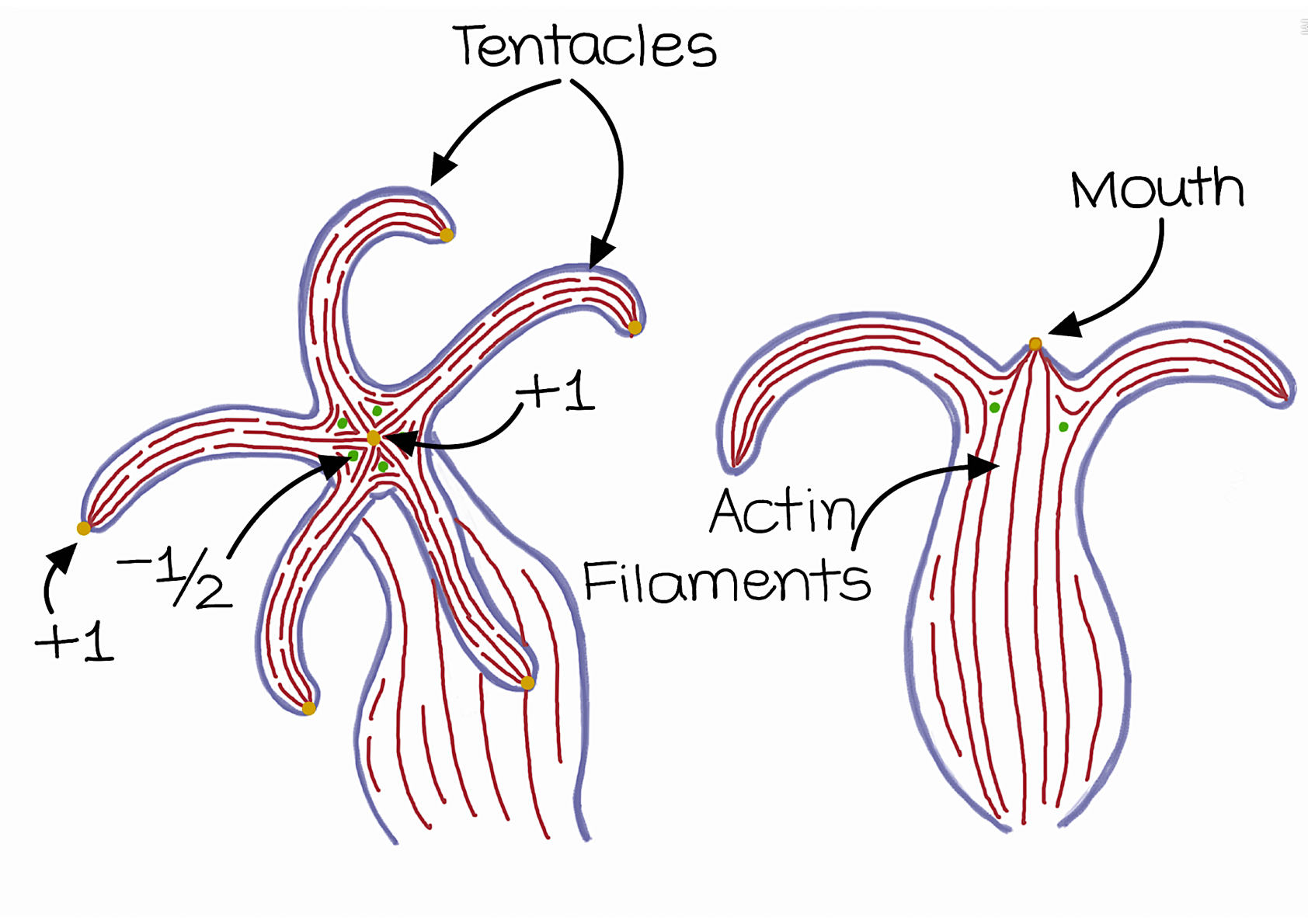}
\caption{Topological defects in an active nematic field formed by actin filaments in a developing Hydra play an important role during morphogenesis. For example, in Hydra, the generation of tentacles is associated with the formation of $+1$ defects (yellow dots) at the future mouth and at the tips of the tentacles, and a pair of $-1/2$ defects (green dots) near the base of each tentacle~\cite{maroudas-sacksTopologicalDefectsNematic2021}.
}
\label{fig:tissue_defects}
\end{figure}

A complementary, and increasingly influential, continuum viewpoint treats the tissue itself as an active nematic. Elongated cells, or the actin fibres within them, define a local axis with no head--tail distinction, so the coarse-grained orientation field carries the topological defects characteristic of nematic order. In two dimensions the relevant defects carry charge $\pm 1/2$, and their motility ($+1/2$ defects are intrinsically self-propelled, whereas $-1/2$ defects are not) makes them natural organizing centres for morphogenesis. In cultured epithelia, $+1/2$ defects have been shown to govern the sites of cell death and extrusion~\cite{saw2017}, and active-nematic behaviour emerges even in monolayers of nominally isotropic cells~\cite{mueller2019emergence}. In the developing \emph{Hydra}, integer ($+1$) defects in the actin-fibre nematic mark the sites of the future mouth and tentacle tips, with pairs of $-1/2$ defects near the tentacle bases (FIG.~\ref{fig:tissue_defects})~\cite{maroudas-sacksTopologicalDefectsNematic2021}. The location of a defect thus becomes a geometric marker of where force is focused and new form is generated, a concrete instance of the broader role of topological and geometric structure in force transmission~\cite{doostmohammadiActiveNematics2018}.

\subsection{Comparing tissue models}

A tissue has degrees of freedom that a single cell does not (cells change shape, exchange neighbours, divide, die, and couple to the extracellular matrix), and no single framework captures all of them. The choice of model is therefore dictated by the question asked and by the measurements available. Vertex models (Section~\ref{sec:vertex_model}) are the natural choice for confluent epithelial monolayers, where cells tile space and share well-defined junctions. When cells instead deform strongly, round up, separate, or migrate through their surroundings, methods that resolve the entire cell outline are more appropriate.

Cellular Potts models represent each cell as a set of sites on a lattice and evolve the configuration by Metropolis-type Monte-Carlo moves whose acceptance is biased by an adhesion-and-area energy~\cite{scianna2013cellular}; they capture differential adhesion, cell sorting, and migration for large numbers of cells, but because their dynamics is  stochastic and energy-based rather than governed by a direct force balance, they are ill-suited to problems in which the actual forces (measured tractions or internal stresses) are the point of comparison. Phase-field models instead describe each cell by a smooth field evolving under relaxational partial differential equations, into which active stresses, motility, and hydrodynamic coupling can be built directly~\cite{mueller2019emergence}; they therefore retain access to forces and stresses and allow large shape changes while penalising overlap between cells. Active-particle models go to the opposite extreme, reducing each cell to a self-propelled point or disc carrying a polarity, and are appropriate when cell motion and interactions matter more than cell shape. A distinct axis of classification is whether momentum is conserved: the vertex and continuum descriptions above are overdamped and exchange momentum with a substrate through friction (Eqs.~\eqref{eq:VM_dynamics}, \eqref{eq:tissue_force_balance}), whereas momentum-conserving hydrodynamics is appropriate only when exchange with substrates, the extracellular matrix, and surrounding media is negligible on the scales of interest. Because different models can reproduce similar flows and shapes, cell-shape statistics, junctional recoil, rearrangement rates, and traction each provide complementary constraints.

A related distinction is between kinematic and dynamic descriptions. Graner-type decompositions~\cite{etournay2015, guirao2015} express an observed tissue deformation as a sum of cell-shape changes, neighbour exchanges, divisions, and extrusions; they are powerful because they identify which cellular events produced the deformation, but they do not by themselves supply the forces or constitutive laws responsible. Active continuum descriptions instead start from fields (velocity, stress, density, cell-shape anisotropy, and polarity) and are suited to asking how active stress, friction, and rheology generate flow. Reconciling the two, by inferring active stresses and material parameters from measured cell-scale events, remains a central open problem.

Whether a tissue responds as a solid, a fluid, or something in between depends on the timescale and the perturbation: elastic stress stored over short times can relax over longer times through T1 transitions, division, apoptosis, or active junctional fluctuations. The language of jamming and unjamming usefully connects tissue rigidity to cell shape, density, adhesion, and activity (Section~\ref{sec:vertex_model}), but it does not by itself fix the molecular origin of that activity or adhesion, which must be tied to measurable quantities such as myosin fluctuations, cadherin dynamics, and cell turnover. Anisotropy is equally central: isotropic mechanics governs area, density, and pressure, whereas anisotropic mechanics governs shear, elongation, convergent extension, and oriented rearrangements. One and the same tissue-scale deformation can arise from planar cell polarity, oriented actomyosin cables, oriented divisions or T1 transitions, external stretch, or aligned matrix.

Finally, planar models are controlled reductions. When curvature is itself central to the phenomenon (as in the buckling that produces intestinal villi and crypts~\cite{hannezo2011instabilities}, or the differential apical--basal contractility that shapes pancreatic ducts~\cite{messal2019tissue}), the out-of-plane degrees of freedom captured by three-dimensional vertex or surface models become essential~\cite{hannezo2014}, and curvature, lateral adhesion, basement-membrane mechanics, and external confinement enter as new variables. Across all of these choices runs one difficulty: because many mechanisms can reproduce the same final morphology, shape alone rarely discriminates between them. Perturbative and time-resolved experiments are therefore decisive: laser ablation for junctional tension, traction-force microscopy for substrate coupling, embedded droplets for internal stress, and myosin, cadherin, or matrix perturbations to separate internally driven from externally imposed deformation. The frontier is to pair such measurements with models that make genuinely different, testable predictions. 

\section{\label{sec:other_examples}Forces in other systems: viral packaging, swimming, and plants}

The cytoskeletal polymerisation, molecular motors, and actomyosin cortex emphasised so far do not exhaust the settings in which force generation organises living matter. Three further domains are collected below, each re-expressing a principle already developed in this review in a different physical regime: the non-equilibrium ratchets and stall forces of Sections~\ref{sec:poly_forces} and~\ref{sec:molecular_motors}, the low-Reynolds-number hydrodynamics of Section~\ref{sec:soft_matter_theory}, and an osmotic or turgor pressure balanced against a deformable boundary. Force \emph{sensing} at the molecular level (through mechanosensitive ion channels and adhesion molecules such as talins, vinculin, integrins and cadherins) has been discussed in Section~\ref{sec:infer_forces}; the emphasis here returns to force generation. The subsections that follow are illustrative rather than exhaustive.

\subsection{Virus assembly and genome packaging}

Mechanical forces play a central role in the assembly, compaction, and delivery of the genome in viruses. In many double-stranded DNA bacteriophages, the genome is driven into a preformed capsid by an ATP-consuming portal motor situated at a unique vertex~\cite{rickgauer2008portal}. This motor is among the most powerful known: single-molecule optical-tweezer measurements on bacteriophage $\phi29$ show that it packages DNA against internal loads of up to $\approx 57$~pN~\cite{smith2001bacteriophage}, two orders of magnitude larger than the $\approx 1$~pN generated by a single polymerising actin filament (Section~\ref{sec:poly_forces}). The resisting force builds up because double-stranded DNA is a stiff, charged polymer forced far below its $\approx 50$~nm persistence length: the work of packaging is set by a competition between bending elasticity (the worm-like-chain energetics of Section~\ref{sec:soft_matter_theory}) and the electrostatic self-repulsion of neighbouring strands, screened over the nanometre Debye length discussed in Section~\ref{sec:force_scales}.

At full packing the confined genome exerts an internal pressure of tens of atmospheres on the capsid, which must be balanced by tension in the protein shell; atomic-force-microscopy indentation places the forces needed to deform such capsids in the range of several hundred piconewtons. On infection, this stored pressure helps to eject the DNA into the host cell, the osmotic-pressure difference driving ejection being of order $10$--$100$ atmospheres~\cite{brandariz2019pressure}. Whether the packaged genome is best described as an ordered inverse spool or a more disordered condensate remains model-dependent; single-molecule force measurements together with osmotic-suppression experiments are the observations that discriminate between these pictures~\cite{zandiVirusGrowthForm2020, bruinsmaPhysicsViralDynamics2021}.

\subsection{Swimming}

Swimming is the whole-cell manifestation of the molecular force generators of Section~\ref{sec:molecular_motors}. Bacteria such as \textit{E.~coli} are propelled by rotary motors (driven by a transmembrane ion-motive force) that turn helical flagella at torques of order $10^{3}$~pN\,nm; bundling of the flagella produces smooth ``runs'' at $\approx 20$--$30~\mu$m/s, and reversal of motor rotation unbundles them, so that the cell tumbles and reorients~\cite{berg2004coli}. Eukaryotic swimmers such as \textit{C.~reinhardtii} instead beat two flagella whose internal force is generated by axonemal dynein (Section~\ref{sec:molecular_motors}).

These motions occur at the low Reynolds numbers of Section~\ref{sec:soft_matter_theory} ($Re \lesssim 10^{-4}$), where inertia is negligible and Stokes flow is time-reversible. A purely reciprocal shape change then produces no net translation (Purcell's scallop theorem), so swimming requires a non-reciprocal, time-irreversible stroke~\cite{purcell1977life}. The far field of a force-free swimmer is that of a force dipole, decaying as $1/r^{2}$ and distinguishing ``pushers'' such as \textit{E.~coli} from ``pullers'' such as \textit{C.~reinhardtii}. Hydrodynamic interactions among many such swimmers generate collective flows and large-scale ``bacterial turbulence'', connecting this problem to the active-matter descriptions that recur throughout this review~\cite{lauga2009hydrodynamics, LaugaBook}.

\subsection{Plants}

Plant cells generate and withstand forces in a regime set apart from animal cells by the stiff cell wall and by turgor: the hydrostatic pressure of the vacuole against the wall is typically $0.3$--$1$~MPa~\cite{niklasPlants, dumaisVegetableDynamicksRole2012}. Because turgor is isotropic, it cannot by itself set the direction of growth; growth acquires direction because the wall yields \emph{anisotropically}, its cellulose microfibrils (laid down under the guidance of cortical microtubules) making it more extensible in one direction than another~\cite{coenMechanicsPlantMorphogenesis2023}. Irreversible expansion sets in only once turgor exceeds a wall-yield threshold~\cite{lockhart1965analysis}, a yield-stress rheology of the kind met in Section~\ref{sec:soft_matter_theory}. Spatial gradients in this growth (\emph{differential growth}) bend and curve an initially straight organ, and are the mechanical basis of tropisms (the bending of shoots and roots toward light or along gravity) and, more broadly, of plant morphogenesis~\cite{moultonMultiscaleIntegrationEnvironmental2020}. Plants also possess their own mechanosensors: stretch-activated ion channels transduce wall and membrane tension into physiological responses, the plant counterpart of the sensing machinery of Section~\ref{sec:infer_forces}~\cite{basu2017plant}.

At the cellular level, many plant cells display cytoplasmic streaming, the bulk circulation of the cytoplasm generated as myosin~XI motors haul organelles and endoplasmic reticulum along actin bundles~\cite{woodhouseCytoplasmicStreamingPlant2013, tominaga2015molecular}. In the giant internodal cells of \textit{Chara} the streaming velocity reaches $\sim 50$--$100~\mu$m/s, among the fastest active intracellular flows known, and is thought to enhance the transport and mixing of nutrients relative to diffusion.

\section{\label{sec:conclusions}Conclusions} 

This review argues that nanoscale forces, originating in local non-equilibrium processes, are central to cellular form and biological function. The settings in which they arise are familiar objects of study in soft condensed matter physics. These ideas provide a natural background, when coupled to existing tools of statistical physics, to the study of a broad class of questions in mechanobiology.

Describing such systems is intrinsically harder than their equilibrium counterparts: with no free energy to minimise, they must be treated at the level of equations of motion and, because the aqueous medium transmits momentum, of coarse-grained hydrodynamics, for which no general far-from-equilibrium theory yet exists, though symmetry-based arguments are a reasonable starting point. What such active systems can do is nonetheless already striking: instabilities of ordered states, giant fluctuations that violate central-limit expectations, and self-tuning to the vicinity of dynamical phase transitions~\cite{menon2010activematter, marchettiHydrodynamicsSoftActive2013}. How these signatures are put to use in biological function remains largely open.

The most distinctive thread running through the review 
is treating mechanics as a distinct signalling channel, complementary to biochemical cascades but with its own capacity, energetic cost, latency, and spatial range: biochemical signals are scalar, diffusion-limited, and local, whereas forces are tensorial, propagate rapidly through elastic and hydrodynamic couplings, and act at a distance. A scalar chemical potential difference $\Delta\mu$ is converted into anisotropic stress only through an extended, oriented structure: the nematic active stress $\overline{\zeta}\,\Delta\mu\,\mathsf{Q}$ of Eq.~\eqref{eq:Q_active_stress}, realized for instance by the flow-aligned actomyosin of the cytokinetic ring (FIG.~\ref{fig:cytokinesis}), is one example, with the cytoskeleton, adhesions, or membrane supplying the tensorial degrees of freedom. This coupling is quantifiable: thermodynamic uncertainty relations bound the precision of force output by the associated dissipation~\cite{barato2015thermodynamic}, and cost-of-sensing arguments bound how accurately such structures read mechanical inputs~\cite{bialek2005physical, govern2014optimal}. This perspective is testable~\cite{parrondo2015thermodynamics}: disrupting the mechanical link should measurably degrade downstream transcriptional readouts, connecting phenomenological activity parameters to molecular information rates.

Force generation also has an evolutionary dimension. The extraction of genetic information by machines such as RNA polymerase was already a mechanochemical, force-requiring process in the earliest cells, so mechanisms transducing chemical energy into information-controlled force were plausibly favoured early by selection. The later advent of large-scale, motor-driven force generation on the cytoskeleton lifted the diffusion limit on cell size and internal organisation, enabling directed cargo transport, the building of organelles such as the nucleus, spindle, and Golgi, and ultimately the collective cell dynamics underlying multicellularity~\cite{zaidel2015contractome, thattai2019contraction, brunet2023cell}. Innovations in force generation could in turn feed back on higher-level control of gene expression, polarity, signalling, and development~\cite{Engler2004, uhlerRegulationGenomeOrganization2017}.

Understanding how biological, biochemical, and biophysical descriptions converge (to describe, and perhaps one day recreate, the functions of a living cell) remains an open and challenging goal, still framed by the question Schr\"{o}dinger posed in \emph{What is Life?}~\cite{schrodinger1948life}. A concrete and demanding instance of this goal is the bottom-up construction of a minimal synthetic cell that can grow, divide, and sustain itself. Reconstituting even the division machinery alone requires orchestrating symmetry breaking, membrane constriction, and abscission (precisely the force-generating and force-transmitting processes discussed throughout this review), making it a stringent test of whether our physical understanding of cell-scale forces is complete~\cite{dekker2026divisome}. What this review adds is the claim that the mechanics of the cell cannot be cleanly separated from the biochemical processing of information: their coupling is a potentially important, and still underexplored, organising principle for cellular form.

\section{Glossary}

\begin{description}[font=\normalfont\small\itshape, leftmargin=0cm, labelindent=0cm, before=\normalfont\small]

\item[Active process] A physico-chemical process held out of equilibrium through the throughput of energy at the microscopic scale, typically via chemical reactions involving energy storing molecules such as ATP.

\item[Actomyosin cortex] Term applied to the dense layer of actin filaments, myosin motors and actin cross-linkers, along with other regulatory molecules, that lie just below the cell membrane.

\item[ATP] Adenosine triphosphate, the principal chemical energy currency of the cell; hydrolysis of one of its phosphate bonds releases about $20~k_BT$ and powers most active processes.

\item[Axoneme] The conserved microtubule-based core of a cilium or flagellum, typically a $9+2$ arrangement of doublets driven by dynein motors, forming an active elastic bundle that beats.

\item[Brownian ratchet] A model that has functioned as a paradigm for how processes involving non-equilibrium energy flow can produce biased motion, allowing forces to be exerted.

\item[Canham--Helfrich energy] The bending energy of a fluid membrane, quadratic in the mean curvature with a bending rigidity $\kappa$, supplemented by surface-tension and Gaussian-curvature terms.

\item[Catch and slip bonds] Molecular bonds whose lifetime respectively increases (catch) or decreases (slip) with applied force; the distinction sets whether an adhesion reinforces or releases under load.

\item[Chromosome] The term applied to DNA in its cellular context, where it is found bound to many other proteins, including histones, chromatin remodellers and molecular motors such as RNA polymerase.

\item[Comoving corotating derivative] accounts for the kinematical changes in a tensor field as seen in a frame being advected and rotated by a flow field.

\item[Constitutive relation] The relation between stresses and strains (or strain rates) that defines the mechanical response of a material that is being deformed.

\item[Curie principle] The symmetry principle that, in an isotropic medium, a scalar cause cannot by itself produce a vectorial effect; directed force generation therefore requires a broken symmetry, such as the polarity of a filament.

\item[Cytokinesis] The cell-scale sequence of processes through which a cell cleaves into two daughter cells.

\item[Cytoskeleton] Term applied to a specific set of polymers (actin, microtubules and intermediate filaments) in the cell, which underlie cell shape, architecture and motility while also provide a scaffolding for several classes of molecular motors.

\item[Dalton] Unit of molecular mass (Da), one-twelfth the mass of a carbon-12 atom; a single amino acid has a mass of about $100~{\rm Da}$.

\item[Debye length] The length scale over which the electric field of a charge is screened by mobile ions in an electrolyte; nanometric in the salty cytoplasm and micron-scale in pure water.

\item[Depletion interaction] An effective, entropy-driven attraction between large particles suspended in a bath of smaller ones: clustering the large particles frees configurational volume for the small ones and raises the total entropy.

\item[Detailed balance] The relationship between forward and backward rates of statistical processes that must exist for macroscopic thermodynamic equilibrium states to result from the dynamics.

\item[Epithelial tissue] The layer of cells that form the interface of organs and are exposed directly to fluid flow as well as external perturbations.

\item[Euchromatin and heterochromatin] The loosely packed, gene-rich, actively transcribed state of chromatin (euchromatin) versus its densely packed, gene-poor, transcriptionally quiescent state (heterochromatin).

\item[Eukaryotic cell] A cell that encloses its DNA within a membrane-bound nucleus and partitions its interior into membrane-bound organelles; the cells of animals, plants, fungi and protists.

\item[Fluctuation-Dissipation theorem] A fundamental relationship between noise strength and damping coefficients that ensures that thermodynamic equilibrium is attained.

\item[Focal adhesion] An integrin-based multiprotein assembly that mechanically couples the extracellular matrix to the actin cytoskeleton, both transmitting and sensing force.

\item[Fokker--Planck equation] The partial differential equation governing the time evolution of the probability density of a continuous stochastic variable; the continuum limit of the master equation and the counterpart of the Langevin description.

\item[Hooke's law] The linear constitutive relation for an isotropic elastic solid: stress is proportional to strain through the shear and bulk moduli.

\item[Hydrodynamic approach] The procedure of replacing a microscopic description of a large number of momenta and positions with a coarse-grained description where quantities of interest (the density and momentum density) are instead represented by their coarse-grained averages, and a local equation for their evolution, at each point in space and time.

\item[Hydrodynamic stress] Force per unit area acting across surfaces, that arise as a consequence of fluid flow, distortions of an order parameter and gradients in other related quantities.

\item[Ion-channels and ion-pumps] Proteins that span the plasma membrane (transmembrane), selectively transferring chemical species from one side to the other, helping maintain potential differences across the cell surface.

\item[Langevin equation] A stochastic equation of motion for a particle, balancing a systematic force against a velocity-dependent drag and a random thermal noise; at equilibrium the noise and drag are tied by the fluctuation-dissipation theorem.

\item[Linear response theory]  The approximation that changes in
observables are linear in sufficiently weak applied perturbations, and 
near equilibrium it underlies linear force-flux relations.

\item[Markov process] Stochastic process connecting allowed states in which the transition probabilities (or rates) depend only on the current state, not on prior history.

\item[Maxwell / Kelvin--Voigt model] The two simplest linear viscoelastic models: a spring and dashpot in series (Maxwell, a fluid at long times) or in parallel (Kelvin--Voigt, a solid at long times), interpolating between elastic and viscous response.

\item[Mechanochemical coupling] The use of chemical reactions to drive mechanical processes and vice-versa, typically shape changes in molecules. 

\item[Mechanosensing] The detection by a cell of applied forces, or of the mechanical properties of its environment, through force-dependent molecular conformations and reaction rates.

\item[Mechanotransduction] The conversion of a mechanical stimulus into a biochemical or transcriptional response.

\item[Metabolism] The term applied to biochemical reactions that convert nutrients into high-energy molecules such as ATP.

\item[Mitotic spindle] An ordered, characteristically shaped structure formed from microtubules that separates the duplicated chromosomes.

\item[Molecular motor] A molecular assembly that consumes energy in the form of ATP molecules, changes conformation, and uses this change of conformation to perform mechanical work.

\item[Navier--Stokes equation] The momentum-balance equation for a Newtonian fluid; at the low Reynolds numbers of cell-scale flows its inertial (nonlinear) terms are negligible and it reduces to the linear Stokes equation.

\item[Nematic order] An axially aligned state formed in systems of fore-aft symmetric anisotropic molecules.

\item[Nonequilibrium steady state] Time-independent macroscopic state of a system which is not governed by the thermodynamic equilibrium distribution, and has non-zero fluxes (of energy or matter) and constant entropy production.

\item[Organelle] A membrane-bound compartment within a eukaryotic cell (e.g., nucleus, mitochondrion, endoplasmic reticulum, Golgi apparatus) that isolates a specialized chemical environment for particular reactions.

\item[Osmotic pressure] The pressure that must be applied to prevent the net flow of solvent across a semipermeable membrane separating solutions of differing solute concentration.

\item[P\'eclet number] A dimensionless number quantifying the balance of directed and diffusive motion.

\item[Persistence length] The length scale over which the tangent vectors along a polymer remain correlated, quantifying filament stiffness (of order $50$~nm for DNA, $10~\mu$m for actin and $1$~mm for microtubules).

\item[Polar order] A locally aligned state formed in systems of fore-aft asymmetric anisotropic molecules.

\item[Polymerization force] Forces exerted by molecules that are allowed to polymerize.

\item[Prokaryotic cell] A cell, such as a bacterium, that lacks a membrane-bound nucleus and internal membrane compartments; evolutionarily more ancient than eukaryotes.

\item[Reynolds number] A dimensionless quantity representing the ratio of inertial forces to viscous forces, which effectively measures the balance of advection and dissipation in a fluid.

\item[Rheology] The study of flow and deformation of materials.

\item[Sarcomere] The basic contractile unit of striated muscle, an array of antiparallel actin filaments and myosin thick filaments bounded by Z-discs.

\item[Self-organized patterns] Spatiotemporal structures that emerge from the dynamics in non-equilibrium states.

\item[Stall force] The opposing load at which a force-generating process (a molecular motor or a polymerizing filament) ceases to advance, its velocity vanishing.

\item[Thermodynamic equilibrium] Macroscopic states with vanishing fluxes (of energy or matter) and no entropy production, wherein the probability of microscopic configurations is governed by the Boltzmann distribution.

\item[Tissue] A collection of cells, and their extracellular matrix, with a specific functionality.

\item[Topological defect] A point or line in an orientationally ordered medium where the order parameter is undefined, characterized by a conserved topological charge (winding number); in active systems some defects self-propel.

\item[Traction force] Force per unit area applied across a surface.

\item[Transcription] The biochemical process by which DNA sequence is read out into an RNA molecule.

\item[Transcription factors] Proteins that regulate gene expression by binding to specific DNA sequences, controlling the rate of transcription.

\item[Translation] The biochemical process by which an RNA molecule is converted into a protein.

\item[Treadmilling] The steady state of a polar filament in which monomers add at one end at the same rate as they are lost from the other, so the filament translocates at fixed length.

\item[Turgor pressure] The hydrostatic pressure exerted by the fluid contents of a walled cell, notably a plant cell, against its cell wall.

\item[Vertex model] A, typically two-dimensional, geometrical model for an epithelial tissue in which changes in cell shapes and sizes are accounted by forces at cell-cell junctions, while incorporating dynamical processes that represent their cellular counterparts.

\item[Viscoelastic fluids] Materials with both viscous and elastic responses whose shear modulus relaxes to zero at long times, so that they flow under sustained stress.

\item[Viscoelastic solids] Materials with both viscous and elastic responses that retain a nonzero shear modulus at long times.

\item[Worm-like chain] A model of a semiflexible polymer as an inextensible elastic curve with a bending energy, characterized by a single persistence length; the standard description of DNA and cytoskeletal filaments.

\end{description}

\bigskip
\begin{acknowledgments}
K.V.K. acknowledges support from the Department of Atomic Energy, Government of India, under Project No.~RTI4001, from the Department of Biotechnology, Government of India, through a Ramalingaswami re-entry fellowship, from the Max Planck Society through a Max-Planck-Partner-Group at ICTS-TIFR, and from the John Templeton Foundation. M.M.I. acknowledges financial support from CEFIPRA-CSRP 6704-5.  P.A.P. acknowledges support through The Wellcome Trust DBT India Alliance (grant~IA/TSG/20/1/600137). G.I.M acknowledges current support from the Centre for Computational Biology and Bioinformatics at Ashoka University supported by the DBT (grant BT/PR40220/BTIS/137/22/2021), as well as from the AXIS Bank. G.I.M is grateful to TIFR, Mumbai for an Adjunct Professorship awarded in the initial stages of this work. 

We thank G. Kumaraswamy, A. Rajput, A. Padmanabhan, R. Parthasarathy, V. Nanjundiah, and S. Ramaswamy for a careful reading of the manuscript and useful feedback.  We especially thank the anonymous referees whose detailed comments significantly improved the manuscript. 

The initial submission was prepared without any AI assistance. All figures were hand-drawn by the authors. In the resubmission, we used Claude (Anthropic) to help streamline the manuscript, and locate relevant literature. All AI-assisted content was critically reviewed, verified, and edited by the authors, who take full responsibility for the accuracy, integrity, and originality of the work.

\end{acknowledgments}

\bibliography{references}

\end{document}